%% file: table-slab-forces.tex
\documentclass[10pt]{iopart}

\usepackage{iopams,amsfonts,amssymb,amsthm} 
\usepackage{latexsym}
\usepackage{rawfonts}
\usepackage{multirow}
\usepackage{rotating}
\usepackage{lscape}
\usepackage{graphicx}\graphicspath{{figs/}}
\usepackage{subfigure}
\usepackage{xcolor}
\usepackage{fancybox}

\input{prepictex}
\input{pictex}
\input{postpictex}

\definecolor{Maroon}{rgb}{0.70,0.0,0.0}
\definecolor{Maroon1}{rgb}{0.40,0.0,0.0}
\definecolor{Brown}{rgb}{0.7,0.3,0}
\definecolor{Navy}{rgb}{0.3,0.0,0.4}
\definecolor{Green}{cmyk}{1,0,1,0.2}
\definecolor{Red}{cmyk}{0,1,1,0}
\definecolor{DarkRed}{cmyk}{0,1,1,0.6}
\definecolor{DarkBlue}{cmyk}{1,1,0,0.5}
\definecolor{DarkGreen}{cmyk}{1,0,1,0.65}
\definecolor{OrangeRed}{cmyk}{0,1,0.87,0}
\definecolor{RedOrange}{cmyk}{0,0.77,0.87,0}
\definecolor{Orange}{cmyk}{0,0.61,0.87,0}
\definecolor{Offwhite}{cmyk}{.07,.15,.15,0}
\definecolor{Offwhite2}{cmyk}{.04,.02,.03,0}

\newcommand{\q}{\quad}

\newcommand{\Ref}[1]{(\ref{#1})}

\newcommand{\RealN}{{\mathbb{R}}}
\newcommand{\IntN}{{\mathbb{Z}}}

\newcommand{\Slab}{{\mathbb{S}}}

\def\L{\left(}
\def\R{\right)}

\def\LC{\left\{}
\def\RC{\right\}}
\def\LA{\left\langle}
\def\RA{\right\rangle}

\def\vert{{\,\hbox{\large$|$}\,}}
\def\Vert{\hbox{\large$|$}}

\def\U#1{\underline{#1}}

\def\C#1{{\cal #1}}

\def\v#1{{v_#1}}

\def\sfrac#1#2{\hbox{\normalsize $\frac{#1}{#2}$}}

\def\fns{\scriptsize}

\begin{document}
\title[Squeezed Lattice Knots]{Lattice Knots in a Slab}
\author{
D. Gasumova$\dagger$, 
E. J. Janse van Rensburg$\dagger$\footnote[3]{To whom 
correspondence should be addressed (\texttt{rensburg@yorku.ca)}}
and A. Rechnitzer$\ddagger$}

\address{$\dagger$Department of Mathematics and Statistics, 
York University\\ Toronto, Ontario M3J~1P3, Canada\\
\texttt{rensburg@yorku.ca}}

\address{$\ddagger$Department of Mathematics, 
The University of British Columbia\\
Vancouver V6T~1Z2, British Columbia , Canada\\
\texttt{andrewr@math.ubc.ca}}

\begin{abstract}
In this paper the number and lengths of minimal length lattice knots 
confined to slabs of width $L$, is determined.  Our data on minimal
length verify the results by Ishihara et. al. \cite{ISDAVS12} for the similar 
problem, expect in a single case, where an improvement is found.  From our
data we construct two models of grafted knotted ring polymers squeezed 
between hard walls, or by an external force.  In each model,
we determine the entropic forces arising when the
lattice polygon is squeezed by externally applied forces.  The
profile of forces and compressibility of several knot types are
presented and compared, and in addition, the total work done on the
lattice knots when it is squeezed to a minimal state is determined. 
\end{abstract}
\maketitle

\section{Introduction}

Chemically identical ring polymers may be knotted and these are
examples of topological isomers which may have chemical and physical
properties determined by their topology.  There has been a sustained interest 
in the effects of knotting and entanglement in polymer physics and chemistry,
and it is known that entanglements may play an important role in the 
chemistry and biological function of DNA \cite{ZKC97}.  For example,
entanglement and knotting are active aspects of the functioning
of DNA and are mediated by topoisomerases \cite{KSH07,LMZC06}, 
while proteins are apparently rarely knotted in their natural active 
state \cite{T00}.

Ring polymers with specified knot type have been chemically synthesized
\cite{DBS89}, but more often, random knotting of ring polymers occurs in
ring closure reactions \cite{SW93,BOS07,MW86}.  In this case, 
a spectrum of knot types are encountered \cite{VKK05,DSB01,SD02}, and 
these are a function of the length of the polymer:  Numerical studies 
show that longer ring polymers are knotted with higher frequency 
and complexity \cite{JvRW90}.

Ring polymers adsorbing in a plane or compressed in a slab also appear 
to have increasing knot probability \cite{MHDKK02,OSV04}, although the 
probability may decrease in very narrow slabs \cite{TJvROW94}.  Similar effects
are seen when a force squeezes a ring polymer in 
a slab \cite{JvROTW08}, and the results of the calculation in 
reference \cite{JvR07} suggest that knotted polygons will exert higher
entropic forces on the walls of a confining slit. More generally,
the phase behaviour of lattice ring polymers confined to slabs
and subjected to external forces were examined in reference \cite{SBA12}.

The entropic force of a knotted ring polymer confined to a
slab between two plates were examined using a bead-spring model
in reference \cite{MLY11}.  In this study it was found that more
complex knot-types in a ring polymer exert higher forces on the
confining walls of the slab (if the slab is narrow).

In this paper we obtain qualitative results on the entropic 
properties of tightly knotted polymers confined to a slab or squeezed 
by a flexible membrane, using minimal length cubic lattice knots.
We will consider two different models.

\begin{figure}[t!]
\centering
 \includegraphics[scale = 0.45] {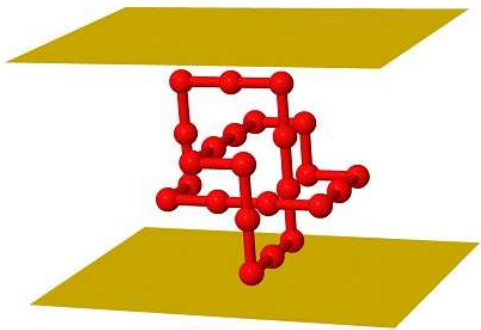}%
\caption{A minimal length lattice knot of type $3_1$ squeezed between
two hard walls. The lattice knot is grafted to the bottom wall -- that
is, it has at least one vertex in this wall. If the two hard walls 
are a distance $L$ apart, then the height of the lattice knot is
$h \leq L$, since the polygon must fit between the two walls.
The number of these lattice knots of length $n$ are denoted $p_n^L (K)$.}
\label{FIG1} 
\end{figure}

The first is a model of a tightly knotted ring polymer of fixed 
length squeezed between two hard walls or plates (see figure \ref{FIG1}).
Self-avoidance introduces steric repulsions between monomers 
which causes (self)-entanglement of the polymer.  Confining
the polymer to a slab results in the loss of configurational
entropy inducing a repulsive force which depends
on the entanglements between the walls of the slab.  
This entropically induced repulsion will be overcome by an externally applied 
force at critical magnitudes, and we shall determine these
critical forces for several knot types in our model.  An external force
will tend to squeeze the walls of the slab together, and at some
critical widths of the slab, the polymer cannot shrink further without 
expanding laterally and increasing its length.  Beyond this critical width
there may also be an elastic energy contribution to the free
energy -- the polymer stretches in length to accommodate the narrow
slab.  We model this with a Hooke energy.

The second model is inspired by the study in reference \cite{GWF06}.
In figure 1 therein, a polymer is grafted to a hard wall and covered
by a soft flexible membrane.  The membrane may be modeled by a hard wall 
as in the first model above.  On the other hand, a pressure difference
across the membrane will induce forces on the monomers in the top
layer of the grafted polymer.  This couples the force to the
height of the monomers in the top layer -- the force is mediated
directed through the membrane and push on the highest monomers of 
the polymer.

\begin{figure}[b!]
\centering
 \includegraphics[scale = 0.55] {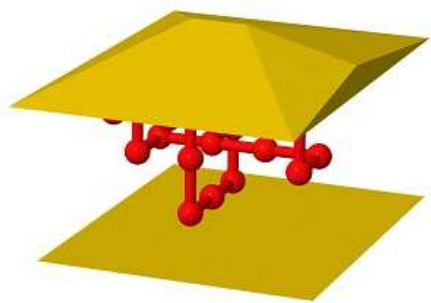}%
\caption{A minimal length lattice knot of type $3_1$ grafted to
the bottom wall and covered by a weightless flexible membrane.
The lattice knot is grafted to the bottom wall (which is hard)
by having at least one vertex in this wall. 
A pressure gradient over the top membrane will exert a force
compressing the knot onto the bottom wall.  The force is mediated
through the membrane onto the monomers in the top layer of the
lattice knot.  If the height of the lattice knot is $h$, then the
number of vertices at height $h$ is conjugate to a force pushing the
lattice knot towards the bottom wall.
The number of these lattice knots of length $n$ is denoted by 
$p_n (K;h)$.}
\label{FIG2} 
\end{figure}

We consider the models presented above and illustrated in 
figures \ref{FIG1} and \ref{FIG2} in turn.  In section 2 we 
present our models and discuss the collection of numerical data.
An implementation of the GAS-algorithm for knotted cubic lattice
polygons in slabs \cite{JvRR09} was used,
and we collected data on the entropy and minimal length of
knotted lattice polygons up to eight crossings.  Our data 
verify similar results obtained in reference \cite{ISDAVS12}.  
In section 3 we discuss the first model and present our data,
and in section 4 we present data on the second model and
discuss our results.  The paper is concluded in section 5.

\section{Models of lattice knots in slabs}

A lattice polygon is a sequence $\omega_n = \{\v0,\v1,\ldots,v_{n-1}\}$ 
of distinct vertices $\v{i}$ such that $\v{i}v_{i+1}$, 
for $i=0,1,\ldots,n-2$ and $v_{n-1}\v0$ are unit length edges in
the cubic lattice $\IntN^3$.  Two polygons $\omega_0$ and $\omega_1$
are equivalent if they are translates of each other.  The 
length of a polygon is its number of edges (or steps).  A polygon is,
by inclusion, a tame embedding of the unit circle in $\RealN^3$ and so 
has a well defined knot type.  A lattice polygon with specified 
knot type is a \textit{lattice knot}.

The number of distinct polygons of length $n$ and knot type $K$ will
be denoted $p_n (K)$.  For example $p_4 (0_1)=3$ while $p_n (0_1) = 0$
if $n<4$ or if $n$ is odd, where $0_1$ is the unknot in standard
knot notation.  It is also known that $p_{24} (3_1^+) = 1664$ 
while $p_n (3_1^+) = 0$ if $n < 24$ \cite{SIADSV09}, where $3_1^+$
is the trefoil knot type.  By symmetry, $p_n(K^+) = p_n(K^-)$ if
$K$ is a chiral knot type.

The \textit{minimum length} of a lattice knot $K$ is the minimum number of 
edges required to realise it as a polygon in $\IntN^3$.  For example, 
the minimal length of knot type $0_1$ is $4$ and of knot type $3_1^+$ 
or $3_1^-$ is $24$ edges.  The minimal length is denoted by $n_K$, 
so that $n_{0_1}=4$ and $n_{3_1^+} = 24$ \cite{D93}.  It also known 
that $n_{4_1} = 30$ ($4_1$ is the figure eight knot) and 
$n_{5_1^+}=34$ \cite{SIADSV09}.  Beyond these, only upper bounds on
$n_K$ are known.

If $v \in \omega_n$ is a vertex in a lattice polygon, then the
Cartesian coordinates of $v$ are $\L X(v),Y(v),Z(v)\R$. 
A polygon is \textit{grafted} to the $XY$ plane $Z=0$ if 
$Z(v) \geq 0$ for all vertices $v$ in $\omega$, and there exists one
vertex, say $\v0$, such that $Z(\v0)=0$.  The \textit{height} of a 
grafted polygon is $h = \max\LC Z(v) \vert v \in \omega_n \RC$.

A grafted polygon $\omega_n$ is said to be confined to a slab $\Slab_L$ 
of width $L$ if $0 \leq Z(v) \leq L$ for each vertex $v\in\omega_n$.
A polygon confined to a slab is illustrated 
in figure \ref{FIG1}, where the \textit{bottom wall} and \textit{top
wall} of the slab are indicated.

Next we define the number of polygons of length $n$, knot type $K$, 
and confined in a slab  of width $L$ by $p_n^L (K)$.  Similarly, 
define the number of polygons of length $n$, knot type $K$, and of 
height $h$ by $p_n (K;h)$. Clearly, 
\begin{equation}
p_n^L (K) = \sum_{h \leq L} p_n (K;h).
\end{equation}

The minimal length of a lattice knot in a slab of width $L$ will be
denoted $n_{L,K}$.  For example, one may deduce that $n_{2,3_1} = 24$
from reference \cite{D93}. However, simulations show that $n_{1,3_1}=26$
 \cite{ISDAVS12}.  In other words, in a slab of width $L=1$, it is necessary
to have a polygon of length $26$ to tie a lattice trefoil, while if $L\geq 2$
then $24$ edges will be sufficient (and necessary) \cite{D93}. 

With these definitions, we define two models of grafted lattice knots.
The first model is that of a grafted lattice knot in a slab with hard walls
(see figure \ref{FIG1}). In this model the confinement of the lattice
knot will decrease its entropy, and this will induce an entropic repulsion
between the top and bottom walls of the slab.  The discrete geometry
in this model implies that the induced entropic force is given by 
free energy differences if the distance between the walls is 
decreased by one step.

The second model is illustrated in figure \ref{FIG2}.  The grafted
lattice knot is covered by a flexible and weightless membrane.
A (positive) pressure difference in the fluid above and below the 
membrane induces a force pushing on the top vertices in the lattice knot.
A negative pressure difference in the fluid results in an effective
pulling force on the vertices in the top layer of the lattice knot.
In this model, the partition function is given by all the states of
grafted lattice knot (including those of any height).  The force $f$ 
induced by the pressure difference will push on the highest vertices
in the polygon as it is mediated by the flexible lightweight membrane
onto these vertices. A (linear) compressibility of the lattice knot 
can be determined by taking second derivatives of the free energy of this model
to the applied force, as we shall show below.

\subsection{Numerical approach}

In this paper we examine the properties of minimal length
lattice knots squeezed between two hard walls, or squeezed 
by an applied force towards a hard wall.  In both these models 
it is necessary to determine the number and length of
lattice knots in slabs of width $L$.  Some data of this kind
were obtained in reference \cite{ISDAVS12}, and we will at the
same time verify in most cases, and improve in one case, on
their results.

Our numerical approach will be the implementation of the
GAS-algorithm for lattice knots \cite{JvRR09,JvRR10,JvRR12} using 
BFACF elementary moves \cite{AdCCF83,BF81}.  The lattice knots will
be confined to slab $\Slab_L$ of width $L$ defined by $\Slab_L = \IntN^2 \times
\{0,1,\ldots,L\}$.  We implement the algorithm by noting that BFACF 
elementary moves on unrooted cubic lattice polygons are known to have 
irreducibility classes which are the knot types of the polygons.  The
proof of this fact can be found in reference \cite{JvRW91}.  Note
that the proof in reference \cite{JvRW91} applies \textit{mutatis mutandis} 
to the model in this paper as well, provided that $L\geq 2$.

We estimate $p_{n_{L,k}}(K;h)$ (the number of polygons of knot type $K$, 
height $h$ and of length $n_{L,K}$) using the GAS algorithm
for knotted polygons. Here $n_{L,K}$ is the minimal
length of grafted lattice knots of type $K$ in a slab of width $L$.
By summing $h\leq L$, one obtains $p_{n_{L,K}}^L (K)$, the number of grafted
lattice knots of type $K$ which can fit in a slab of width $L$, 
of length $n_{L,K}$ (and thus of height $h \leq L$).  Our results
are not rigorous, and in a strict sense the results of $n_{L,K}$ 
are upper bounds while $p_{n_{L,K}} (K;h)$ are lower bounds. 
Since the GAS algorithm can be implemented as a flat histogram method, 
it is efficient at rare event sampling and thus at finding knotted
polygons of minimal length. A comparison of our results with the 
data in reference \cite{ISDAVS12} makes us confident that our results 
are exact in almost all cases.

Data on lattice knots in slabs $\Slab_L$,
with $L\geq 2$ can be collected as in reference \cite{JvRR10}.
The case $L=1$ requires further scrutiny.  Data in this ensemble were
collected by generating lattice knots in $\Slab_2$, and sieving out
lattice polygons which fit into $\Slab_1$.  By biasing the sampling
to favour polygons which fit into $\Slab_1$, we were successful
in generating lists of lattice knots in $\Slab_1$.  We are reasonably
confident that in most cases our lists of knotted polygons are complete.

We display our data on the minimal length of lattice knots
in $\Slab_L$ in appendix A in tables \ref{Lengths} and 
\ref{Lengths-C} (for some compound knots).  
The data in table \ref{Lengths} agree for all knot types, except for 
$8_{18}$, with the data in reference \cite{ISDAVS12}.  We improved on 
the estimate for the minimal length of $8_{18}$ in $\Slab_1$ by
finding states at length $n=70$, compared to $72$ in that reference.

As one might expect, we observe a steady increase of $n_{L,K}$ with decreasing 
$L$ for each knot type, and also down table \ref{Lengths}.  Decreasing $L$ 
squeezes lattice knots in narrower slabs, and at critical values of
$L$ there is an increase in the minimal length.  For example, for
the trefoil knot $3_1^+$, there are realizations of lattice knots with
this knot type at $n=24$ edges for $L\geq 2$.  However, if $L=1$,
then $26$ edges are needed.  In figure \ref{FIG1X} the minimal lengths
of lattice knots in $\Slab_1$ are compared to the minimal lengths
of lattice knots in the bulk lattice.  In this scatter plot each
knot type has coordinates $(n_{1,K}, n_K)$.  We found that 
$n_{1,K} > n_K$ for all non-trivial knot types, but the data do cluster
along a line showing a strong correlation between these two quantities.
For slabs $\Slab_L$ our data are displayed in table \ref{Lengths},
showing that $n_{1,K} \geq n_{L,K} \geq n_K$ generally.

Naturally, $n_{L,K}$ is a non-increasing function of $L$ for a given
knot type $K$.  In some cases there is a large increase in $n_{L,K}$
with decreasing $L$. For example, $8_{18}$ increases from $52$ at $L=4$
through $56$, $60$ and to $70$ as $L$ decreases through $3$, $2$ and $1$.

\begin{figure}[t!]
\input{ScatterData.tex}
\caption{A scatter plot of minimal lengths $n_{1,K}$ in $\Slab_1$ 
($X$-axis) and minimal lengths $n_K$ ($Y$-axis).}
\label{FIG1X} 
\end{figure}
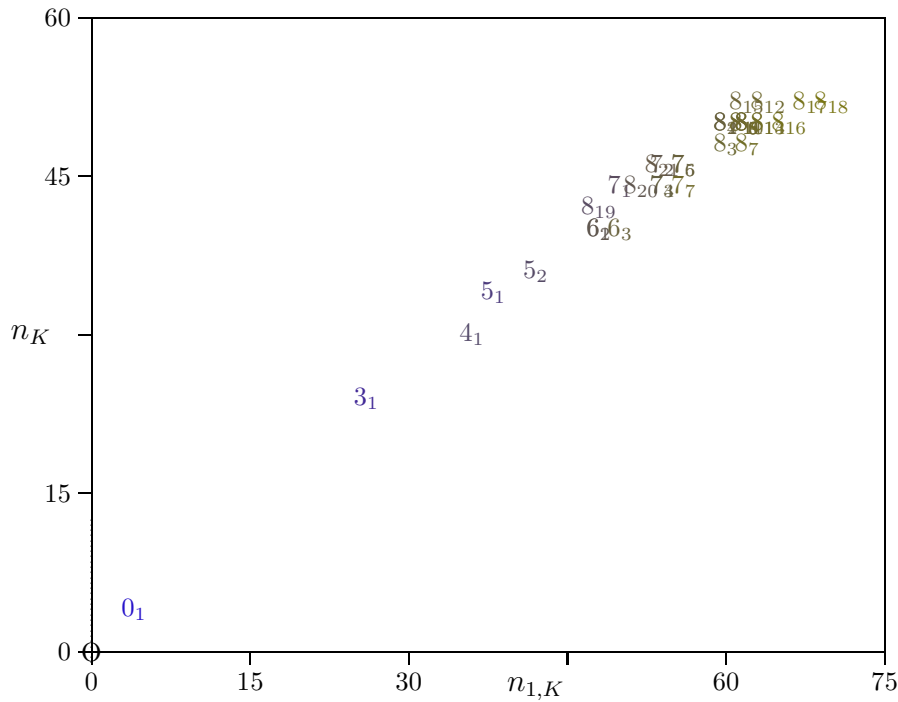

\begin{figure}[t!]
\input{ScatterData2.tex}
\caption{A scatter plot of the number of minimal length lattice knots 
$p_{n_{1,K}}(K)$ in $\Slab_1$ ($X$-axis) and the number of minimal 
length lattice knots $p_{n_K}(K)$ ($Y$-axis) on a log-log scale.}
\label{FIG2X} 
\end{figure}
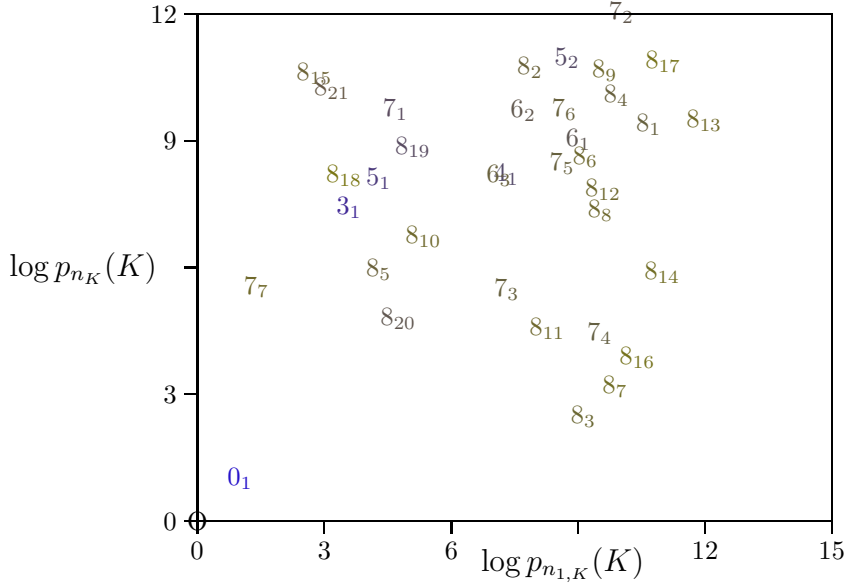

The number of lattice knots of minimal length in $\Slab_L$ are displayed
in tables \ref{Entropy} and \ref{Entropy-C} (for some compound knots)
in the appendix.  
We display some of these results in figure \ref{FIG2X}, where we plot 
the number of minimal length lattice knots against the number of 
minimal length lattice knots in $\Slab_1$ for different knot types 
on a log-log scale.  The data scatter in the plot, showing that 
the number of minimal length lattice knots may change significantly 
with a decrease in $L$.

Lattice knots are partitioned into symmetry classes due to invariance under 
rotations or (in the case of amphichireal knots) reflections which respect 
the orientation of $\Slab_L$.  Thus, the sets of lattice 
knots enumerated in tables \ref{Entropy} and \ref{Entropy-C}
partition into symmetry classes.  These classes
are listed in tables \ref{S1} and \ref{S2} in the appendix.  
The symmetry classes are denoted by $2^{a}4^b6^c8^d12^e16^f$ for each
set of lattice knots.  For example, for the knot type $3_1^+\# 3_1^+$
in table \ref{S1} and for $L=1$, the symmetry classes are $2^3 4^{36} 8^{285}$
meaning that there are $3$ classes with two members (each member is a 
lattice knot), $36$ classes with $4$ members and $285$ with $8$ members.

The data show that the entropy decreases with decreasing $L$, 
if he the minimum length of the polygon does not change.  In cases where 
the minimum length increases with decreasing $L$, it may be accompanied 
by a large decrease or increase in entropy.  See for example the data 
for $8_5^+$ and $8_6^+$ at $L=1$ and $L=2$ in table \ref{Entropy}.

In what follows, we will use the data in these tables to determine the
response of the lattice knots when forces are applied to squeeze them
in slabs with hard walls, or in a model where the forces are mediated
via a flexible membrane to the highest vertices in the lattice
knots.

\section{A grafted lattice knot between hard walls}

The free energy of grafted lattice knots in a slab of width $L$ is
given by
\begin{equation}
\C{F}_L  = \hbox{Energy} - T\cdot\hbox{ Entropy}.
\end{equation}
The lattice knots have fixed length (this is the canonical ensemble),
and we assume in this model that the length is fixed at the
minimal length in the slab $\Slab_L$. 

In this model, the entropy should be given by $\log p_{n_{L,K}}^L (K)$,
where $n_{L,K}$ is the minimal length of the lattice knot of type
$K$ in a slab of width $L$.

We assign an energy to the lattice knot as follows: Compressing
a minimal length lattice knot in a slab of width $L$ will generally
reduce its entropy, but at a minimum value of $L$, no further compression
can take place because a minimal length lattice knot cannot be realised
in a narrower slab.  Instead, further increase of pressure on the $\Slab_L$
will induce forces along the edges of the polygon, and at a critical
value of the force, these induced forces will overcome the elastic or tensile
strength of the edges composing the lattice knot.  The result is that the 
lattice knot will either stretch in length to fit in a narrower slab, 
or it will break apart and be destroyed.  We assume the former case
(our data show that in most cases the level of stretching is less than
10\% of the rest length of the lattice polygon).  Thus, assign a 
Hooke energy to the polygon, with rest length
equal to $n_{K}$.  The energy is then given by
\begin{equation}
\hbox{Energy} = k \cdot(n_{L,K}-n_{K})^2,
\end{equation}
where $n_{L,K}$ is the minimum length to accommodate the lattice knot in a 
slab of width $L$.  There is no Hooke energy in the event that $n_{L,K} = n_{K}$.

With the above in mind, we define the free energy as follows
\begin{equation}
\C{F}_L = k  (n_{L,K}-n_{K})^2 - T \log p_{n_{L,K}}^L (K) .
\label{eqnF} 
\end{equation}
For example, in the case of the unknot $0_1$, one may determine
directly that $n_{0_1}=4$ and $p_{n_{0,0_1}} = 1$ while $p_{n_{1,0_1}}=3$.
This shows that $\C{F}_0 = - T\log 1 = 0$ and $\C{F}_1 = -T \log 3$.

It is important to note that this free energy is a low temperature or a 
stiff Hooke spring approximation.  It is low temperature because
thermal fluctuations in the length of the lattice knot are not modeled
(that is, the lattice knot is always in the shortest possible conformations
in $\Slab_L$), and it is a stiff Hooke spring approximation for that same
reason: the energy barrier to stretch the polygon to $n_{L,K}+2$ edges
in length is too big, and those states do not make a measurable contribution
to the free energy.

Free energy differences as a result change in entropy and the Hooke term
induces entropic forces pushing against the walls of the slab.  
These forces should push the walls apart, both decreasing the length of 
the lattice knot and increasing its entropy. They are given by
\begin{equation}
F_L = \Delta_1 \C{F}_L = \C{F}_{L} - \C{F}_{L-1}.
\end{equation}
If an externally applied force $f$ squeezing the walls of $\Slab_{L}$ together
exceeds $F_L$ (that is, if $|f| > |F_L|$), then the force $f$ will
overcome the entropic and Hooke terms in the free energy and squeeze the
lattice knot into $\Slab_{L-1}$.  Thus, the critical values of an applied
force pushing against $F_L$ are given when
\begin{equation}
f_L = - F_L
\end{equation}
and if $|f|>|f_L|$ then the walls of $\Slab_L$ will be squeezed together
to compress the lattice knot.  We call $f_L$ the \textit{critical force}
of the model.

Compressing the lattice knot between two hard walls performs work on the
knot, and conversely, if a lattice knot is placed in a narrow slab and the
slab expands as a result, then the lattice knot performs work on the walls
of the slab.  The maximum amount of useful work that can be extracted from
this process is given by
\begin{equation}
\C{W}_K = \sum_L f_L \cdot \delta L
\end{equation}
assuming that the expansion is isothermic and reversible.  In our geometry, 
$\delta L = 1$. Thus, $\C{W}_K$ reduces to $\sum_L f_L$.

For example, compressing a minimal length unknotted polygon between 
two hard walls reduces the entropy of the unknot polygon only when 
$L$ transitions from $1$ to $0$.  That is, the top wall in 
figure \ref{FIG1} can be pushed down without encountering resistance 
until $L=1$.  Further compressing to $L=0$ reduces the free energy by
\begin{equation}
F_1 = \Delta_1 \C{F}_1 = \L \C{F}_1 - \C{F}_0 \R = - T \log 3 .
\end{equation}
This shows that the critical force is $f_1 = T \log 3$ in this model.
In this case, the unknot lattice polygon can do at most $\C{W}_{0_1}
= T \log 3$ units of work, assuming that it is placed in $\Slab_0$ and 
allowed to expand the slab.

\subsection{Squeezing minimal length lattice trefoils between two planes}

These ideas can be extended to lattice knots, using the data in
tables \ref{Lengths} and \ref{Entropy}.

In the case of the trefoil knot $3_1^+$, it follows from table \ref{Lengths} 
that $n_{L,3_1^+}=24$ if $L \geq 2$ but that $n_{1,3_1^+} = 26$.  In other 
words, the length of the lattice knot increases from $24$ to $26$ 
if it is squeezed by a force into a slab $\Slab_1$.  This stretching
of the polygon stores work done by the compressing force in the form of
elastic energy, which we indicate by the Hooke term in equation \Ref{eqnF}.

If the compressing force is removed, the lattice knot will rebound to 
length $24$, and expand the slab to width $L=2$, performing work while
doing so.  In $L=2$ the lattice knot is not stretched, but it still 
suffers a reduction in entropy.  Further expansion of the slab width
to $L=3$ increases entropy, and there is thus an entropic force
pushing the hard walls apart until the lattice knot enters a state 
of maximum entropy for values of $L$ large enough.

With this in mind, the free energies of minimal length lattice knots 
of type $3_1^+$ as a function of $L$ may be obtained from the
data in tables \ref{Lengths} and \ref{Entropy}.  The results are
\begin{eqnarray*}
\C{F}_1 (3_1^+)&=& 4\,k - T\log 36; \\
\C{F}_2 (3_1^+)&=&  - T\log 152 ;\\
\C{F}_3 (3_1^+)&=&  - T\log 1660; \\
\C{F}_{\geq 4} (3_1^+)&=&  - T\log 1664 .
\end{eqnarray*}
Observe the steady increase in the entropy of the lattice knot with
increasing $L$.  At $L=2$ the knot cannot be compressed to $L=1$ without
increasing its length to $26$ edges, and the Hooke term $4\,k$
appears in $\C{F}_1 (3_1^+)$.

From the above data one may compute the critical forces for 
lattice knots of type $3_1^+$.  These are
\begin{equation}
f_L = \cases{
\infty , & \hbox{if $L=1$}; \\
 4k + T \log \L 38/9 \R,  & \hbox{if $L=2$}; \\
 T \log \L 415/38 \R,  & \hbox{if $L=3$};  \\
 T \log \L 416/415 \R,  & \hbox{if $L=4$}; \\
0,  & \hbox{if $L\geq 5$}.
}
\label{eqn7}
\end{equation}
The results for $f_L$ above show that there is no entropy loss with decreasing
$L$ until $L=4$.  Thereafter, compressing the lattice knot results in entropy
loss, and the critical forces are given above.  At $L=2$ the knot
stretches to accommodate conformations in $L=1$, with the result that
a Hooke term appears.  Observe that $f_1 = \infty$, since a non-trivial
lattice knot cannot be squeezed into $\Slab_0$, unless the compressing
force overcomes the strength of the edges and breaks the polygon apart.

\begin{figure}[t!]
\centering
\input{Bargraphs/Bargraph3-1k1.tex}
\caption{A profile of critical forces $f_L$ for the lattice knot 
of type $3_1^+$.  Compressing the knot between two plates 
encounters no resistance for $L \geq 5$.  If $L=4$, then there is
a small resistance (not visible on this scale), 
and for $L\leq 3$ a larger resistance.  
Bars in red indicate that the critical forces are due to entropy
reduction alone, and blue bars denote a Hooke contribution to 
the critical force.  In this example, $k=1/4$ and $T=1$.}
\label{FIG3} 
\end{figure}
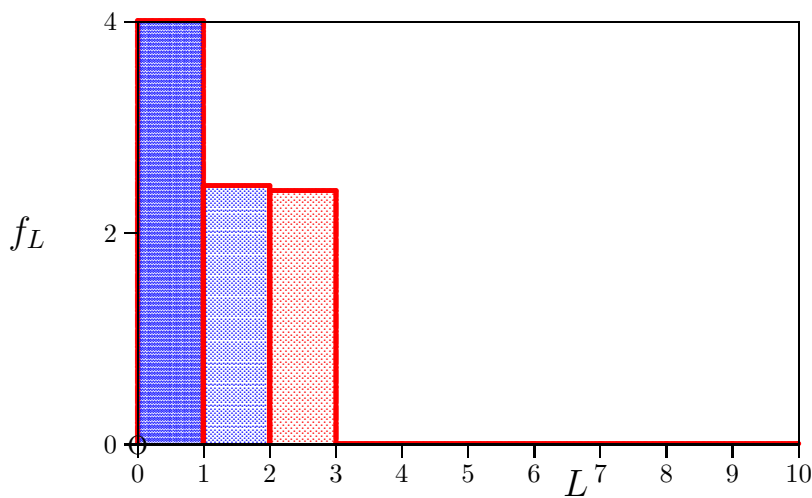

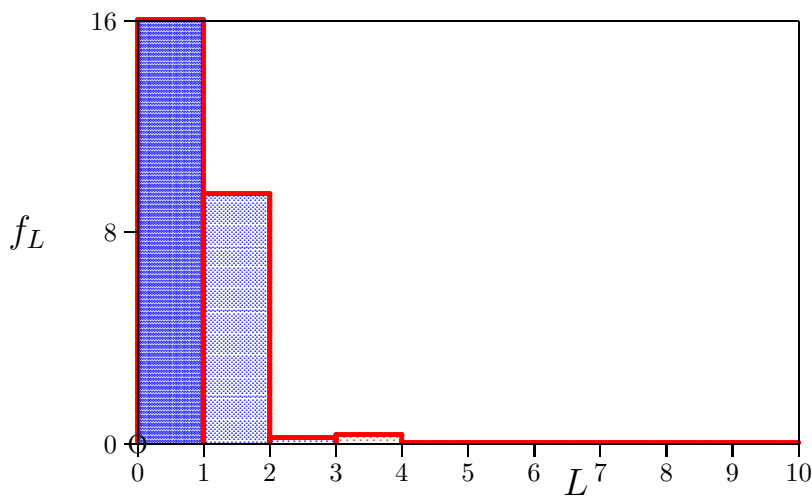
\begin{figure}[t!]
\centering
\input{Bargraphs/Bargraph4-1k1.tex}
\caption{A profile of critical forces $f_L$ for the lattice knot 
of type $4_1$.  Compressing the knot between two plates 
encounters no resistance for $L \geq 5$.  If $L=4$, then there is
a small resistance, and for $L\leq 3$ a larger resistance.  
Bars in red indicate that the critical forces are due to entropy
reduction alone, and blue bars denote a Hooke contribution to 
the critical force.  In this example, $k_1=k_2=1/4$ and $T=1$.}
\label{FIG4} 
\end{figure}
\begin{figure}[t!]
\subfigure[$5_1^+$]{
   \input{Bargraphs/Bargraph5-1k1.tex}
   \label{FIG5-1}
 }
 \subfigure[$5_2^+$]{
   \input{Bargraphs/Bargraph5-2k1.tex}
   \label{FIG5-2}
 }
\caption{A profile of critical forces $f_L$ for lattice knots 
of types $5_1^+$ and $5_2^+$.  There is no resistance to compression until 
$L = 5$.  There are small resistances if $L=4$, and this decreases 
even further for $L=3$ in the case of $5_1^+$.  In the case of $5_2^+$ 
the resistance increases with decreasing $L$. Bars in red indicate that 
the critical forces are due to entropy reduction alone, and blue bars 
denote a Hooke contribution to the critical force.  In this example, 
$k=1/4$ and $T=1$.}
\label{FIG5} 
\end{figure}
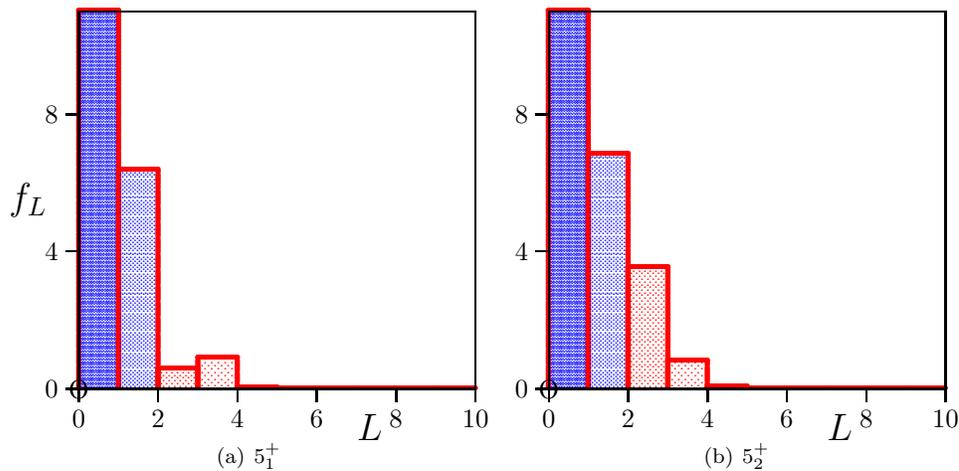
\begin{figure}[h!]
\subfigure[$6_1^+$]{
   \input{Bargraphs/Bargraph6-1k1.tex}
   \label{FIG6-1}
 }
 \subfigure[$6_2^+$]{
   \input{Bargraphs/Bargraph6-2k1.tex}
   \label{FIG6-2}
 }
\subfigure[$6_3$]{
   \input{Bargraphs/Bargraph6-3k1.tex}
   \label{FIG6-3}
 }
\caption{A profile of critical forces $f_L$ for the lattice knots 
of types $6_1^+$, $6_2^+$ and $6_3$.  There is no resistance 
to compression until $L = 5$ in all cases.  There are small resistances 
if $L=4$.  The negative bar between $L=2$ and $L=3$ for $6_1^+$ shows
that a gain in entropy overwhelms the Hooke forces when the lattice
knot is compressed from $L=3$ to $L=2$ -- in fact, no force is necessary,
as the entropic force pulls the walls together. Bars in red indicate 
that the critical forces are due to entropy reduction alone, and blue 
bars denote a Hooke contribution to the critical force.  In this example, 
$k=1/4$ and $T=1$.}
\label{FIG6} 
\end{figure}
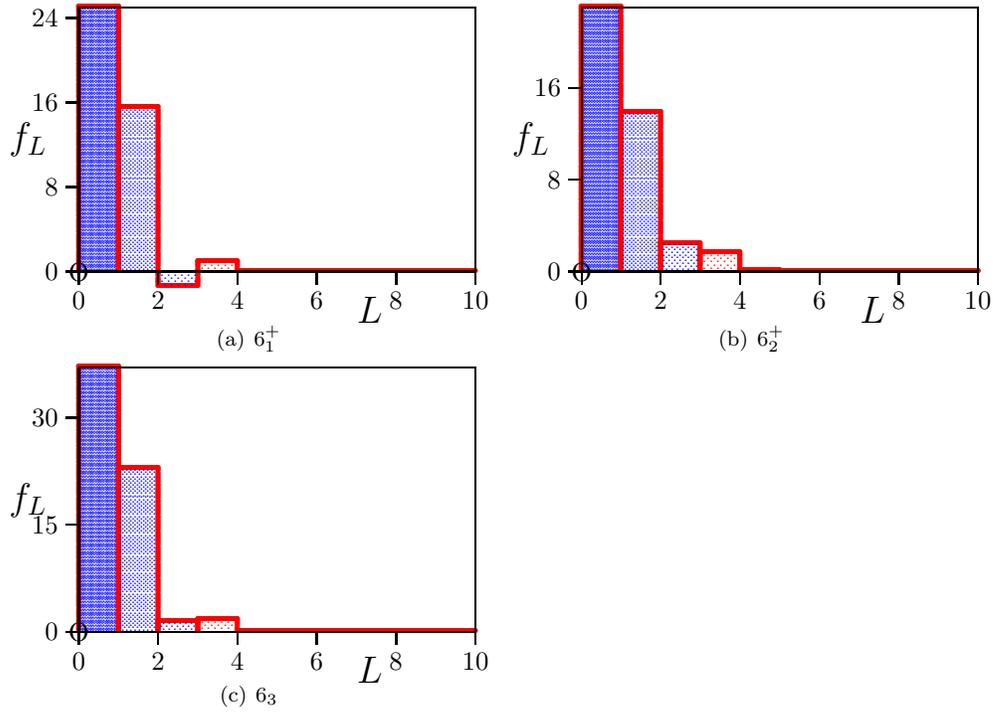
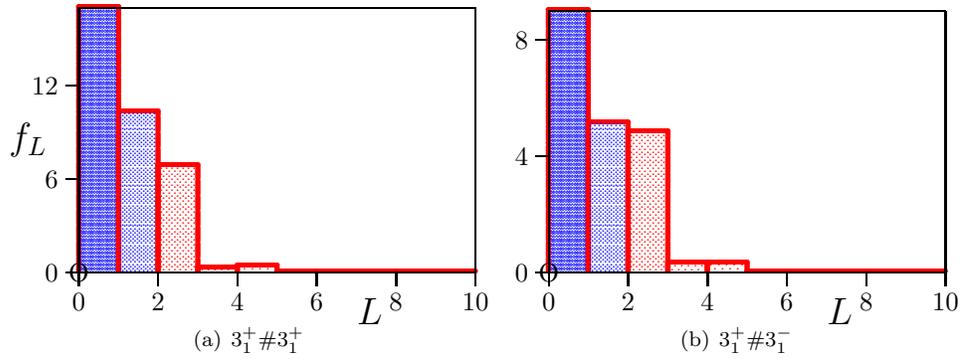
\begin{figure}[t!]
\subfigure[$3_1^+\# 3_1^+$]{
   \input{Bargraphs/Bargraph3-1p+3-1pk1.tex}
   \label{FIG7-1}
 }
 \subfigure[$3_1^+\# 3_1^-$]{
   \input{Bargraphs/Bargraph3-1p+3-1mk1.tex}
   \label{FIG7-2}
 }
\caption{A profile of critical forces $f_L$ for the compound
lattice knots of types $3_1^+\# 3_1^+$ and $3_1^+\# 3_1^-$.  
There is no resistance to compression until $L = 5$ in all cases.  
Bars in red indicate that the critical forces are due to entropy 
reduction alone, and blue bars denote a Hooke contribution to the 
critical force.  In this example, $k=1/4$ and $T=1$.}
\label{FIG7} 
\end{figure}
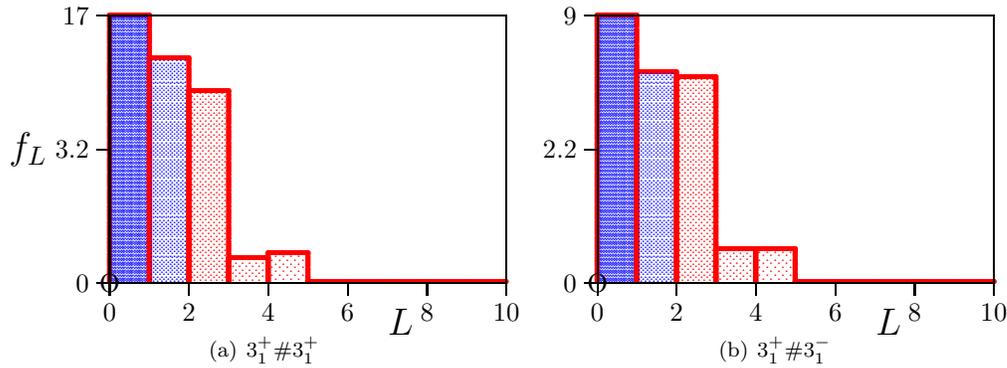
\begin{figure}[ht!]
\subfigure[$3_1^+\# 3_1^+$]{
   \input{Bargraphs/Bargraph3-1p+3-1pk1L.tex}
   \label{FIG8-1}
 }
 \subfigure[$3_1^+\# 3_1^-$]{
   \input{Bargraphs/Bargraph3-1p+3-1mk1L.tex}
   \label{FIG8-2}
 }
\caption{The same data as in figure \ref{FIG7}, but with the
vertical axes scaled logarithmically to enhance the data at larger
values of $L$.}
\label{FIG8} 
\end{figure}

The above expressions for the critical forces gives a compression
profile for squeezing the lattice trefoil between two planes.  We
illustrate this profile as a bargraph in figure \ref{FIG3} -- 
where we put $k=1/4$ and $T=1$.  In this case one may also
compute $\C{W}_{3_1^+} = 4k + T\log (416/9)$.  The choice of $k=1/4$
in figure \ref{FIG3} gives a Hooke contribution of one to the
free energy if the polygon should stretch by two edges.  At this
level, the Hooke energy does not dominate the free energy.

Similar data can be obtained for the figure eight knot $4_1$, and its
critical forces are give by
\begin{equation}
f_L = \cases{
\infty , & \hbox{if $L=1$}; \\
32k + T \log \L 758/185 \R,  & \hbox{if $L=2$}; \\
4k - T \log \L 379/170 \R,  & \hbox{if $L=3$};  \\
 T \log \L 114/85 \R,  & \hbox{if $L=4$}; \\
0,  & \hbox{if $L\geq 5$},
}
\end{equation}
displayed in figure \ref{FIG4}.

In this knot type, the lattice knot increases in length both
in the transition from $L=3$ to $L=2$, and then again to $L=1$,
as seen in table \ref{Lengths}. While one will generally expect
$f_L \geq f_{L+1}$ (that is, a larger force is necessary to 
compress the knot in narrower slabs), there is an interplay between
entropy and the Hooke terms in the free energy, and in some cases
$f_L < f_{L+1}$.  This is for example seen in figure \ref{FIG4} in
the data for $L=3$ and $L=4$ for the knot type $4_1$.  One may also 
verify that $\C{W}_{4_1} = 36k + T \log (456/185)$.

The results for five crossing knots $5_*^+$ and six crossing knots
are illustrated in figures \ref{FIG5} and \ref{FIG6}.

Our numerical data on seven and eight crossings knots in tables \ref{Lengths}
and \ref{Entropy} were used to compute critical forces and $\C{W}$ for
each of those knot types.  Data on compound knots up to eight crossings
are listed in tables \ref{Lengths-C} and \ref{Entropy-C}, and we
similarly determined critical forces and $\C{W}$ for those knot types.
The results are shown in tables \ref{Forces7} and \ref{Forces8}.

\subsection{Discussion}

Compressing a lattice knot between two hard walls decreases the entropy
of the knot, until the walls are close enough together.  Then the lattice
knot expands laterally and in length as it finds conformations which can be 
accommodated in even narrower slabs.  

For example, a lattice trefoil loses entropy in $\Slab_L$ as $L$ is 
reduced from $L=4$ to $L=2$, but then has to increase in length by $2$ 
if it is compressed into $\Slab_1$.  This stretching of the lattice knot 
to a longer length changes its entropic properties, and may even increase 
its entropy, as the longer lattice polygon may be able to explore more states. 
However, the Hooke energy involved in stretching the lattice knot increases the free
energy, and also increases the critical force necessary to compress the knot
into a narrower slab.

The critical forces of the trefoil knot $3_1^+$ are listed in equation
\Ref{eqn7}.  These forces are induced by entropy loss when $L >2$,
but at $L=2$ a Hooke energy contribution also appears.   The sum of these
forces gives the total amount of work in an isothermic compression
of the lattice knot.  We found this to be $\C{W}_{3_1^+} = 4k+T\log(416/9)$.
There are two contributions to $\C{W}_{3_1^+}$, namely an entropic 
and a Hooke contribution.  These contributions are equal in magnitude
at a critical value of $k$ (or equivalently, a critical value of $T$).  
In the case of the trefoil knot this critical value of $k$ is $k_{3_1^+} 
= [T\log(416/9) ]/4 \approx (0.95836\ldots)T$. If $k>k_c$, then the 
Hooke term dominates $\C{W}_{3_1^+}$ and if $k<k_c$, then the work done 
has a larger contribution from entropy reduction in the process.

In the case of the figure eight knot, one may similarly determine
the critical value of $k$: $k_{4_1}=(0.02505\ldots)T$.  In this case,
$k_c$ is very small, and the total amount of work done in compressing
the knot to $L=2$ is dominated by the Hooke term even for relatively
small values of $k$.  The two five crossing knots have
$k_{5_1^+}=(0.23974\ldots)T$ and $k_{5_2^+}=(0.06167\ldots)T$.

Data on other knots are listed in tables \ref{Forces7} and \ref{Forces8}.  
The critical values of $k$ can be determined by solving for $k$ from the 
last column in these tables.  Observe that some knot types have negative 
values of $k_c$, for example, $k_{6_1^+} = -0.015025 T$.
This implies that a large entropy gain occurs when the knot 
stretches in length to fit in $\Slab_1$, and this can only be matched by 
a negative Hooke constant.

\begin{landscape}
\begin{table}
\begin{center}
 \begin{tabular}{||c||c|c|c|c|c|c|c|c|c||}
  \hline  
   & $L=0$ & $L=1$ & $L=2$ & $L=3$ & $L=4$ & $L=5$ & $L=6$ & $L=7$ & $\C{W}$ \\
   \hline
  $0_1$ &\fns $\log 3$ &\fns $0$ &\fns $0$ &\fns $0$ &\fns $0$ &\fns $0$ &\fns $0$ &\fns $0$ &\fns $\log 3$  \\
$3_1^+$ &\fns $\infty$ &\fns $4\,k+\log \left( {\frac {38}{9}} \right) $ &\fns $\log \left( {\frac {415}{38}} \right) $ &\fns $\log \left( {\frac {416}{415}} \right) $ &\fns $0$ &\fns $0$ &\fns $0$ &\fns $0$ &\fns $4\,k+\log \left( {\frac {416}{9}} \right) $ \\

$4_1$   &\fns $\infty$ &\fns $32\,k+\log \left( {\frac {758}{185}} \right) $ &\fns $
4\,k+\log \left( {\frac {170}{379}} \right) $ &\fns $
\log \left( {\frac {114}{85}} \right) $ &\fns $
0$ &\fns $
0$ &\fns $
0$ &\fns $
0$ &\fns $
36\,k+\log \left( {\frac {456}{185}} \right) $ \\

$5_1^+$ &\fns  $\infty$ &\fns $16\,k+\log \left( {\frac {95}{9}} \right) $ &\fns $
\log \left( {\frac {34}{19}} \right) $ &\fns $
\log \left( {\frac {206}{85}} \right) $ &\fns $
\log \left( {\frac {417}{412}} \right) $ &\fns $
0$ &\fns $
0$ &\fns $
0$ &\fns $
16\,k+\log \left( {\frac {139}{3}} \right) $ \\

$5_2^+$ &\fns  $\infty$ &\fns $36\,k+\log \left( {\frac {3}{26}} \right) $ &\fns $
\log \left( {\frac {1033}{30}} \right) $ &\fns $
\log \left( {\frac {6917}{3099}} \right) $ &\fns $
\log \left( {\frac {7182}{6917}} \right) $ &\fns $
0$ &\fns $
0$ &\fns $
0$ &\fns $
36\,k+\log \left( {\frac {1197}{130}} \right) 
$ \\

$6_1^+$ &\fns  $\infty$ &\fns $60\,k+\log \left( {\frac {3176}{2009}} \right) $ &\fns $
4\,k+\log \left( {\frac {73}{794}} \right) $ &\fns $
\log \left( {\frac {375}{146}} \right) $ &\fns $
\log \left( {\frac {128}{125}} \right) $ &\fns $
0$ &\fns $
0$ &\fns $
0$ &\fns $
64\,k+\log \left( {\frac {768}{2009}} \right) 
$ \\

$6_2^+$ &\fns  $\infty$ &\fns $60\,k+\log \left( {\frac {15}{46}} \right) $ &\fns $
4\,k+\log \left( {\frac {373}{90}} \right) $ &\fns $
\log \left( {\frac {1829}{373}} \right) $ &\fns $
\log \left( {\frac {2052}{1829}} \right) $ &\fns $
0$ &\fns $
0$ &\fns $
0$ &\fns $
64\,k+\log \left( {\frac {171}{23}} \right) $ \\

$6_3$   &\fns  $\infty$ &\fns $84\,k+\log \left( {\frac {157}{26}} \right) $ &\fns $
16\,k+\log \left( {\frac {37}{471}} \right) $ &\fns $
\log \left( {\frac {213}{37}} \right) $ &\fns $
\log \left( {\frac {74}{71}} \right) $ &\fns $
0$ &\fns $
0$ &\fns $
0$ &\fns $
100\,k+\log \left( {\frac {37}{13}} \right) $ \\

$7_1^+$   &\fns  $\infty$ &\fns $36\,k+\log \left( {\frac {100}{9}} \right) $ &\fns $
\log \left( {\frac {1417}{300}} \right) $ &\fns $
\log \left( {\frac {2890}{1417}} \right) $ &\fns $
\log \left( {\frac {422}{289}} \right) $ &\fns $
\log \left( {\frac {849}{844}} \right) $ &\fns $
0$ &\fns $
0$ &\fns $
36\,k+\log \left( {\frac {1415}{9}} \right) $ \\

$7_2^+$   &\fns  $\infty$ &\fns $60\,k+\log \left( {\frac {62507}{5607}} \right) $ &\fns $
4\,k+\log \left( {\frac {12924}{62507}} \right) $ &\fns $
\log \left( {\frac {7688}{3231}} \right) $ &\fns $
\log \left( {\frac {41069}{30752}} \right) $ &\fns $
\log \left( {\frac {42045}{41069}} \right) $ &\fns $
0$ &\fns $
0$ &\fns $
64\,k+\log \left( {\frac {14015}{1869}} \right) $ \\

$7_3^+$   &\fns  $\infty$ &\fns $84\,k+\log \left( {\frac {1202}{31}} \right) $ &\fns $
16\,k+\log \left( {\frac {5}{1803}} \right) $ &\fns $
0$ &\fns $
\log \left( \frac{3}{2} \right) $ &\fns $
0$ &\fns $
0$ &\fns $
0$ &\fns $
100\,k+\log \left( {\frac {5}{31}} \right) $ \\

$7_4^+$   &\fns  $\infty$ &\fns $96\,k+\log \left( {\frac {100}{3361}} \right) $ &\fns $
4\,k+\log \left( \frac{1}{20} \right) $ &\fns $
\log \left( {\frac {16}{5}} \right) $ &\fns $
\log \left( {\frac {21}{16}} \right) $ &\fns $
0$ &\fns $
0$ &\fns $
0$ &\fns $
100\,k+\log \left( {\frac {21}{3361}} \right) $ \\

$7_5^+$   &\fns  $\infty$ &\fns $96\,k+\log \left( {\frac {52}{115}} \right) $ &\fns $
4\,k+\log \left( {\frac {5}{104}} \right) $ &\fns $
\log \left( {\frac {581}{15}} \right) $ &\fns $
\log \left( {\frac {591}{581}} \right) $ &\fns $
0$ &\fns $
0$ &\fns $
0$ &\fns $
100\,k+\log \left( {\frac {197}{230}} \right) $ \\

$7_6^+$   &\fns  $\infty$ &\fns $96\,k+\log \left( {\frac {220}{731}} \right) $ &\fns $
4\,k+\log \left( \frac{4}{11} \right) $ &\fns $
\log \left( {\frac {323}{16}} \right) $ &\fns $
\log \left( {\frac {2127}{1615}} \right) $ &\fns $
0$ &\fns $
0$ &\fns $
0$ &\fns $
100\,k+\log \left( {\frac {2127}{731}} \right) $ \\

$7_7^+$   &\fns  $\infty$ &\fns $108\,k+\log \left( 2248 \right) $ &\fns $
32\,k+\log \left( {\frac {161}{1124}} \right) $ &\fns $
4\,k+\log \left( {\frac {7}{46}} \right) $ &\fns $
\log \left( {\frac {9}{7}} \right) $ &\fns $
0$ &\fns $
0$ &\fns $
0$ &\fns $
144\,k+\log \left( 63 \right) $ \\

\hline

$3_1^+\#3_1^+$   &\fns  $\infty$ &\fns $
64\,k+\log \left( {\frac {4}{1215}} \right) $ &\fns $
\log \left( 975 \right) $ &\fns $
\log \left( {\frac {2623}{1950}} \right) $ &\fns $
\log \left( {\frac {3826}{2623}} \right) $ &\fns $
0$ &\fns $
\log \left( {\frac {1914}{1913}} \right) $ &\fns $
0$ &\fns $
64\,k+\log \left( {\frac {2552}{405}} \right) $ \\

$3_1^+\#3_1^-$   &\fns  $\infty$ &\fns $
16\,k+\log \left( {\frac {28}{9}} \right) $ &\fns $
\log \left( {\frac {1011}{8}} \right) $ &\fns $
\log \left( {\frac {3274}{2359}} \right) $ &\fns $
\log \left( {\frac {6839}{4911}} \right) $ &\fns $
0$ &\fns $
\log \left( {\frac {13679}{13678}} \right) $ &\fns $
\log \left( {\frac {13680}{13679}} \right) $ &\fns $
16\,k+\log \left( 760 \right) $ \\

$3_1^+\#4_1$   &\fns  $\infty$ &\fns $
60\,k+\log \left( {\frac {3434}{1509}} \right) $ &\fns $
4\,k+\log \left( {\frac {3559}{1717}} \right) $ &\fns $
\log \left( {\frac {16181}{7118}} \right) $ &\fns $
\log \left( {\frac {21026}{16181}} \right) $ &\fns $
\log \left( {\frac {11241}{10513}} \right) $ &\fns $
0$ &\fns $
0$ &\fns $
64\,k+\log \left( {\frac {7494}{503}} \right) $ \\

$3_1^+\#5_1^+$   &\fns  $\infty$ &\fns $
64\,k+\log \left( {\frac {10}{27}} \right) $ &\fns $
\log \left( 395 \right) $ &\fns $
\log \left( {\frac {53}{25}} \right) $ &\fns $
\log \left( {\frac {4464}{4187}} \right) $ &\fns $
\log \left( {\frac {4183}{2976}} \right) $ &\fns $
\log \left( {\frac {4187}{4183}} \right) $ &\fns $
0$ &\fns $
64\,k+\log \left( {\frac {4187}{9}} \right) $ \\

$3_1^+\#5_1^-$   &\fns  $\infty$ &\fns $
36\,k+\log \left( {\frac {197}{18}} \right) $ &\fns $
\log \left( {\frac {3374}{197}} \right) $ &\fns $
\log \left( {\frac {12431}{3374}} \right) $ &\fns $
\log \left( {\frac {15328}{12431}} \right) $ &\fns $
\log \left( {\frac {35447}{30656}} \right) $ &\fns $
\log \left( {\frac {35547}{35447}} \right) $ &\fns $
0$ &\fns $
36\,k+\log \left( {\frac {11849}{12}} \right) 
$ \\

$3_1^+\#5_2^+$   &\fns  $\infty$ &\fns $
64\,k+\log \left( {\frac {235}{18802}} \right) $ &\fns $
\log \left( {\frac {91191}{235}} \right) $ &\fns $
\log \left( {\frac {273449}{91191}} \right) $ &\fns $
\log \left( {\frac {398758}{273449}} \right) $ &\fns $
\log \left( {\frac {454641}{398758}} \right) $ &\fns $
\log \left( {\frac {153271}{151547}} \right) $ &\fns $
0$ &\fns $
64\,k+\log \left( {\frac {459813}{18802}} \right) $ \\

$3_1^+\#5_2^-$   &\fns  $\infty$ &\fns $
96\,k-\log \left( 246 \right) $ &\fns $
4\,k+\log \left( {\frac {110}{7}} \right) $ &\fns $
\log \left( {\frac {113}{55}} \right) $ &\fns $
\log \left( {\frac {165}{113}} \right) $ &\fns $
0$ &\fns $
0$ &\fns $
0$ &\fns $
100\,k+\log \left( {\frac {55}{287}} \right) 
$ \\

$4_1\#4_1$   &\fns  $\infty$ &\fns $
60\,k+\log \left( {\frac {5477}{100}} \right)$ &\fns $ 
4\,k+\log \left( {\frac {5076}{5477}} \right) $ &\fns $
\log \left( {\frac {21601}{5076}} \right) $ &\fns $
\log \left( {\frac {38653}{21601}} \right) $ &\fns $
\log \left( {\frac {41515}{38653}} \right) $ &\fns $
\log \left( {\frac {41853}{41515}} \right) $ &\fns $
0$ &\fns $
64\,k+\log \left( {\frac {41853}{100}} \right) $ \\

\hline
\end{tabular}
\end{center}
 \caption{Critical forces $f_L$ for knots up to seven crossings and
for compound knot types up to eight crossings.  The last
column is the maximum amount of useful work which can be extracted if the
lattice knot expands against the hard walls of the slab, pushing them
apart from $L=1$.  In all these cases we set $T=1$.}
  \label{Forces7}  
\end{table}
\end{landscape}

\begin{landscape}
\begin{table}
\begin{center}
 \begin{tabular}{||c||c|c|c|c|c|c|c|c|c||}
  \hline  
   & $L=0$ & $L=1$ & $L=2$ & $L=3$ & $L=4$ & $L=5$ & $L=6$ & $L=7$ & $\C{W}$ \\
   \hline

$8_1^+$   &\fns  $\infty$ &\fns $96\,k+\log \left( {\frac {254}{2145}} \right) $ &\fns $
4\,k+\log \left( {\frac {274}{635}} \right) $ &\fns $
\log \left( {\frac {1003}{274}} \right) $ &\fns $
\log \left( {\frac {2787}{2006}} \right) $ &\fns $
\log \left( {\frac {989}{929}} \right) $ &\fns $
0$ &\fns $
0$ &\fns $
100\,k+\log \left( {\frac {989}{3575}} \right) $  \\

$8_2^+$   &\fns  $\infty$ &\fns $
84\,k+\log \left( {\frac {73714}{649}} \right) $ &\fns $
16\,k+\log \left( {\frac {1291}{36857}} \right) $ &\fns $
\log \left( {\frac {3248}{1291}} \right) $ &\fns $
\log \left( {\frac {5263}{3248}} \right) $ &\fns $
\log \left( {\frac {5730}{5263}} \right) $ &\fns $
0$ &\fns $
0$ &\fns $
100\,k+\log \left( {\frac {11460}{649}} \right) $ \\

$8_3$   &\fns  $\infty$ &\fns $
128\,k+\log \left( {\frac {500}{2273}} \right) $ &\fns $
16\,k+\log \left( {\frac {1}{250}} \right) $ &\fns $
0$ &\fns $
\log \left( \frac{3}{2} \right) $ &\fns $
0$ &\fns $
0$ &\fns $
0$ &\fns $
144\,k+\log \left( {\frac {3}{2273}} \right) 
$ \\

$8_4^+$   &\fns  $\infty$ &\fns $
96\,k+\log \left( {\frac {140}{297}} \right) $ &\fns $
4\,k+\log \left( {\frac {23}{119}} \right) $ &\fns $
\log \left( {\frac {933}{115}} \right) $ &\fns $
\log \left( {\frac {2933}{1866}} \right) $ &\fns $
\log \left( {\frac {2991}{2933}} \right) $ &\fns $
0$ &\fns $
0$ &\fns $
100\,k+\log \left( {\frac {1994}{1683}} \right) $ \\

$8_5^+$   &\fns  $\infty$ &\fns $
64\,k+\log \left( {\frac {102047}{18}} \right) $ &\fns $
32\,k+\log \left( {\frac {1236}{102047}} \right) $ &\fns $
4\,k+\log \left( {\frac {8}{103}} \right) $ &\fns $
\log \left( \frac{3}{2} \right) $ &\fns $
0$ &\fns $
0$ &\fns $
0$ &\fns $
100\,k+3\,\log \left( 2 \right) $ \\

$8_6^+$   &\fns  $\infty$ &\fns $
140\,k-\log \left( 1205 \right) $ &\fns $
\log \left( 3023 \right) $ &\fns $
4\,k+\log \left( {\frac {460}{3023}} \right) $ &\fns $
\log \left( \frac{3}{2}\right) $ &\fns $
0$ &\fns $
0$ &\fns $
0$ &\fns $
144\,k+\log \left( {\frac {138}{241}} \right) 
$ \\

$8_7^+$   &\fns  $\infty$ &\fns $
160\,k+\log \left( {\frac {6172}{2429}} \right) $ &\fns $
36\,k+\log \left( {\frac {1}{6172}} \right) $ &\fns $
\log \left( 2 \right) $ &\fns $
\log \left( \frac{3}{2}\right) $ &\fns $
0$ &\fns $
0$ &\fns $
0$ &\fns $
196\,k+\log \left( {\frac {3}{2429}} \right) 
$ \\

$8_8^+$   &\fns  $\infty$ &\fns $
128\,k+\log \left( {\frac {347}{1696}} \right) $ &\fns $
12\,k+\log \left( {\frac {5035}{347}} \right) $ &\fns $
4\,k+\log \left( {\frac {26}{1007}} \right) $ &\fns $
\log \left( \frac{3}{2}\right) $ &\fns $
0$ &\fns $
0$ &\fns $
0$ &\fns $
144\,k+\log \left( {\frac {195}{1696}} \right) $ \\

$8_9$   &\fns  $\infty$ &\fns $
128\,k+\log \left( {\frac {355}{318}} \right) $ &\fns $
16\,k+\log \left( {\frac {20}{213}} \right) $ &\fns $
\log \left( {\frac {147}{10}} \right) $ &\fns $
\log \left( \frac{3}{2}\right) $ &\fns $
\log \left( {\frac {3541}{2940}} \right) $ &\fns $
0$ &\fns $
0$ &\fns $
144\,k+\log \left( {\frac {3541}{1272}} \right) $ \\

$8_{10}^+$   &\fns  $\infty$ &\fns $
108\,k+\log \left( {\frac {2982}{13}} \right) $ &\fns $
32\,k+\log \left( {\frac {221}{1988}} \right) $ &\fns $
4\,k+\log \left( {\frac {70}{663}} \right) $ &\fns $
\log \left( \frac{3}{2}\right) $ &\fns $
0$ &\fns $
0$ &\fns $
0$ &\fns $
144\,k+\log \left( {\frac {105}{26}} \right) 
$ \\

$8_{11}^+$   &\fns  $\infty$ &\fns $
128\,k+\log \left( {\frac {453}{488}} \right) $ &\fns $
12\,k+\log \left( {\frac {5066}{453}} \right) $ &\fns $
4\,k+\log \left( {\frac {4}{2533}} \right) $ &\fns $
\log \left( \frac{3}{2}\right) $ &\fns $
0$ &\fns $
0$ &\fns $
0$ &\fns $
144\,k+\log \left( {\frac {3}{122}} \right) 
$ \\

$8_{12}$   &\fns  $\infty$ &\fns $
128\,k+\log \left( {\frac {2897}{183}} \right) $ &\fns $
16\,k+\log \left( {\frac {18}{14485}} \right) $ &\fns $
\log \left( 6 \right) $ &\fns $
\log \left( \frac{3}{2}\right) $ &\fns $
0$ &\fns $
0$ &\fns $
0$ &\fns $
144\,k+\log \left( {\frac {54}{305}} \right) $ \\

$8_{13}^+$   &\fns  $\infty$ &\fns $
180\,k+\log \left( {\frac {95}{9987}} \right) $ &\fns $
16\,k-\log \left( 19 \right) $ &\fns $
\log \left( {\frac {537}{5}} \right) $ &\fns $
\log \left( {\frac {802}{537}} \right) $ &\fns $
\log \left( {\frac {408}{401}} \right) $ &\fns $
0$ &\fns $
0$ &\fns $
196\,k+\log \left( {\frac {272}{3329}} \right) $ \\

$8_{14}^+$   &\fns  $\infty$ &\fns $
180\,k+\log \left( {\frac {172}{1475}} \right) $ &\fns $
12\,k+\log \left( {\frac {77}{86}} \right) $ &\fns $
4\,k+\log \left( {\frac {3}{77}} \right) $ &\fns $
\log \left( \frac{3}{2}\right) $ &\fns $
0$ &\fns $
0$ &\fns $
0$ &\fns $
196\,k+\log \left( {\frac {9}{1475}} \right) $ \\

$8_{15}^+$   &\fns  $\infty$ &\fns $
84\,k+\log \left( 55 \right) $ &\fns $
16\,k+\log \left( {\frac {282}{55}} \right) $ &\fns $
\log \left( {\frac {3047}{564}} \right) $ &\fns $
\log \left( {\frac {5013}{3047}} \right) $ &\fns $
0$ &\fns $
0$ &\fns $
0$ &\fns $
100\,k+\log \left( {\frac {5013}{2}} \right) 
$ \\

$8_{16}^+$   &\fns  $\infty$ &\fns $
220\,k+\log \left( {\frac {73}{2045}} \right) $ &\fns $
32\,k+\log \left( {\frac {3}{73}} \right) $ &\fns $
4\,k+\log \left( 2/3 \right) $ &\fns $
\log \left( \frac{3}{2}\right) $ &\fns $
0$ &\fns $
0$ &\fns $
0$ &\fns $
256\,k+\log \left( {\frac {3}{2045}} \right) 
$ \\

$8_{17}$   &\fns  $\infty$ &\fns $
220\,k+\log \left( {\frac {518}{3771}} \right) $ &\fns $
32\,k+\log \left( {\frac {231}{74}} \right) $ &\fns $
4\,k+\log \left( {\frac {557}{539}} \right) $ &\fns $
\log \left( {\frac {1098}{557}} \right) $ &\fns $
\log \left( {\frac {554}{549}} \right) $ &\fns $
0$ &\fns $
0$ &\fns $
256\,k+\log \left( {\frac {1108}{1257}} \right) $ \\

$8_{18}$   &\fns  $\infty$ &\fns $
260\,k+\log \left( {\frac {189}{8}} \right) $ &\fns $
48\,k+\log \left( {\frac {17000}{189}} \right) $ &\fns $
16\,k+\log \left( {\frac {37}{2125}} \right) $ &\fns $
\log \left( 3 \right) $ &\fns $
0$ &\fns $
0$ &\fns $
0$ &\fns $
324\,k+\log \left( 111 \right) $ \\

$8_{19}^+$   &\fns  $\infty$ &\fns $
32\,k+\log \left( {\frac {1656}{163}} \right) $ &\fns $
4\,k+\log \left( {\frac {283}{414}} \right) $ &\fns $
\log \left( {\frac {1677}{283}} \right) $ &\fns $
\log \left( {\frac {583}{559}} \right) $ &\fns $
0$ &\fns $
0$ &\fns $
0$ &\fns $
36\,k+\log \left( {\frac {6996}{163}} \right) $ \\

$8_{20}^+$   &\fns  $\infty$ &\fns $
60\,k+\log \left( {\frac {318}{29}} \right) $ &\fns $
4\,k+\log \left( {\frac {5}{159}} \right) $ &\fns $
\log \left( 2 \right) $ &\fns $
\log \left( \frac{3}{2}\right) $ &\fns $
0$ &\fns $
0$ &\fns $
0$ &\fns $
64\,k+\log \left( {\frac {30}{29}} \right) $ \\

$8_{21}^+$   &\fns  $\infty$ &\fns $
60\,k+\log \left( 30 \right) $ &\fns $
4\,k+\log \left( {\frac {857}{180}} \right) $ &\fns $
\log \left( {\frac {6375}{857}} \right) $ &\fns $
\log \left( {\frac {467}{425}} \right) $ &\fns $
0$ &\fns $
0$ &\fns $
0$ &\fns $
64\,k+\log \left( {\frac {2335}{2}} \right) 
$ \\

\hline
\end{tabular}
\end{center}
 \caption{Critical forces $f_L$ for knots with eight crossings.  The last
column is the maximum amount of useful work which can be extracted if the
lattice knot expands against the hard walls of the slab, pushing them
apart from $L=1$.  In all these cases we set $T=1$.}
  \label{Forces8}  
\end{table}
\end{landscape}

\section{A grafted lattice knot pushed by a force}

In this section we consider the model inspired by figure \ref{FIG2}:
An external force $f$ compresses a grafted lattice knot onto a hard wall.
If $p_n(h,K)$ is the number of such lattice knots with highest vertices 
at height $h$ above the bottom wall and length $n_{h;K}$, then the 
partition function is given by
\begin{equation}
Z (f) = \sum_{n,h=0}^\infty p_{n_{h;K}} (h,K) e^{-E_h-fh}
\end{equation}
where $n_{h;K} = \min_{L\leq h}\{ n_{L,K}\}$ is the minimal length of the
polygons in $\Slab_L$ for all $L\leq h$. For example, $n_{1;3_1^+} = 26$
and $n_{h;3_1^+} = 24$ for all $h>1$.

In the partition function, positive values of $f$ mean that the vertices
in the top layer are pushed towards the bottom wall, and negative
values $f$ mean that the force is pulling the vertices in the 
top layer from the bottom wall.  Observe that we put the Boltzman factor 
$k_B T=1$ in this definition, and so use lattice units throughout.

The function $E_h$ is an energy of the lattice knot.  We shall again
use a Hooke energy for the polygons, namely
\begin{equation}
E_h = k (n_{h;K} - n_K)^2 .
\end{equation}
In addition, we assume a low temperature approximation, namely that
only polygons of the shortest length in $\Slab_L$ contribute to $Z(f)$
for all $L$. That is, in the case of $3_1^+$, only polygons of lengths
$24$ and $26$ (when $h=1$) contribute.  This approximation is also
valid in the regime that the Hooke constant $k$ is large. 

In other word, the low temperature and large Hooke constant
approximation of the partition function is
\begin{equation}
Z^* (f) = \sum_{h=0}^\infty p_{n_{h,K}} (h,K) e^{-k(n_{h,K} - n_K)^2 - fh}.
\end{equation}
For $f<0$ we observe that the polygon is pulled by its highest
vertices from the bottom wall, and that it will stretch in length if the
forces overcome the tensile strength of the edges.  Stretching the
edges in this way will cause the Hooke energy term to increase quadratically
in $n \propto h$, while the force $f$ couples only linearly in with $h$.  Thus,
this regime will be a purely Hooke regime, provided that $k$ is large enough.

Thus, we obtain a model of fixed length lattice knots, which may stretch
to longer states if pushed against the bottom wall by large forces to
accommodate itself into a conformation with small $h$.

The (extensive) free energy in this model is given by
\begin{equation}
\C{F}_f = \log Z^* (f)
\end{equation}
and its derivatives give the thermodynamic observables of the model.
For example, the mean height of the grafted lattice knot is
\begin{equation}
\LA h \RA_K = - \frac{d\C{F}_f}{df} 
= \frac{\sum_{h=0}^\infty h\, p_n(K;h) e^{-k(n_{h,K} - n_K)^2-fh}}{Z^* (f)},
\end{equation}
while the second derivative
\begin{equation}
\kappa_{K} = - \frac{d\log \LA h \RA_K}{df} 
= - \frac{1}{\LA h \RA_K} \frac{d \LA h \RA_K}{df} 
\end{equation}
is the fractional rate of change in mean height with $f$, and is a 
measure of the linear compressibility of the lattice
knot due to a force acting on its highest vertices.

\begin{figure}[t!]
\centering
\input{Compressgraphs/Compress0-1.tex}
\caption{The mean height $\LA h \RA_{0_1}$ (red curve) and compressibility
$\kappa_{0_1}$ (blue curve) as a function of $f$ for the unknot $0_1$.  
For negative values of $f$ (pulling forces) the mean height is $1$, 
and as the lattice knot is compressed into pushing forces, 
its mean height decreases until it approaches zero.  Note that
$\LA h \RA = 2/3$ if $f=0$.}
\label{FIG9} 
\end{figure}
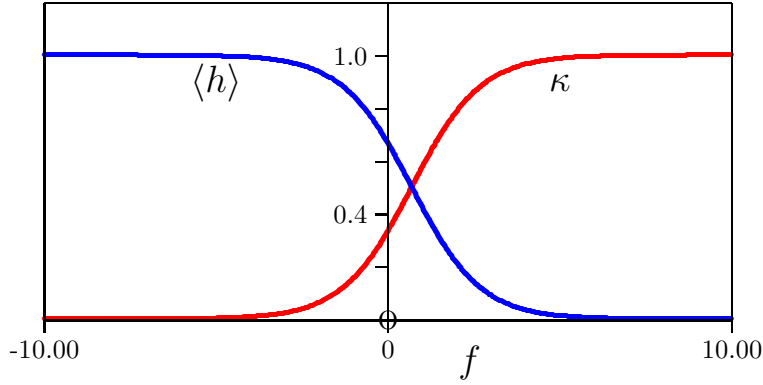

The data in tables \ref{Lengths} and \ref{Entropy} can be used to determine
$\kappa_K$.  For example, for the unknot grafted to the bottom wall,
one has $n_{h;K} = n_{L,K}=4$ if $L=h=0$ or $L=h=1$. Thus, we determine
the partition function in this case to be
\begin{equation}
Z^*_{0_1} (f) = 1 + 2\,e^{-f}
\end{equation}
since $E_h=0$ for both $h=0$ and $h=1$ in this case.  Observe that
$p_4(0,0_1)=1$ and $p_4(1,0_1)=2$ in this model.

One may now compute the mean height and $\kappa_{0_1}$ for the unknot
directly from the above.  The results are
\begin{equation}
\LA h \RA_{0_1} = \frac{2\,e^{-f}}{1+2\,e^{-f}},\quad 
\kappa_{0_1} = \frac{1}{1+2\,e^{-f}}  
\end{equation}
and these are plotted in figure \ref{FIG9}. If this lattice knot 
is released from its maximum compressed state in a slab of width $L=0$, 
and allowed to expand, then work can be extracted from the expansion.  
The maximum amount of work which may be extracted is given by the 
free energy differences between the $L=0$ slab and the free energy 
at $f=0$.  This is given by
\begin{equation}
\C{W}_{0_1} = \C{F}_f \Vert_{f=0} = \C{F}_f \Vert_{L=0}
= \log 3 - \log 1 = \log 3.
\end{equation}
This is the same value obtained in the first model.

\subsection{Compressing minimal length lattice trefoil knots}

By noting that $n_{h;3_1^+}=24$ if $h\geq 2$ and $n_{1;3_1^+}=26$
in table \ref{Lengths}, one may determine $p_{n(h,3_1^+)} (h,3_1^+)$ by
examining the data in table \ref{Entropy}.

In particular, it is apparent that $p_{n(1,3_1^+)}(1,3_1^+) = 36$ and
$p_{n(2,3_1^+)}(2,3_1^+) = 152$.  However, these 152 lattice knots of length
$n=24$ in $\Slab_2$ are also counted in $\Slab_3$, and so must be 
subtracted from the data in column $L=3$ to obtain $p_{n(3,3_1^+)}(3,3_1^+)$.
In particular, it follows that $p_{n(3,3_1^+)}(3,3_1^+) = 1660-152=1508$.

Similarly, one may show that $p_{n(4;3_1^+)}(4,3_1^+) = 4$ 
and $p_{n(\geq5;3_1^+)}(\geq 5,3_1^+) = 0$.
\begin{figure}[t!]
\subfigure[The mean height of $3_1^+$]{
   \input{Compressgraphs/Compress3-1L.tex}
   \label{FIG10-1}
 }
 \subfigure[The compressibility of $3_1^+$]{
   \input{Compressgraphs/Compress3-1F.tex}
   \label{FIG10-2}
 }
\caption{The mean height $\LA h \RA_{3_1^+}$ and compressibility
$\kappa_{3_1^+}$ of grafted lattice knots of type $3_1^+$.  Negative
forces are pulling forces, stretching the lattice knot from the 
bottom plane.  Positive forces are pushing forces.  The mean height
decreases in steps from a maximum of about $4$ to about $3$
at $f=0$ and then to $1$ for large positive $f$.  There are two peaks
in $\kappa_{3_1^+}$.  The highest peak at positive $f$ corresponds
to the pushing force overcoming the Hooke term in $\C{F}$,
increasing the length of the lattice from $24$ to $26$ and pushing
it into a slab of width $L=1$.  In this example, $k=1/4$.}
\label{FIG10} 
\end{figure}
This shows that
\begin{equation}
Z^*(3_1^+) = 36\,e^{-4k-f} + 152\,e^{-2f} + 1508\,e^{-3f} + 4\,e^{-4f}.
\end{equation}
The mean height of the lattice knot is
\begin{equation}
\LA h \RA_{3_1^+} = 
{\frac {9\,{{\rm e}^{-4\,k}}+76\,{{\rm e}^{-\,f}}
+1131\,{{\rm e}^{-2\,f}}+4\,{{\rm e}^{-3\,f}}}{9\,{{\rm e}^{-4\,k}}
+38\,{{\rm e}^{-\,f}}+377\,{{\rm e}^{-2\,f}}+{{\rm e}^{-3\,f}}}}
\end{equation}
The (linear) compressibility of the lattice knot of type $3_1^+$ is a 
more complicated expression, given by
\begin{equation*}
\fl
\kappa_{3_1^+} =
{\frac {\L 342 +13572\,{{\rm e}^{-f}}+81\,{{\rm e}^{-2\,f}}\R e^{-4\,k-f}+\L 14326+152\,{{\rm e}^{-f}}+377\,{{\rm e}^{-2\,f}}\R 
{\rm e}^{-3\,f}}
{ \left( 9\,{{\rm e}^{-4\,k}}+76\,{{\rm e}^{-f}}+1131\,{{\rm e}^{-2\,f}}
+4\,{{\rm e}^{-3\,f}}
 \right)  \left( 9\,{{\rm e}^{-4\,k}}+38\,{
{\rm e}^{-f}}+377\,{{\rm e}^{-2\,f}}+{{\rm e}^{-3\,f}} \right) }} .
\end{equation*}
In figures \ref{FIG10} and \ref{FIG11} the mean height and $\kappa_{3_1^+}$
are plotted as a function of the force for $k=1/4$ (figure \ref{FIG10})
and $k=4/3$ (figure \ref{FIG11}).  In figure \ref{FIG10} the mean
height decreases to $L=1$ in two steps, the first at negative (pulling)
forces, and the second a step from a height of roughly $h=3$ to $h=1$.
In this step (which shows up as a peak in figure \ref{FIG10-2}) the
polygon is squeezed into $\Slab_1$ as the force overcomes both
the entropy reduction and Hooke term.

Increasing the Hooke constant $k$ produces graphs similar to figure
\ref{FIG11}.  There are now three peaks in $\kappa$, each corresponding
to a reduction of the lattice knot from height $h$ to height $h-1$ as
the applied force first overcomes entropy and then the
Hooke energy to push the lattice knot into $\Slab_1$.

The total amount of work done by letting the lattice polygon expand
at zero force from its maximal compressed state in $L=1$ is given by
$\C{W}_{3_1^+} = \C{F}_f\Vert_{L=1} - \C{F}_f \Vert_{f=0}$.
This gives
\begin{equation}
\C{W}_{3_1^+} =  \log \L 1664+36 e^{-4k} \R  -  \log \L 36 e^{-4k} \R = 
\log \L {\sfrac {416}{9}}\,{{\rm e}^{4\,k}}+1 \R .
\end{equation}

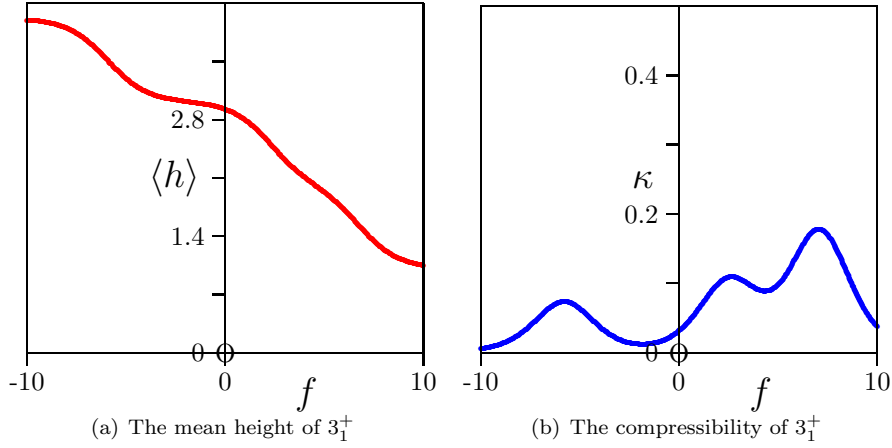
\begin{figure}[t!]
\subfigure[The mean height of $3_1^+$]{
   \input{Compressgraphs/Compress3-1LA.tex}
   \label{FIG11-1}
 }
 \subfigure[The compressibility of $3_1^+$]{
   \input{Compressgraphs/Compress3-1FA.tex}
   \label{FIG11-2}
 }
\caption{The similar plots to figure \ref{FIG10}, but with $k=4/3$.
This larger Hooke energy requires a larger force to push the
lattice knots from a $L=2$ slab into $\Slab_1$.  This shows
up as a third peak in the $\kappa$-graph above.}
\label{FIG11} 
\end{figure}

\subsection{Compressing minimal length lattice figure eight knots}

The minimal length of a lattice polygon of knot type $4_1$ (the
figure eight knot) is $36$ in $\Slab_1$, $32$ in $\Slab_2$ and
$30$ in $\Slab_L$ with $L\geq 3$.  In other words, states 
with heights $1$ or $2$ will have a Hooke energy, as they have been
stretched in length to squeeze into slabs with small height.

\begin{figure}[t!]
\subfigure[The mean height of $4_1$]{
   \input{Compressgraphs/Compress4-1L.tex}
   \label{FIG12-1}
 }
 \subfigure[The compressibility of $4_1$]{
   \input{Compressgraphs/Compress4-1F.tex}
   \label{FIG12-2}
 }
\caption{The mean height $\LA h \RA_{4_1}$ and compressibility
$\kappa_{4_1}$ of grafted lattice knots of type $4_1$.  Negative
forces are pulling forces, stretching the lattice knot from the 
bottom plane.  Positive forces are pushing forces.  The mean height
decreases in steps from a maximum of $4$ to about $2.8$
at $f=0$ and then to $1$ for large positive $f$.  There are two peaks
in $\kappa_{4_1}$.  Both peaks correspond to a reduction in the
slab width $L$ as the force first overcomes the Hooke term from
$L=3$ to $L=2$, and then again at larger pushing forces, the
Hooke term from $L=2$ to $L=1$. In this example, $k=1/4$.}
\label{FIG12} 
\end{figure}
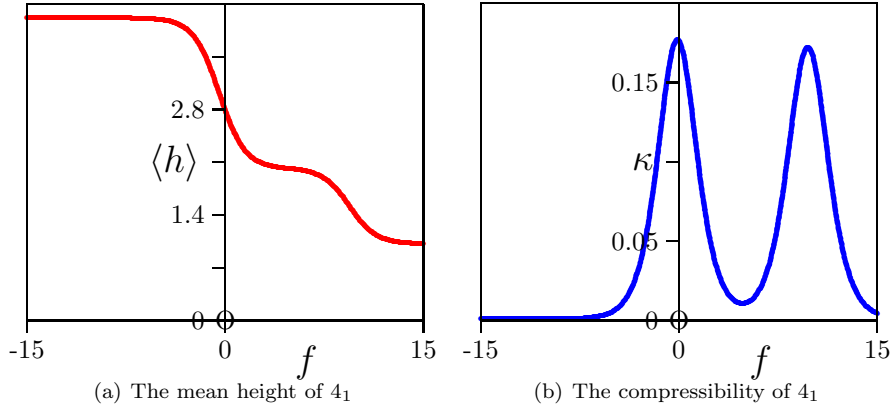

By consulting the data in tables \ref{Lengths}
and \ref{Entropy}, the partition function of this model
can be determined, and it is given by
\begin{equation}
Z^*(4_1) = 
1480\,{{\rm e}^{-f-36\,k}}+6064\,{{\rm e}^{-2\,f-4\,k}}+2720\,{{\rm e}
^{-3\,f}}+928\,{{\rm e}^{-4\,f}} .
\end{equation}
The mean height of this lattice knot is given by
\begin{equation}
\LA h \RA_{4_1} = 
{\frac {185\,{{\rm e}^{-36\,k}}+1516\,{{\rm e}^{-f-4\,k}}+1020\,{
{\rm e}^{-2\,f}}+464\,{{\rm e}^{-3\,f}}}{185\,{{\rm e}^{-36\,k}}+758\,
{{\rm e}^{-f-4\,k}}+340\,{{\rm e}^{-2\,f}}+116\,{{\rm e}^{-3\,f}}}}
\end{equation}
and it is plotted in figure \ref{FIG12-1} for $k=1/4$.  Observe that 
the mean height decreases in steps with increasing $f$, from height
$4$ to height $2$, before it is squeezed into $\Slab_1$ for sufficiently
large values of $f$. 

The compressibility is plotted in figure \ref{FIG12-2}, and the two
peaks correspond to the decreases in $\LA h\RA_{4_1}$ in steps.
At the peaks, the lattice knot has a maximum response to changes in $f$.
The expression for the compressibility is lengthy and will not be
reproduced here. 

Finally, the total amount of work done by releasing the knot from
$\Slab_1$ at zero pressure and letting it expand isothermically, is
given by
\begin{equation}
\C{W}_{4_1} = 
\log \left( {\sfrac {456}{185}}\,{{\rm e}^{36\,k}}+{\sfrac {758}{185}}\,
{{\rm e}^{32\,k}}+1 \right) 
\end{equation}

In the case that $k=0$, $\C{W}_{3_1^+} \approx 3.855
> 2.023 \approx \C{W}_{4_1}$.  In other words, more work is done by the
trefoil knot.  However, if $k=1/4$ (and a Hooke term is present),
then the relation is the opposite: 
$\C{W}_{3_1^+} \approx 4.841 < 10/379 \approx \C{W}_{4_1}$.
Equality is obtained when $k=0.06588\ldots$.

\subsection{Compressing minimal length lattice knots of 
types $5_1^+$ and $5_2^+$}

The partition functions of minimal lattice knots of types $5_1^+$
and $5_2^+$ are given by
\begin{eqnarray}
\fl
Z^*(5_1^+) &
= 72\,{{\rm e}^{-f-16\,k}}+760\,{{\rm e}^{-2\,f}}
+600\,{{\rm e}^{-3\,f}}+1936\,{{\rm e}^{-4\,f}}+40\,{{\rm e}^{-5\,f}}; \cr
\fl
Z^*(5_2^+) &
= 6240\,{{\rm e}^{-f-36\,k}}+720\,{{\rm e}^{-2\,f}}
+24072\,{{\rm e}^{-3\,f}}+30544\,{{\rm e}^{-4\,f}}+2120\,{{\rm e}^{-5\,f}}.
\end{eqnarray}
These expressions show that the maximum height is $5$ while there are 
Hooke terms for the transition from $h=2$ to $h=1$.

The mean heights can be computed, and are given by
\begin{eqnarray}
\fl 
\LA h \RA_{5_1^+} &
= {\frac {9\,{{\rm e}^{-16\,k}}+190\,{{\rm e}^{-f}}+225\,{{\rm e}^{-2\,f}}
+968\,{{\rm e}^{-3\,f}}+25\,{{\rm e}^{-4\,f}}}
{9\,{{\rm e}^{-16\,k}}+95\,{{\rm e}^{-f}}+75\,{{\rm e}^{-2\,f}}
+242\,{{\rm e}^{-3\,f}}+5\,{{\rm e}^{-4\,f}}}}
 , \cr
\fl
\LA h \RA_{5_2^+} &
= {\frac {780\,{{\rm e}^{-36\,k}}+180\,{{\rm e}^{-f}}+9027\,{{\rm e}^{-2\,f}}
+15272\,{{\rm e}^{-3\,f}}+1325\,{{\rm e}^{-4\,f}}}
{780\,{{\rm e}^{-36\,k}}+90\,{{\rm e}^{-f}}+3009\,{{\rm e}^{-2\,f}}
+3818\,{{\rm e}^{-3\,f}}+265\,{{\rm e}^{-4\,f}}}} .
\end{eqnarray}

$\C{W}_K$ are similar given by
\begin{equation}
\C{W}_{5_1^+}
= \log \left( {\sfrac {139}{3}}\,{{\rm e}^{16\,k}}+1 \right) 
\q\hbox{and}\q
\C{W}_{5_2^+}
= \log \left( {\sfrac {1197}{130}}\,{{\rm e}^{36\,k}}+1 \right) .
\end{equation}

\begin{figure}[t!]
\subfigure[The mean height of $5$-crossing knots]{
   \input{Compressgraphs/Compress5-12L.tex}
   \label{FIG13-1}
 }
 \subfigure[The compressibility of $5$-crossing knots]{
   \input{Compressgraphs/Compress5-12F.tex}
   \label{FIG13-2}
 }
\caption{The mean height $\LA h \RA_{5_*}$ and compressibility
$\kappa_{5_*}$ of grafted lattice knots of types $5_1^+$ (red curves)
and $5_2^+$ (blue curves).  The mean height decreases in steps from a 
maximum of $5$ to $1$ for large positive $f$. The transition for
$5_2^+$ is smoother than for $5_1^+$, which exhibits peaks in $\kappa$
corresponding to critical forces overcoming the Hooke terms and
entropy.   There is one low and are two prominent peaks
in $\kappa_{5_1^+}$.  In this example, $k=1/4$.}
\label{FIG13} 
\end{figure}
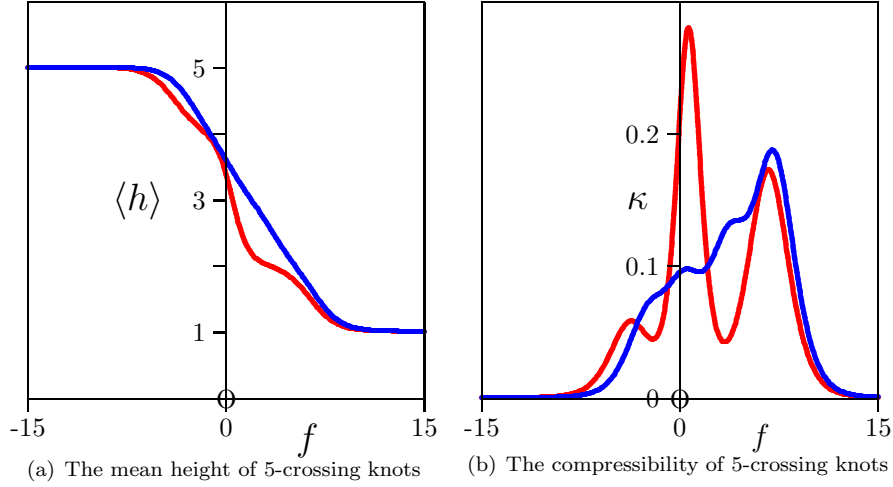
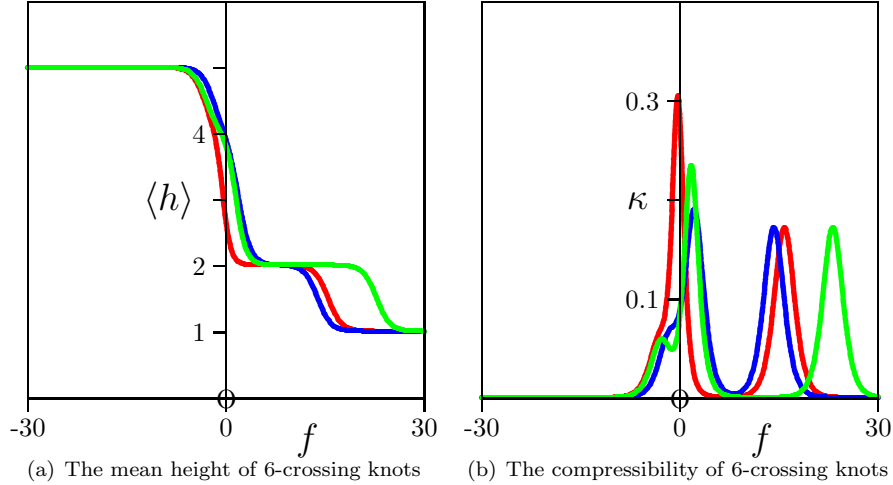
\begin{figure}[b!]
\subfigure[The mean height of $6$-crossing knots]{
   \input{Compressgraphs/Compress6-123L.tex}
   \label{FIG14-1}
 }
 \subfigure[The compressibility of $6$-crossing knots]{
   \input{Compressgraphs/Compress6-123F.tex}
   \label{FIG14-2}
 }
\caption{The mean height $\LA h \RA_{6_*}$ and compressibility
$\kappa_{6_*}$ of grafted lattice knots of types $6_1^+$ (red curves), 
$6_2^+$ (blue curves) and $6_3$ (green curves).  The mean height 
decreases in steps from a maximum of $5$ to $1$ for large positive $f$. 
These knot types follow a very similar pattern, with $6_3$ posing
the most resistance to compression.  Each knot also exhibits two
peaks in $\kappa$ of similar height. In this example, $k=1/4$.}
\label{FIG14} 
\end{figure}

The results for the $5$-crossing knots are plotted in figure \ref{FIG13}.
The curves for the mean height $\LA h \RA$ are very similar , with
$5_2^+$ undergoing a smoother compression with increasing
$f$.  The knot type $5_1^+$ shows more variation, and this is very
visible in the plots for $\kappa$ in figure \ref{FIG13-2}, where there
are several sharp peaks for $5_1^+$, but less pronounced changes
for $5_2^+$.  In both knot types, the Hooke constant is $k=1/4$.

\subsection{Compressing minimal length lattice knots of 
types $6_1^+$, $6_2^+$ and $6_3$}

The results for $6$-crossing knot types are displayed in figure \ref{FIG14}.  
These knot types exhibit more similar behaviour than the two 
$5$-crossing knot types.


\begin{table}[t!]
\begin{center}
 \begin{tabular}{||c||l||}
  \hline  
Knot & Work \\
   \hline
{$0_1$}   &\fns $\log 3$ \\
{$3_1^+$} &\fns $\log  \left( {\sfrac {416}{9}}\,{{\rm e}^{4\,k}}+1 \right) $ \\
{$4_1$}   &\fns $\log  \left( {\sfrac {456}{185}}\,{{\rm e}^{36\,k}}+{\sfrac {758}{185}}\,{{\rm e}^{32\,k}}+1 \right) $ \\
{$5_1^+$} &\fns $\log  \left( {\sfrac {139}{3}}\,{{\rm e}^{16\,k}}+1 \right) $ \\
{$5_2^+$} &\fns $\log  \left( {\sfrac {1197}{130}}\,{{\rm e}^{36\,k}}+1 \right) $ \\
{$6_1^+$} &\fns $\log  \left( {\sfrac {768}{2009}}\,{{\rm e}^{64\,k}}+{\sfrac {3176}{2009}}\,{{\rm e}^{60\,k}}+1 \right) $ \\
{$6_2^+$} &\fns $\log  \left( {\sfrac {171}{23}}\,{{\rm e}^{64\,k}}+{\sfrac {15}{46}}\,{{\rm e}^{60\,k}}+1 \right) $ \\
{$6_3$}   &\fns $\log  \left( {\sfrac {37}{13}}\,{{\rm e}^{100\,k}}+{\sfrac {157}{26}}\,{{\rm e}^{84\,k}}+1 \right) $ \\
{$7_1^+$} &\fns $\log  \left( {\sfrac {1415}{9}}\,{{\rm e}^{36\,k}}+1 \right) $ \\
{$7_2^+$} &\fns $\log  \left( {\sfrac {14015}{1869}}\,{{\rm e}^{64\,k}}+{\sfrac {62507}{5607}}\,{{\rm e}^{60\,k}}+1 \right) $ \\
{$7_3^+$} &\fns $\log  \left( {\sfrac {5}{31}}\,{{\rm e}^{100\,k}}+{\sfrac {1202}{31}}\,{{\rm e}^{84\,k}}+1 \right) $ \\
{$7_4^+$} &\fns $\log  \left( {\sfrac {21}{3361}}\,{{\rm e}^{100\,k}}+{\sfrac {100}{3361}}\,{{\rm e}^{96\,k}}+1 \right) $ \\
{$7_5^+$} &\fns $\log  \left( {\sfrac {197}{230}}\,{{\rm e}^{100\,k}}+{\sfrac {52}{115}}\,{{\rm e}^{96\,k}}+1 \right) $ \\
{$7_6^+$} &\fns $\log  \left( {\sfrac {2127}{731}}\,{{\rm e}^{100\,k}}+{\sfrac {220}{731}}\,{{\rm e}^{96\,k}}+1 \right) $ \\
{$7_7^+$} &\fns $\log  \left( 63\,{{\rm e}^{144\,k}}+322\,{{\rm e}^{140\,k}}+2248\,{{\rm e}^{108\,k}}+1 \right) $ \\
{$8_1^+$} &\fns $\log  \left( {\sfrac {989}{3575}}\,{{\rm e}^{100\,k}}+{\sfrac {254}{2145}}\,{{\rm e}^{96\,k}}+1 \right) $ \\
{$8_2^+$} &\fns $\log  \left( {\sfrac {11460}{649}}\,{{\rm e}^{100\,k}}+{\sfrac {73714}{649}}\,{{\rm e}^{84\,k}}+1 \right) $ \\
{$8_3$} &\fns $\log  \left( {\sfrac {3}{2273}}\,{{\rm e}^{144\,k}}+{\sfrac {500}{2273}}\,{{\rm e}^{128\,k}}+1 \right) $ \\
{$8_4^+$} &\fns $\log  \left( {\sfrac {1994}{1683}}\,{{\rm e}^{100\,k}}+{\sfrac {140}{297}}\,{{\rm e}^{96\,k}}+1 \right) $ \\
{$8_5^+$} &\fns $\log  \left( 8\,{{\rm e}^{100\,k}}+{\sfrac {206}{3}}\,{{\rm e}^{96\,k}}+{\sfrac {102047}{18}}\,{{\rm e}^{64\,k}}+1 \right) $ \\
{$8_6^+$} &\fns $\log  \left( {\sfrac {138}{241}}\,{{\rm e}^{144\,k}}+{\sfrac {1}{1205}}\,{{\rm e}^{140\,k}}+1 \right) $ \\
{$8_7^+$} &\fns $\log  \left( {\sfrac {3}{2429}}\,{{\rm e}^{196\,k}}+{\sfrac {6172}{2429}}\,{{\rm e}^{160\,k}}+1 \right) $ \\
{$8_8^+$} &\fns $\log  \left( {\sfrac {195}{1696}}\,{{\rm e}^{144\,k}}+{\sfrac {95}{32}}\,{{\rm e}^{140\,k}}+{\sfrac {347}{1696}}\,{{\rm e}^{128\,k}}+1 \right) $ \\
{$8_9^+$} &\fns $\log  \left( {\sfrac {245}{106}}\,{{\rm e}^{144\,k}}+{\sfrac {355}{318}}\,{{\rm e}^{128\,k}}+1 \right)$ \\
{$8_{10}^+$} &\fns $\log  \left( {\sfrac {105}{26}}\,{{\rm e}^{144\,k}}+{\sfrac {51}{2}}\,{{\rm e}^{140\,k}}+{\sfrac {2982}{13}}\,{{\rm e}^{108\,k}}+1 \right) $ \\
{$8_{11}^+$} &\fns $\log  \left( {\sfrac {3}{122}}\,{{\rm e}^{144\,k}}+{\sfrac {2533}{244}}\,{{\rm e}^{140\,k}}+{\sfrac {453}{488}}\,{{\rm e}^{128\,k}}+1 \right) $ \\
{$8_{12}^+$} &\fns $\log  \left( {\sfrac {54}{305}}\,{{\rm e}^{144\,k}}+{\sfrac {2897}{183}}\,{{\rm e}^{128\,k}}+1 \right) $ \\
{$8_{13}^+$} &\fns $\log  \left( 1632\,{{\rm e}^{144\,k}}+190\,{{\rm e}^{128\,k}}+1 \right) $ \\
{$8_{14}^+$} &\fns $\log  \left( {\sfrac {9}{1475}}\,{{\rm e}^{196\,k}}+{\sfrac {154}{1475}}\,{{\rm e}^{192\,k}}+{\sfrac {172}{1475}}\,{{\rm e}^{180\,k}}+1 \right) $ \\
{$8_{15}^+$} &\fns $\log  \left( {\sfrac {5013}{2}}\,{{\rm e}^{100\,k}}+55\,{{\rm e}^{84\,k}}+1 \right) $ \\
{$8_{16}^+$} &\fns $\log  \left( {\sfrac {3}{2045}}\,{{\rm e}^{256\,k}}+{\sfrac {3}{2045}}\,{{\rm e}^{252\,k}}+{\sfrac {73}{2045}}\,{{\rm e}^{220\,k}}+1 \right) $ \\
{$8_{17}^+$} &\fns $\log  \left( {\sfrac {1108}{1257}}\,{{\rm e}^{256\,k}}+{\sfrac {539}{1257}}\,{{\rm e}^{252\,k}}+{\sfrac {518}{3771}}\,{{\rm e}^{220\,k}}+1 \right) $ \\
{$8_{18}^+$} &\fns $\log  \left( 111\,{{\rm e}^{324\,k}}+2125\,{{\rm e}^{308\,k}}+{\sfrac {189}{8}}\,{{\rm e}^{260\,k}}+1 \right) $ \\
{$8_{19}^+$} &\fns $\log  \left( {\sfrac {6996}{163}}\,{{\rm e}^{36\,k}}+{\sfrac {1656}{163}}\,{{\rm e}^{32\,k}}+1 \right) $ \\
{$8_{20}^+$} &\fns $\log  \left( {\sfrac {30}{29}}\,{{\rm e}^{64\,k}}+{\sfrac {318}{29}}\,{{\rm e}^{60\,k}}+1 \right) $ \\
{$8_{21}^+$} &\fns $\log  \left( {\sfrac {2335}{2}}\,{{\rm e}^{64\,k}}+30\,{{\rm e}^{60\,k}}+1 \right) $ \\
\hline
\end{tabular}
\end{center}
 \caption{The work done by releasing the lattice knot in $\Slab_1$ and letting
it expand to equilibrium at zero pressure.}
  \label{Work-Mod2}  
\end{table}

\begin{table}[t!]
\begin{center}
 \begin{tabular}{||c||l||}
  \hline  
Knot & Work \\
   \hline
{$3_1^+\#3_1^+$} &\fns $\log  \left( {\sfrac {2552}{405}}\,{{\rm e}^{64\,k}}+1 \right) $ \\
{$3_1^+\#3_1^-$} &\fns $\log  \left( 760\,{{\rm e}^{16\,k}}+1 \right) $ \\
{$3_1^+\#4_1$} &\fns $\log  \left( {\sfrac {7494}{503}}\,{{\rm e}^{64\,k}}+{\sfrac {3434}{1509}}\,{{\rm e}^{60\,k}}+1 \right) $ \\
{$3_1^+\#5_1^+$} &\fns $\log  \left( {\sfrac {4187}{9}}\,{{\rm e}^{64\,k}}+1 \right) $ \\
{$3_1^+\#5_1^-$} &\fns $\log  \left( {\sfrac {11849}{12}}\,{{\rm e}^{36\,k}}+1 \right) $ \\
{$3_1^+\#5_2^+$} &\fns $\log  \left( {\sfrac {459813}{18802}}\,{{\rm e}^{64\,k}}+1 \right) $ \\
{$3_1^+\#5_2^-$} &\fns $\log  \left( {\sfrac {55}{287}}\,{{\rm e}^{100\,k}}+{\sfrac {1}{246}}\,{{\rm e}^{96\,k}}+1 \right) $ \\
{$4_1\#4_1$} &\fns $\log  \left( {\sfrac {41853}{100}}\,{{\rm e}^{64\,k}}+{\sfrac {5477}{100}}\,{{\rm e}^{60\,k}}+1 \right) $ \\
\hline
\end{tabular}
\end{center}
 \caption{The work done by releasing the lattice knot in $\Slab_1$ and letting
it expand to equilibrium at zero pressure.}
  \label{Work-Mod2A}  
\end{table}

\subsection{Discussion}

The compression of lattice knots typically show a few peaks in the
compressibility.  Since the compressibility is the defined as the
fractional change in the width of the lattice knot with incrementing
force, a high compressibility corresponds to a ``soft" lattice knot,
and a low compressibility to a ``hard" lattice knot.  The peaks observed
in figures from figure \ref{FIG10} to figure \ref{FIG14} correspond
to values of $f$ where the lattice knots are soft.  

Our general observation is that the compressibility of lattice
knots in this model is not monotonic:  That is, the lattice knot
does not become increasingly more resistant to further compression 
with increasing force.  Instead, it may alternately become more
or less resistant to compression, as the compressibility  goes
through peaks (``soft regimes") and troughs (``hard regimes").

The work done by releasing the lattice knot in $\Slab_1$ and letting
it expand at zero pressure was computed above for knots up to $5$-crossings.
For example, if the lattice unknot is placed in $\Slab_0$ and allowed to expand 
isothermically to equilibrium at $f=0$, then $\C{W}_{0_1}=\log 3$.

A similar calculation gave $\C{W}_{3_1^+} = \log\L \sfrac{416}{9} e^{4k} + 1\R$
for expansion from $\Slab_1$, and expressions were also determined for other 
knot types.  The results are given in tables \ref{Work-Mod2} and \ref{Work-Mod2A} 
for the remaining knot types.

Observe that these expressions are different from those in the model in 
section 3:  There the data were obtained by computing the sum over all 
the critical forces.  In each case the minimal values of the forces required 
to overcome resistance by the squeezed polygon were determined.  
This is a different process to the above, where the lattice knots are placed 
in $\Slab_1$ and then allowed to expand isothermically to equilibrium.  The
total work was obtained by computing the free energy difference between
the initial and final states.

\begin{figure}[t!]
\input{Compressgraphs/ScatterW.tex}
\caption{A scatter plot of $\C{W}$ with $k=1/4$ (horizontal axis)
and $k=1$, (vertical axis).  The data for prime knot types accumulate
along a band, indicating that the Hooke term makes a significant contribution
to the free energy.}
\label{FIG15} 
\end{figure}
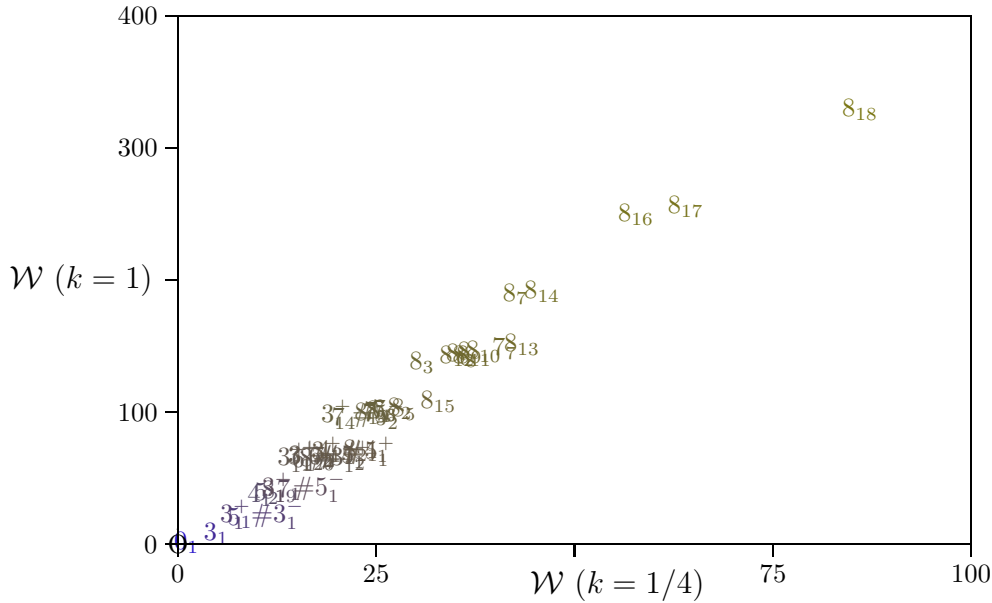

Finally, we examined the effect of the Hooke parameter $k$ by comparing
$\C{W}$ for two different values of $k$, namely $k=1/4$ and $k=1$.  For
each knot type, the point $(\C{W}_{k=1/4},\C{W}_{k=1})$ is plotted as a
scatter plot in figure \ref{FIG15}. These points line up along a band,
indicating that the Hooke term makes a significant contribution to the
free energy.

\section{Conclusions}

In this paper we presented data on minimal length lattice polygons
grafted to the bottom walls of slabs $\Slab_L$.  We collected data on 
these polygons by implementing the GAS-algorithm for knotted lattice polygons 
in a slab in the lattice, and sieving out minimal length polygons.
Our data were presented in tables \ref{Lengths} and \ref{Entropy}
for prime knot types, and tables \ref{Lengths-C} and \ref{Entropy-C}
for compound knot types.  Our results verify, up to single exceptions,
the data obtained in reference \cite{ISDAVS12}.

We determined the compressibility properties of minimal lattice
polygons in two ensembles.  The first was a model of a ring polymer
grafted to the bottom wall of a slab with hard walls, and squeezed by
the top wall into a narrower slab.

The second was a model of a ring polymer grafted to a hard wall,
and then pushed by a force $f$ towards the bottom wall.

In these models we determined free energies, from which thermodynamic
quantities were obtained.  In the first model we computed
the critical forces which squeeze the lattice polygons into 
narrower slabs, and we computed the total amount of work that would be
done if the lattice polygon was squeezed into $\Slab_1$.  These
data are displayed in tables \ref{Forces7} and \ref{Forces8}.

The profiles of critical forces in this model are dependent on knot 
types, and generally increases with decreasing values of $L$.  
The dependency on knot type shows that entanglements in the
lattice knot, in addition to the Hooke term and the entropy, plays
a role in determining the resistance of the lattice knot to be
squeezed in ever thinner slabs.

We examined the mean height and compressibility profiles
for some simple knot types in section 4.  The amount of work done by
expanding expanding lattice knots from maximal compression at zero force was
also determined, and the results were listed in table \ref{Work-Mod2}
and \ref{Work-Mod2A}.

In contrast with the results in section 3, we computed the compressibility
of lattice knots as a function of the applied force in section 4.
Our results show that the compressibility is dependent
on the knot type and the Hooke energy and is not a monotonic
function of the applied force.  In many cases there are peaks in
the compressibility where the lattice knot becomes (relatively) 
softer (more compressible) with increasing force.
This may feature may be dependent on either the entanglements, or the
loss of entropy, or the Hooke term, or on a combination of these. These
observations indicate that such peaks may be seen in some conditions
when knotted ring polymers are compressed.

Finally, the inclusion of other energy terms, for example a bending
term, or a binding term to to the walls of the slab, can be done.
We expect such additional terms to have an effect on our results.  For
example, a bending energy will make the lattice knot more rigid and
less compressible, increasing the critical forces computed in 
section 3, and reducing the heights of the peaks seen in the
compressibility in section 4.  A further generalisation would
be to simulate lattice knots in other confined spaces, for example
in pores or in channels.  Such models pose unique questions, since
the algorithm is not known to be irreducible.

\vspace{1cm}
\section*{Acknowledgements}
EJJvR and AR acknowledge support in the form of NSERC Discovery Grants from
the Government of Canada.

\vspace{1cm}
\section*{Bibliography}
\bibliographystyle{plain}
\bibliography{table-slab-forces}

\vspace{1cm}

\section*{Appendix: Numerical Results}

Our raw data on the estimates of the exact minimal lengths and number
of minimal length lattice knots in $\Slab_L$ are displayed in tables
\ref{Lenghts}, \ref{Lengths-C}, \ref{Entropy} and \ref{Entropy-C}.

Data on the symmetry classes of lattice knots confined to slabs $\Slab_L$.
Data are presented in the format $A^aB^bC^c\ldots$ denoting $a$ symmetry
classes of polygons, each with $A$ members, $b$ symmetry classes of polygons,
each with $B$ members, and so on. 

\begin{table}[t!]
\begin{center}
 \begin{tabular}{||c||c|c|c|c|c|c|c|c||}
 \hline
  Knot & \multicolumn{8}{|c||}{$n_{L,K}$} \\
  \hline  
     & $L=1$ & $L=2$ & $L=3$ & $L=4$ & $L=5$ & $L=6$ & $L=7$ & $L=8$ \\
   \hline
$0_1$ & $4$ & $4$ & $4$ & $4$ & $4$ & $4$ & $4$ & $4$ \\
$3_1^+$ & \U{$26$} & $24$ & $24$ & $24$ & $24$& $24$& $24$& $24$ \\
$4_1$ & \U{$36$} & \U{$32$} & $30$ & $30$ & $30$ & $30$ & $30$ & $30$ \\
$5_1^+$ & \U{$38$} & $34$ & $34$ & $34$ & $34$ & $34$ & $34$ & $34$ \\
$5_2^+$ & \U{$42$} & $36$ & $36$ & $36$ & $36$ & $36$ & $36$ & $36$ \\
$6_1^+$ & \U{$48$} & \U{$42$} & $40$ & $40$ & $40$ & $40$ & $40$ & $40$ \\
$6_2^+$ & \U{$48$} & \U{$42$} & $40$ & $40$ & $40$ & $40$ & $40$ & $40$ \\
$6_3$ & \U{$50$} & \U{$44$} & $40$ & $40$ & $40$ & $40$ & $40$ & $40$ \\
$7_1^+$ & \U{$50$} & \U{$44$} & $44$ & $44$ & $44$ & $44$ & $44$ & $44$ \\
$7_2^+$ & \U{$54$} & \U{$48$} & $46$ & $46$ & $46$ & $46$ & $46$ & $46$ \\
$7_3^+$ & \U{$54$} & \U{$48$} & $44$ & $44$ & $44$ & $44$ & $44$ & $44$ \\
$7_4^+$ & \U{$54$} & \U{$46$} & $44$ & $44$ & $44$ & $44$ & $44$ & $44$ \\
$7_5^+$ & \U{$56$} & \U{$48$} & $46$ & $46$ & $46$ & $46$ & $46$ & $46$ \\
$7_6^+$ & \U{$56$} & \U{$48$} & $46$ & $46$ & $46$ & $46$ & $46$ & $46$ \\
$7_7^+$ & \U{$56$} & \U{$50$} & \U{$46$} & $44$ & $44$ & $44$ & $44$ & $44$ \\
$8_1^+$ & \U{$60$} & \U{$52$} & $50$ & $50$ & $50$ & $50$ & $50$ & $50$ \\
$8_2^+$ & \U{$60$} & \U{$54$} & $50$ & $50$ & $50$ & $50$ & $50$ & $50$ \\
$8_3$ & \U{$60$} & \U{$52$} & $48$ & $48$ & $48$ & $48$ & $48$ & $48$ \\
$8_4^+$ & \U{$60$} & \U{$52$} & $50$ & $50$ & $50$ & $50$ & $50$ & $50$ \\
$8_5^+$ & \U{$60$} & \U{$56$} & \U{$52$} & $50$ & $50$ & $50$ & $50$ & $50$ \\
$8_6^+$ & \U{$62$} & \U{$52$} & \U{$52$} & $50$ & $50$ & $50$ & $50$ & $50$ \\
$8_7^+$ & \U{$62$} & \U{$54$} & $48$ & $48$ & $48$ & $48$ & $48$ & $48$ \\
$8_8^+$ & \U{$62$} & \U{$54$} & \U{$52$} & $50$ & $50$ & $50$ & $50$ & $50$ \\
$8_9$ & \U{$62$} & \U{$54$} & $50$ & $50$ & $50$ & $50$ & $50$ & $50$ \\
$8_{10}^+$ & \U{$62$} & \U{$56$} & \U{$52$} & $50$ & $50$ & $50$ & $50$ & $50$ \\
$8_{11}^+$ & \U{$62$} & \U{$54$} & \U{$52$} & $50$ & $50$ & $50$ & $50$ & $50$ \\
$8_{12}$ & \U{$64$} & \U{$56$} & $52$ & $52$ & $52$ & $52$ & $52$ & $52$ \\
$8_{13}^+$ & \U{$64$} & \U{$54$} & $50$ & $50$ & $50$ & $50$ & $50$ & $50$ \\
$8_{14}^+$ & \U{$64$} & \U{$54$} & \U{$52$} & $50$ & $50$ & $50$ & $50$ & $50$ \\
$8_{15}^+$ & \U{$62$} & \U{$56$} & $52$ & $52$ & $52$ & $52$ & $52$ & $52$ \\
$8_{16}^+$ & \U{$66$} & \U{$56$} & \U{$52$} & $50$ & $50$ & $50$ & $50$ & $50$ \\
$8_{17}$ & \U{$68$} & \U{$58$} & \U{$54$} & $52$ & $52$ & $52$ & $52$ & $52$ \\
$8_{18}$ & \U{$70$} & \U{$60$} & \U{$56$} & $52$ & $52$ & $52$ & $52$ & $52$ \\
$8_{19}^+$ & \U{$48$} & \U{$44$} & $42$ & $42$ & $42$ & $42$ & $42$ & $42$ \\
$8_{20}^+$ & \U{$52$} & \U{$46$} & $44$ & $44$ & $44$ & $44$ & $44$ & $44$ \\
$8_{21}^+$ & \U{$54$} & \U{$48$} & $46$ & $46$ & $46$ & $46$ & $46$ & $46$ \\
\hline
 \end{tabular}
\end{center}
 \caption{The minimal length of lattice knots of prime knot type to eight
crossings confined to slabs of width $L$.
Cases with an increase in minimal length are underlined.}
  \label{Lengths} 
\end{table}

\begin{table}[t!]
\begin{center}
 \begin{tabular}{||c||c|c|c|c|c|c|c|c||}
 \hline
  Knot & \multicolumn{8}{|c||}{$p_{n_{L,K}} (K)$} \\
  \hline  
     & $L=1$ & $L=2$ & $L=3$ & $L=4$ & $L=5$ & $L=6$ & $L=7$ & $L=8$ \\
\hline
$0_1$ & $3$ & $3$ & $3$ & $3$ & $3$ & $3$ & $3$ & $3$ \\
$3_1^+$ & $36$ & $152$ & $1660$ & $1664$ & $1664$ & $1664$ & $1664$ & $1664$ \\
$4_1$ & $1480$ & $6064$ & $2720$ & $3648$ & $3648$& $3648$& $3648$& $3648$ \\
$5_1^+$ & $72$ & $760$ & $1360$ & $3296$ & $3336$ & $3336$& $3336$& $3336$ \\
$5_2^+$ & $6240$ & $720$ & $24792$ & $55336$ & $57456$ & $57456$ & $57456$& $57456$ \\
$6_1^+$ & $8036$ & $12704$ & $1168$ & $3000$ & $3072$ & $3072$ & $3072$& $3072$ \\
$6_2^+$ & $2208$ & $720$ & $2984$ & $14632$ & $16416$ & $16416$& $16416$& $16416$ \\
$6_3$ & $1248$ & $7536$ & $592$ & $3408$ & $3552$ & $3552$& $3552$& $3552$ \\
$7_1^+$ & $108$ & $1200$ & $5668$ & $11560$ & $16880$ & $16980$ & $16980$& $16980$ \\
$7_2^+$ & $22428$ & $250028$ & $51696$ & $123008$ & $164276$ & $168180$ & $168180$ & $168180$ \\
$7_3^+$ & $1488$ & $57696$ & $160$ & $160$ & $240$ & $240$ & $240$ & $240$ \\
$7_4^+$ & $13444$ & $400$ & $20$ & $64$ & $84$ & $84$ & $84$ & $84$ \\
$7_5^+$ & $5520$ & $2496$ & $120$ & $4648$ & $4728$ & $4728$ & $4728$ & $4728$ \\
$7_6^+$ & $5848$ & $1760$ & $640$ & $12920$ & $17016$ & $17016$ & $17016$ & $17016$ \\
$7_7^+$ & $4$ & $8992$ & $1288$ & $196$ & $252$ & $252$ & $252$ & $252$ \\
$8_1^+$ & $42900$ & $5080$ & $2192$ & $8024$ & $11148$ & $11868$ & $11868$ & $11868$ \\
$8_2^+$ & $2596$ & $294856$ & $10328$ & $25984$ & $42104$ & $45840$ & $45840$ & $45840$ \\
$8_3$ & $9092$ & $2000$ & $8$ & $8$ & $12$ & $12$ & $12$ & $12$ \\
$8_4^+$ & $20196$ & $9520$ & $1840$ & $14928$ & $23464$ & $23928$ & $23928$ & $23928$ \\
$8_5^+$ & $72$ & $408180$ & $4944$ & $384$ & $384$ & $384$ & $384$ & $384$ \\
$8_6^+$ & $9640$ & $8$ & $24184$ & $3680$ & $5520$ & $5520$ & $5520$ & $5520$ \\
$8_7^+$ & $19432$ & $49376$ & $8$ & $16$ & $24$ & $24$ & $24$ & $24$ \\
$8_8^+$ & $13568$ & $2776$ & $40280$ & $1040$ & $1560$ & $1560$ & $1560$ & $1560$ \\
$8_9$ & $15264$ & $17040$ & $4920$ & $30584$ & $42492$ & $42492$ & $42492$ & $42492$ \\
$8_{10}^+$ & $208$ & $47712$ & $5304$ & $560$ & $840$ & $840$ & $840$ & $840$ \\
$8_{11}^+$ & $3904$ & $3624$ & $40528$ & $64$ & $96$ & $96$ & $96$ & $96$ \\
$8_{12}$ & $14640$ & $231760$ & $288$ & $1728$ & $2592$ & $2592$ & $2592$ & $2592$ \\
$8_{13}^+$ & $159792$ & $1520$ & $80$ & $8592$ & $12832$ & $13056$ & $13056$ & $13056$ \\
$8_{14}^+$ & $59000$ & $6880$ & $6160$ & $240$ & $360$ & $360$ & $360$ & $360$ \\
$8_{15}^+$ & $16$ & $880$ & $4512$ & $24376$ & $40104$ & $40104$ & $40104$ & $40104$ \\
$8_{16}^+$ & $32720$ & $1168$ & $48$ & $32$ & $48$ & $48$ & $48$ & $48$ \\
$8_{17}$ & $60336$ & $8288$ & $25872$ & $26736$ & $52704$ & $53184$ & $53184$ & $53184$ \\
$8_{18}$ & $32$ & $756$ & $68000$ & $1184$ & $3552$ & $3552$ & $3552$ & $3552$ \\
$8_{19}^+$ & $163$ & $1656$ & $1132$ & $6708$ & $6996$ & $6996$ & $6996$ & $6996$ \\
$8_{20}^+$ & $116$ & $1272$ & $40$ & $80$ & $120$ & $120$ & $120$ & $120$ \\
$8_{21}^+$ & $24$ & $720$ & $3428$ & $25500$ & $28020$ & $28020$ & $28020$ & $28020$ \\
\hline
 \end{tabular}
\end{center}
 \caption{The number of lattice knots of prime knot type to eight
crossings confined to slabs of width $L$.}
  \label{Entropy}
\end{table}

\begin{table}[h!]
\begin{center}
 \begin{tabular}{||c||c|c|c|c|c|c|c|c||}
 \hline
  Knot & \multicolumn{8}{|c||}{$n_{L,K}$} \\
  \hline  
     & $L=1$ & $L=2$ & $L=3$ & $L=4$ & $L=5$ & $L=6$ & $L=7$ & $L=8$ \\
   \hline
$3_1^+3_1^+$ & \U{$48$} & $40$ & $40$ & $40$ & $40$ & $40$ & $40$ & $40$ \\
$3_1^+3_1^-$ & \U{$44$} & $40$ & $40$ & $40$ & $40$ & $40$ & $40$ & $40$ \\
$3_1^+4_1$   & \U{$54$} & \U{$48$} & $46$ & $46$ & $46$ & $46$ & $46$ & $46$ \\
$3_1^+5_1^+$ & \U{$58$} & $50$ & $50$ & $50$ & $50$ & $50$ & $50$ & $50$ \\
$3_1^+5_1^-$ & \U{$56$} & $50$ & $50$ & $50$ & $50$ & $50$ & $50$ & $50$ \\
$3_1^+5_2^+$ & \U{$60$} & $52$ & $52$ & $52$ & $52$ & $52$ & $52$ & $52$ \\
$3_1^+5_2^-$ & \U{$60$} & \U{$52$} & $50$ & $50$ & $50$ & $50$ & $50$ & $50$ \\
$4_14_1$     & \U{$60$} & \U{$54$} & $52$ & $52$ & $52$ & $52$ & $52$ & $52$ \\
\hline
 \end{tabular}
\end{center}
 \caption{The minimal length of lattice knots of compound knot type
to eight crossings confined to slabs of width $L$. Cases with an 
increase in minimal length are underlined.}
  \label{Lengths-C}
\end{table}
\begin{table}[h!]
\begin{center}
 \begin{tabular}{||c||c|c|c|c|c|c|c|c||}
 \hline
  Knot & \multicolumn{8}{|c||}{$p_{n_{L,K}} (K)$} \\
  \hline  
     & $L=1$ & $L=2$ & $L=3$ & $L=4$ & $L=5$ & $L=6$ & $L=7$ & $L=8$ \\
\hline
$3_1^+3_1^+$ & $2430$ & $8$ & $7800$ & $10492$ & $15304$ & $15304$ & $15312$ & $15312$ \\
$3_1^+3_1^-$ & $144$ & $448$ & $56616$ & $78576$ & $109424$ & $109424$ & $109432$ & $109440$ \\
$3_1^+4_1$ & $12072$ & $27472$ & $56944$ & $129448$ & $168208$ & $179856$ & $179856$ & $179856$ \\
$3_1^+5_1^+$ & $216$ & $80$ & $31600$ & $66992$ & $71424$ & $100392$ & $100488$ & $100488$ \\
$3_1^+5_1^-$ & $288$ & $3152$ & $53984$ & $198896$ & $245248$ & $283576$ & $284376$ & $284376$ \\
$3_1^+5_2^+$ & $150416$ & $1880$ & $729528$ & $2187592$ & $3190064$ & $3637128$ & $3678504$ & $3678504$ \\
$3_1^+5_2^-$ & $13776$ & $56$ & $880$ & $1808$ & $2640$ & $2640$ & $2640$ & $2640$ \\
$4_14_1$ & $800$ & $43816$ & $40608$ & $172808$ & $309224$ & $332120$ & $334824$ & $334824$ \\
\hline
 \end{tabular}
\end{center}
 \caption{The number of lattice knots of compound knot type
to eight crossings confined to slabs of width $L$.}
  \label{Entropy-C}
\end{table}

\vfill\eject

\begin{landscape}
\begin{table}[h!]
 \begin{tabular}{||c||c|c|c|c|c|c|c|c||}
 \hline
  Knot & \multicolumn{8}{|c||}{Symmetry Classes} \\
  \hline  
     & $L=1$ & $L=2$ & $L=3$ & $L=4$ & $L=5$ & $L=6$ & $L=7$ & $L=8$ \\
   \hline
$0_1$ &\fns ${3}^{1}$&\fns ${3}^{1}$&\fns ${3}^{1}$&\fns ${3}^{1}$&\fns ${3}^{1}$&\fns ${3}^{1}$&\fns ${3}^{1}$&\fns ${3}^{1}$\\
$3_1^+$ &\fns ${4}^{3}{8}^{3}$&\fns ${8}^{19}$&\fns ${4}^{1}{8}^{3}{12}^{6}{24}^{65}$&\fns ${4}^{1}{8}^{2}{12}^{7}{24}^{65}$&\fns ${4}^{1}{8}^{2}{12}^{7}{24}^{65}$&\fns ${4}^{1}{8}^{2}{12}^{7}{24}^{65}$&\fns ${4}^{1}{8}^{2}{12}^{7}{24}^{65}$&\fns ${4}^{1}{8}^{2}{12}^{7}{24}^{65}$\\
$4_1$ &\fns ${4}^{22}{8}^{174}$&\fns ${8}^{750}{16}^{4}$&\fns ${16}^{116}{24}^{36}$&\fns ${24}^{152}$&\fns ${24}^{152}$&\fns ${24}^{152}$&\fns ${24}^{152}$&\fns ${24}^{152}$\\
$5_1^+$ &\fns ${4}^{6}{8}^{6}$&\fns ${8}^{95}$&\fns ${4}^{6}{8}^{105}{16}^{31}$&\fns ${12}^{6}{16}^{5}{24}^{131}$&\fns ${12}^{6}{24}^{136}$&\fns ${12}^{6}{24}^{136}$&\fns ${12}^{6}{24}^{136}$&\fns ${12}^{6}{24}^{136}$\\
$5_2^+$ &\fns ${4}^{30}{8}^{765}$&\fns ${8}^{90}$&\fns ${4}^{4}{8}^{1687}{16}^{705}$&\fns ${12}^{4}{16}^{265}{24}^{2127}$&\fns ${12}^{4}{24}^{2392}$&\fns ${12}^{4}{24}^{2392}$&\fns ${12}^{4}{24}^{2392}$&\fns ${12}^{4}{24}^{2392}$\\
$6_1^+$ &\fns ${4}^{43}{8}^{983}$&\fns ${8}^{1588}$&\fns ${8}^{112}{16}^{17}$&\fns ${12}^{2}{16}^{9}{24}^{118}$&\fns ${12}^{2}{24}^{127}$&\fns ${12}^{2}{24}^{127}$&\fns ${12}^{2}{24}^{127}$&\fns ${12}^{2}{24}^{127}$\\
$6_2^+$ &\fns ${4}^{18}{8}^{267}$&\fns ${8}^{90}$&\fns ${8}^{365}{16}^{4}$&\fns ${16}^{223}{24}^{461}$&\fns ${24}^{684}$&\fns ${24}^{684}$&\fns ${24}^{684}$&\fns ${24}^{684}$\\
$6_3$ &\fns ${8}^{156}$&\fns ${8}^{942}$&\fns ${8}^{38}{16}^{18}$&\fns ${16}^{18}{24}^{130}$&\fns ${24}^{148}$&\fns ${24}^{148}$&\fns ${24}^{148}$&\fns ${24}^{148}$\\
$7_1^+$ &\fns ${4}^{9}{8}^{9}$&\fns ${8}^{150}$&\fns ${4}^{1}{8}^{666}{16}^{21}$&\fns ${8}^{7}{12}^{2}{16}^{674}{24}^{29}$&\fns ${8}^{5}{12}^{4}{16}^{10}{24}^{693}$&\fns ${12}^{9}{24}^{703}$&\fns ${12}^{9}{24}^{703}$&\fns ${12}^{9}{24}^{703}$\\
$7_2^+$ &\fns ${4}^{41}{8}^{2783}$&\fns ${4}^{9}{8}^{31249}$&\fns ${8}^{4600}{16}^{931}$&\fns ${8}^{127}{16}^{5403}{24}^{1481}$&\fns ${12}^{7}{16}^{488}{24}^{6516}$&\fns ${12}^{7}{24}^{7004}$&\fns ${12}^{7}{24}^{7004}$&\fns ${12}^{7}{24}^{7004}$\\
$7_3^+$ &\fns ${8}^{186}$&\fns ${8}^{7212}$&\fns ${16}^{10}$&\fns ${16}^{10}$&\fns ${24}^{10}$&\fns ${24}^{10}$&\fns ${24}^{10}$&\fns ${24}^{10}$\\
$7_4^+$ &\fns ${2}^{6}{4}^{86}{8}^{1636}$&\fns ${4}^{10}{8}^{45}$&\fns ${4}^{1}{8}^{2}$&\fns ${8}^{1}{16}^{2}{24}^{1}$&\fns ${12}^{1}{24}^{3}$&\fns ${12}^{1}{24}^{3}$&\fns ${12}^{1}{24}^{3}$&\fns ${12}^{1}{24}^{3}$\\
$7_5^+$ &\fns ${8}^{690}$&\fns ${8}^{312}$&\fns ${8}^{13}{16}^{1}$&\fns ${16}^{10}{24}^{187}$&\fns ${24}^{197}$&\fns ${24}^{197}$&\fns ${24}^{197}$&\fns ${24}^{197}$\\
$7_6^+$ &\fns ${8}^{731}$&\fns ${8}^{220}$&\fns ${8}^{80}$&\fns ${16}^{512}{24}^{197}$&\fns ${24}^{709}$&\fns ${24}^{709}$&\fns ${24}^{709}$&\fns ${24}^{709}$\\
$7_7^+$ &\fns ${4}^{1}$&\fns ${4}^{32}{8}^{1108}$&\fns ${4}^{10}{8}^{156}$&\fns ${12}^{1}{16}^{7}{24}^{3}$&\fns ${12}^{1}{24}^{10}$&\fns ${12}^{1}{24}^{10}$&\fns ${12}^{1}{24}^{10}$&\fns ${12}^{1}{24}^{10}$\\
  \hline
$3_1^+3_1^+$ &\fns ${2}^{3}{4}^{36}{8}^{285}$&\fns ${4}^{2}$&\fns ${8}^{303}{16}^{336}$&\fns ${8}^{2}{12}^{1}{16}^{600}{24}^{36}$&\fns ${12}^{2}{16}^{1}{24}^{636}$&\fns ${12}^{2}{16}^{1}{24}^{636}$&\fns ${12}^{2}{24}^{637}$&\fns ${12}^{2}{24}^{637}$\\
$3_1^+3_1^-$ &\fns ${8}^{18}$&\fns ${8}^{56}$&\fns ${8}^{2050}{16}^{2500}{24}^{9}$&\fns ${8}^{1}{16}^{3856}{24}^{703}$&\fns ${8}^{1}{24}^{4559}$&\fns ${8}^{1}{24}^{4559}$&\fns ${16}^{1}{24}^{4559}$&\fns ${24}^{4560}$\\
$3_1^+4_1$ &\fns ${8}^{1509}$&\fns ${8}^{3434}$&\fns ${8}^{4758}{16}^{1180}$&\fns ${8}^{282}{16}^{5737}{24}^{1475}$&\fns ${16}^{1456}{24}^{6038}$&\fns ${24}^{7494}$&\fns ${24}^{7494}$&\fns ${24}^{7494}$\\
$3_1^+5_1^+$ &\fns ${8}^{27}$&\fns ${8}^{10}$&\fns ${8}^{3756}{16}^{97}$&\fns ${8}^{36}{16}^{4115}{24}^{36}$&\fns ${16}^{3633}{24}^{554}$&\fns ${16}^{12}{24}^{4175}$&\fns ${24}^{4187}$&\fns ${24}^{4187}$\\
$3_1^+5_1^-$ &\fns ${8}^{36}$&\fns ${8}^{394}$&\fns ${8}^{6006}{16}^{371}$&\fns ${8}^{470}{16}^{9745}{24}^{1634}$&\fns ${16}^{4891}{24}^{6958}$&\fns ${16}^{100}{24}^{11749}$&\fns ${24}^{11849}$&\fns ${24}^{11849}$\\
$3_1^+5_2^+$ &\fns ${8}^{18802}$&\fns ${8}^{235}$&\fns ${8}^{74813}{16}^{8189}$&\fns ${8}^{37388}{16}^{111588}{24}^{4295}$&\fns ${16}^{61055}{24}^{92216}$&\fns ${16}^{5172}{24}^{148099}$&\fns ${24}^{153271}$&\fns ${24}^{153271}$\\
$3_1^+5_2^-$ &\fns ${8}^{1722}$&\fns ${8}^{7}$&\fns ${8}^{110}$&\fns ${16}^{104}{24}^{6}$&\fns ${24}^{110}$&\fns ${24}^{110}$&\fns ${24}^{110}$&\fns ${24}^{110}$\\
$4_14_1$ &\fns ${8}^{100}$&\fns ${4}^{118}{8}^{5418}$&\fns ${8}^{4466}{16}^{305}$&\fns ${8}^{6873}{16}^{6551}{24}^{542}$&\fns ${12}^{30}{16}^{3200}{24}^{10736}$&\fns ${12}^{30}{16}^{338}{24}^{13598}$&\fns ${12}^{30}{24}^{13936}$&\fns ${12}^{30}{24}^{13936}$\\
 \hline
 \end{tabular}
 \caption{The symmetry classes of prime lattice knots upto seven crossings, 
and compound lattice knots upto eight crossings, confined to $\Slab_L$ with 
$L\leq 8$.}
  \label{S1}
\end{table}
\end{landscape}

\begin{landscape}
\begin{table}[h!]
 \begin{tabular}{||c||c|c|c|c|c|c|c|c||}
 \hline
  Knot & \multicolumn{8}{|c||}{Symmetry Classes} \\
  \hline  
     & $L=1$ & $L=2$ & $L=3$ & $L=4$ & $L=5$ & $L=6$ & $L=7$ & $L=8$ \\
   \hline
$8_1^+$ &\fns ${4}^{55}{8}^{5335}$&\fns ${8}^{635}$&\fns ${8}^{274}$&\fns ${8}^{1}{16}^{480}{24}^{14}$&\fns ${12}^{1}{16}^{90}{24}^{404}$&\fns ${12}^{1}{24}^{494}$&\fns ${12}^{1}{24}^{494}$&\fns ${12}^{1}{24}^{494}$\\
$8_2^+$ &\fns ${4}^{23}{8}^{313}$&\fns ${8}^{36857}$&\fns ${8}^{1271}{16}^{10}$&\fns ${8}^{594}{16}^{1294}{24}^{22}$&\fns ${16}^{467}{24}^{1443}$&\fns ${24}^{1910}$&\fns ${24}^{1910}$&\fns ${24}^{1910}$\\
$8_3$ &\fns ${4}^{3}{8}^{1135}$&\fns ${8}^{250}$&\fns ${8}^{1}$&\fns ${8}^{1}$&\fns ${12}^{1}$&\fns ${12}^{1}$&\fns ${12}^{1}$&\fns ${12}^{1}$\\
$8_4^+$ &\fns ${4}^{63}{8}^{2493}$&\fns ${8}^{1190}$&\fns ${8}^{230}$&\fns ${8}^{147}{16}^{831}{24}^{19}$&\fns ${16}^{58}{24}^{939}$&\fns ${24}^{997}$&\fns ${24}^{997}$&\fns ${24}^{997}$\\
$8_5^+$ &\fns ${2}^{2}{4}^{5}{8}^{6}$&\fns ${4}^{57}{8}^{50994}$&\fns ${4}^{10}{8}^{613}$&\fns ${16}^{24}$&\fns ${16}^{24}$&\fns ${16}^{24}$&\fns ${16}^{24}$&\fns ${16}^{24}$\\
$8_6^+$ &\fns ${8}^{1205}$&\fns ${8}^{1}$&\fns ${8}^{3013}{16}^{5}$&\fns ${16}^{230}$&\fns ${24}^{230}$&\fns ${24}^{230}$&\fns ${24}^{230}$&\fns ${24}^{230}$\\
$8_7^+$ &\fns ${8}^{2429}$&\fns ${8}^{6172}$&\fns ${8}^{1}$&\fns ${16}^{1}$&\fns ${24}^{1}$&\fns ${24}^{1}$&\fns ${24}^{1}$&\fns ${24}^{1}$\\
$8_8^+$ &\fns ${8}^{1696}$&\fns ${8}^{347}$&\fns ${8}^{4959}{16}^{38}$&\fns ${16}^{65}$&\fns ${24}^{65}$&\fns ${24}^{65}$&\fns ${24}^{65}$&\fns ${24}^{65}$\\
$8_9$ &\fns ${8}^{1908}$&\fns ${8}^{2130}$&\fns ${8}^{239}{16}^{134}{24}^{36}$&\fns ${8}^{1}{16}^{1488}{24}^{282}$&\fns ${12}^{1}{24}^{1770}$&\fns ${12}^{1}{24}^{1770}$&\fns ${12}^{1}{24}^{1770}$&\fns ${12}^{1}{24}^{1770}$\\
$8_{10}^+$ &\fns ${8}^{26}$&\fns ${8}^{5964}$&\fns ${8}^{663}$&\fns ${16}^{35}$&\fns ${24}^{35}$&\fns ${24}^{35}$&\fns ${24}^{35}$&\fns ${24}^{35}$\\
$8_{11}^+$ &\fns ${8}^{488}$&\fns ${8}^{453}$&\fns ${8}^{5056}{16}^{5}$&\fns ${16}^{4}$&\fns ${24}^{4}$&\fns ${24}^{4}$&\fns ${24}^{4}$&\fns ${24}^{4}$\\
$8_{12}$ &\fns ${8}^{1830}$&\fns ${8}^{28970}$&\fns ${8}^{36}$&\fns ${16}^{108}$&\fns ${24}^{108}$&\fns ${24}^{108}$&\fns ${24}^{108}$&\fns ${24}^{108}$\\
$8_{13}^+$ &\fns ${8}^{19974}$&\fns ${8}^{190}$&\fns ${8}^{10}$&\fns ${8}^{121}{16}^{316}{24}^{107}$&\fns ${16}^{28}{24}^{516}$&\fns ${24}^{544}$&\fns ${24}^{544}$&\fns ${24}^{544}$\\
$8_{14}^+$ &\fns ${8}^{7375}$&\fns ${8}^{860}$&\fns ${8}^{748}{16}^{11}$&\fns ${16}^{15}$&\fns ${24}^{15}$&\fns ${24}^{15}$&\fns ${24}^{15}$&\fns ${24}^{15}$\\
$8_{15}^+$ &\fns ${4}^{2}{8}^{1}$&\fns ${8}^{110}$&\fns ${4}^{10}{8}^{559}$&\fns ${4}^{7}{8}^{350}{12}^{3}{16}^{1259}{24}^{57}$&\fns ${12}^{10}{24}^{1666}$&\fns ${12}^{10}{24}^{1666}$&\fns ${12}^{10}{24}^{1666}$&\fns ${12}^{10}{24}^{1666}$\\
$8_{16}^+$ &\fns ${4}^{14}{8}^{4083}$&\fns ${8}^{146}$&\fns ${8}^{6}$&\fns ${16}^{2}$&\fns ${24}^{2}$&\fns ${24}^{2}$&\fns ${24}^{2}$&\fns ${24}^{2}$\\
$8_{17}$ &\fns ${8}^{7542}$&\fns ${8}^{1036}$&\fns ${8}^{3234}$&\fns ${8}^{1154}{16}^{998}{24}^{64}$&\fns ${16}^{60}{24}^{2156}$&\fns ${24}^{2216}$&\fns ${24}^{2216}$&\fns ${24}^{2216}$\\
$8_{18}$ &\fns $4^48^2$&\fns ${2}^{6}{4}^{6}{8}^{90}$&\fns ${8}^{8500}$&\fns ${8}^{148}$&\fns ${24}^{148}$&\fns ${24}^{148}$&\fns ${24}^{148}$&\fns ${24}^{148}$\\
$8_{19}^+$ &\fns ${1}^{1}{2}^{3}{4}^{9}{8}^{15}$&\fns ${8}^{207}$&\fns ${4}^{13}{8}^{75}{16}^{30}$&\fns ${8}^{12}{12}^{13}{16}^{30}{24}^{249}$&\fns ${12}^{25}{24}^{279}$&\fns ${12}^{25}{24}^{279}$&\fns ${12}^{25}{24}^{279}$&\fns ${12}^{25}{24}^{279}$\\
$8_{20}^+$ &\fns ${4}^{3}{8}^{13}$&\fns ${8}^{159}$&\fns ${8}^{5}$&\fns ${16}^{5}$&\fns ${24}^{5}$&\fns ${24}^{5}$&\fns ${24}^{5}$&\fns ${24}^{5}$\\
$8_{21}^+$ &\fns ${4}^{2}{8}^{2}$&\fns ${8}^{90}$&\fns ${4}^{1}{8}^{428}$&\fns ${12}^{1}{16}^{315}{24}^{852}$&\fns ${12}^{1}{24}^{1167}$&\fns ${12}^{1}{24}^{1167}$&\fns ${12}^{1}{24}^{1167}$&\fns ${12}^{1}{24}^{1167}$\\
\hline
 \end{tabular}
 \caption{The symmetry classes of prime lattice knots up to eight crossings, confined 
to $\Slab_L$ with $L\leq 8$.}
  \label{S2}
\end{table}
\end{landscape}

\end{document}

%% file: ScatterData.tex
\beginpicture

 \setcoordinatesystem units <4.000 pt, 4.000 pt>

\color[rgb]{0.147438,0.069532,0.791067}  \put {$0_1$} at  4 4 
\color[rgb]{0.259873,0.164787,0.593611}  \put {$3_1$} at  26 24 
\color[rgb]{0.329163,0.290140,0.412418}  \put {$4_1$} at  36 30
\color[rgb]{0.303336,0.235363,0.488992}  \put {$5_1$} at  38 34 
\color[rgb]{0.336295,0.293266,0.401883}  \put {$5_2$} at  42 36 
\color[rgb]{0.367999,0.339272,0.324074}  \put {$6_1$} at  48 40 
\color[rgb]{0.378547,0.343357,0.308775}  \put {$6_2$} at  48 40 
\color[rgb]{0.408879,0.381064,0.238564}  \put {$6_3$} at  50 40 
\color[rgb]{0.356717,0.299817,0.372936}  \put {$7_1$} at  50 44 
\color[rgb]{0.380563,0.343405,0.306492}  \put {$7_2$} at  54 46 
\color[rgb]{0.405128,0.378595,0.245173}  \put {$7_3$} at  54 44
\color[rgb]{0.391214,0.375768,0.262634}  \put {$7_4$} at  54 44
\color[rgb]{0.405633,0.380050,0.243164}  \put {$7_5$} at  56 46 
\color[rgb]{0.408942,0.381084,0.238473}  \put {$7_6$} at  56 46 
\color[rgb]{0.441151,0.412178,0.170452}  \put {$7_7$} at  56 44  
\color[rgb]{0.401865,0.379073,0.248220}  \put {$8_1$} at  60 50 
\color[rgb]{0.414260,0.382590,0.230937}  \put {$8_2$} at  60 50
\color[rgb]{0.421303,0.406285,0.199123}  \put {$8_3$} at  60 48
\color[rgb]{0.406518,0.380326,0.241917}  \put {$8_4$} at  60 50
\color[rgb]{0.415706,0.382054,0.229741}  \put {$8_5$} at  62 50
\color[rgb]{0.431273,0.409654,0.184736}  \put {$8_6$} at  62 50 
\color[rgb]{0.443129,0.429960,0.150192}  \put {$8_7$} at  62 48
\color[rgb]{0.432691,0.408994,0.183769}  \put {$8_8$} at  62 50 
\color[rgb]{0.433896,0.410401,0.180964}  \put {$8_9$} at  62 50 
\color[rgb]{0.436966,0.410755,0.176972}  \put {$8_{10}$} at  62 50 
\color[rgb]{0.434802,0.409123,0.181154}  \put {$8_{11}$} at  62 50 
\color[rgb]{0.430878,0.409018,0.185830}  \put {$8_{12}$} at  64 52 
\color[rgb]{0.444373,0.413785,0.164842}  \put {$8_{13}$} at  64 50
\color[rgb]{0.447782,0.430635,0.144241}  \put {$8_{14}$} at  64 50
\color[rgb]{0.425947,0.386570,0.213014}  \put {$8_{15}$} at  62 52
\color[rgb]{0.460330,0.447641,0.111650}  \put {$8_{16}$} at  66 50 
\color[rgb]{0.465286,0.449123,0.104056}  \put {$8_{17}$} at  68 52 
\color[rgb]{0.477281,0.462492,0.074740}  \put {$8_{18}$} at  70 52 
\color[rgb]{0.348582,0.296885,0.384948}  \put {$8_{19}$} at  48 42
\color[rgb]{0.376618,0.340847,0.313292}  \put {$8_{20}$} at  52 44 
\color[rgb]{0.399164,0.350045,0.278569}  \put {$8_{21}$} at  54 46 


\put { } at -15 0 
 \color{black}
\put {O} at 0 0 
\setplotarea x from 0 to 75, y from 0 to 60
\axis top /
\axis right /
\axis bottom 
 label {{\large \vspace{2mm} \hspace{1cm} $n_{1,K}$}} /
\axis bottom 
 ticks withvalues 0 15 30 {} 60 75 / quantity 6 /
\axis left 
 label {{\large $n_{K}$\hspace{2mm}}} /
\axis left 
 ticks out withvalues 0 15 {} 45 60 / quantity 5 /
\setdots <2pt>
\plot -0.04 0 0.4 0 /
\plot 0 -1 0 13 /
\normalcolor
\endpicture

%% file: ScatterData2.tex
\beginpicture

 \setcoordinatesystem units <16.000 pt, 16.000 pt>

\color[rgb]{0.147438,0.069532,0.791067}  \put {$0_1$} at  1.01 1.01 
\color[rgb]{0.259873,0.164787,0.593611}  \put {$3_1$} at  3.58  7.42 
\color[rgb]{0.329163,0.290140,0.412418}  \put {$4_1$} at  7.30 8.20
\color[rgb]{0.303336,0.235363,0.488992}  \put {$5_1$} at  4.28 8.11
\color[rgb]{0.336295,0.293266,0.401883}  \put {$5_2$} at  8.74 10.96 
\color[rgb]{0.367999,0.339272,0.324074}  \put {$6_1$} at  9.00 9.03 
\color[rgb]{0.378547,0.343357,0.308775}  \put {$6_2$} at  7.70 9.71 
\color[rgb]{0.408879,0.381064,0.238564}  \put {$6_3$} at  7.13 8.18 
\color[rgb]{0.356717,0.299817,0.372936}  \put {$7_1$} at  4.68 9.74
\color[rgb]{0.380563,0.343405,0.306492}  \put {$7_2$} at  10.02 12.03 
\color[rgb]{0.405128,0.378595,0.245173}  \put {$7_3$} at  7.31 5.48
\color[rgb]{0.391214,0.375768,0.262634}  \put {$7_4$} at  9.51 4.43
\color[rgb]{0.405633,0.380050,0.243164}  \put {$7_5$} at  8.62 8.46
\color[rgb]{0.408942,0.381084,0.238473}  \put {$7_6$} at  8.67 9.74
\color[rgb]{0.441151,0.412178,0.170452}  \put {$7_7$} at  1.39 5.54
\color[rgb]{0.401865,0.379073,0.248220}  \put {$8_1$} at  10.67 9.38 
\color[rgb]{0.414260,0.382590,0.230937}  \put {$8_2$} at  7.86 10.73
\color[rgb]{0.421303,0.406285,0.199123}  \put {$8_3$} at  9.12 2.48
\color[rgb]{0.406518,0.380326,0.241917}  \put {$8_4$} at  9.91 10.08
\color[rgb]{0.415706,0.382054,0.229741}  \put {$8_5$} at  4.28 5.95
\color[rgb]{0.431273,0.409654,0.184736}  \put {$8_6$} at  9.17 8.62 
\color[rgb]{0.443129,0.429960,0.150192}  \put {$8_7$} at  9.87 3.18
\color[rgb]{0.432691,0.408994,0.183769}  \put {$8_8$} at  9.52 7.35
\color[rgb]{0.433896,0.410401,0.180964}  \put {$8_9$} at  9.63 10.66
\color[rgb]{0.436966,0.410755,0.176972}  \put {$8_{10}$} at  5.34 6.73
\color[rgb]{0.434802,0.409123,0.181154}  \put {$8_{11}$} at  8.27 4.56
\color[rgb]{0.430878,0.409018,0.185830}  \put {$8_{12}$} at  9.59 7.86
\color[rgb]{0.444373,0.413785,0.164842}  \put {$8_{13}$} at  11.98 9.48
\color[rgb]{0.447782,0.430635,0.144241}  \put {$8_{14}$} at  10.99 5.89
\color[rgb]{0.425947,0.386570,0.213014}  \put {$8_{15}$} at  2.77 10.60
\color[rgb]{0.460330,0.447641,0.111650}  \put {$8_{16}$} at  10.40 3.87
\color[rgb]{0.465286,0.449123,0.104056}  \put {$8_{17}$} at  11.01 10.88
\color[rgb]{0.477281,0.462492,0.074740}  \put {$8_{18}$} at  3.47 8.18
\color[rgb]{0.348582,0.296885,0.384948}  \put {$8_{19}$} at  5.09 8.85
\color[rgb]{0.376618,0.340847,0.313292}  \put {$8_{20}$} at  4.75 4.79
\color[rgb]{0.399164,0.350045,0.278569}  \put {$8_{21}$} at  3.18 10.24


\put { } at -4 0 
 \color{black}
\put {O} at 0 0 
\setplotarea x from 0 to 15, y from 0 to 12
\axis top /
\axis right /
\axis bottom 
 label {{\large \vspace{2mm} \hspace{1cm} $\log p_{n_{1,K}}(K)$}} /
\axis bottom 
 ticks withvalues 0 3 6 {} 12 15 / quantity 6 /
\axis left 
 label {{\large $\log p_{n_{K}}(K)$\hspace{2mm}}} /
\axis left 
 ticks out withvalues 0 3 {} 9 12 / quantity 5 /
\setdots <2pt>
\normalcolor
\endpicture

%% file: Bargraphs/Bargraph3-1k1.tex
\beginpicture

 \setcoordinatesystem units <25.000 pt,40.000 pt>

 \color{red}

\setplotsymbol ({\LARGE $\cdot$})

\plot 0.000 0.000 0.000 4.000 1.000 4.000 1.000 0.000   /

 \color{blue}
\setshadegrid span <0.500000pt>
\hshade 0.000 0.000 1.000 4.000 0.000 1.000  /

 \color{red}

\plot 1.000 0.000 1.000 2.440 2.000 2.440 2.000 0.000   /

 \color{blue}
\setshadegrid span <0.793701pt>
\hshade 0.000 1.000 2.000 2.440 1.000 2.000  /

 \color{red}

\plot 2.000 0.000 2.000 2.391 3.000 2.391 3.000 0.000   /

 \color{red}
\setshadegrid span <1.040042pt>
\hshade 0.000 2.000 3.000 2.391 2.000 3.000  /

 \color{red}

\plot 3.000 0.000 3.000 0.002 4.000 0.002 4.000 0.000   /

\plot 4.000 0.000 4.000 0.000 5.000 0.000 5.000 0.000   /

\plot 5.000 0.000 5.000 0.000 6.000 0.000 6.000 0.000   /

\plot 6.000 0.000 6.000 0.000 7.000 0.000 7.000 0.000   /

\plot 7.000 0.000 7.000 0.000 8.000 0.000 8.000 0.000   /

\plot 8.000 0.000 8.000 0.000 9.000 0.000 9.000 0.000   /

\plot 9.000 0.000 9.000 0.000 10.000 0.000 10.000 0.000   /

 \setcoordinatesystem units <25.000 pt,40.000 pt>

 \color{black}
\put {O} at 0 0 
\setplotarea x from 0 to 10.000000, y from 0 to 4.000000
\axis top /
\axis right /
\axis bottom 
 label {{\Large \hspace{2.7cm} $L$}} /
\axis bottom 
 ticks numbered from 0 to 10 by 1 /
\axis left 
 label {{\Large $f_L \hbox{$\qquad$}$}} /
\axis left 
 ticks numbered from 0 to 4 by 2 /

\normalcolor
\endpicture

%% file: Bargraphs/Bargraph4-1k1.tex
\beginpicture

 \setcoordinatesystem units <25.000 pt,10.000 pt>

 \color{red}

\setplotsymbol ({\LARGE $\cdot$})

\plot 0.000 0.000 0.000 16.000 1.000 16.000 1.000 0.000   /

 \color{blue}
\setshadegrid span <0.500000pt>
\hshade 0.000 0.000 1.000 16.000 0.000 1.000  /

 \color{red}

\plot 1.000 0.000 1.000 9.410 2.000 9.410 2.000 0.000   /

 \color{blue}
\setshadegrid span <0.793701pt>
\hshade 0.000 1.000 2.000 9.410 1.000 2.000  /

 \color{red}

\plot 2.000 0.000 2.000 0.198 3.000 0.198 3.000 0.000   /

 \color{blue}
\setshadegrid span <1.040042pt>
\hshade 0.000 2.000 3.000 0.198 2.000 3.000  /

 \color{red}

\plot 3.000 0.000 3.000 0.294 4.000 0.294 4.000 0.000   /

 \color{red}
\setshadegrid span <1.259921pt>
\hshade 0.000 3.000 4.000 0.294 3.000 4.000  /

 \color{red}

\plot 4.000 0.000 4.000 0.000 5.000 0.000 5.000 0.000   /

\plot 5.000 0.000 5.000 0.000 6.000 0.000 6.000 0.000   /

\plot 6.000 0.000 6.000 0.000 7.000 0.000 7.000 0.000   /

\plot 7.000 0.000 7.000 0.000 8.000 0.000 8.000 0.000   /

\plot 8.000 0.000 8.000 0.000 9.000 0.000 9.000 0.000   /

\plot 9.000 0.000 9.000 0.000 10.000 0.000 10.000 0.000   /

 \setcoordinatesystem units <25.000 pt,10.00 pt>

 \color{black}
\put {O} at 0 0 
\setplotarea x from 0 to 10.000000, y from 0 to 16.000000
\axis top /
\axis right /
\axis bottom 
 label {{\Large \hspace{2.7cm} $L$}} /
\axis bottom 
 ticks numbered from 0 to 10 by 1 /
\axis left 
 label {{\Large $f_L \hbox{$\qquad$}$}} /
\axis left 
 ticks numbered from 0 to 16 by 8 /

\normalcolor
\endpicture

%% file: Bargraphs/Bargraph5-1k1.tex
\beginpicture

 \setcoordinatesystem units <15.000 pt, 13pt>

 \color{red}

\setplotsymbol ({\LARGE $\cdot$})

\plot 0.000 0.000 0.000 11.000 1.000 11.000 1.000 0.000   /

 \color{blue}
\setshadegrid span <0.500000pt>
\hshade 0.000 0.000 1.000 11.000 0.000 1.000  /

 \color{red}

\plot 1.000 0.000 1.000 6.357 2.000 6.357 2.000 0.000   /

 \color{blue}
\setshadegrid span <0.793701pt>
\hshade 0.000 1.000 2.000 6.357 1.000 2.000  /

 \color{red}

\plot 2.000 0.000 2.000 0.582 3.000 0.582 3.000 0.000   /

 \color{red}
\setshadegrid span <1.040042pt>
\hshade 0.000 2.000 3.000 0.582 2.000 3.000  /

 \color{red}

\plot 3.000 0.000 3.000 0.885 4.000 0.885 4.000 0.000   /

 \color{red}
\setshadegrid span <1.259921pt>
\hshade 0.000 3.000 4.000 0.885 3.000 4.000  /

 \color{red}

\plot 4.000 0.000 4.000 0.012 5.000 0.012 5.000 0.000   /

 \color{red}
\setshadegrid span <1.462009pt>
\hshade 0.000 4.000 5.000 0.012 4.000 5.000  /

 \color{red}

\plot 5.000 0.000 5.000 0.000 6.000 0.000 6.000 0.000   /

\plot 6.000 0.000 6.000 0.000 7.000 0.000 7.000 0.000   /

\plot 7.000 0.000 7.000 0.000 8.000 0.000 8.000 0.000   /

\plot 8.000 0.000 8.000 0.000 9.000 0.000 9.000 0.000   /

\plot 9.000 0.000 9.000 0.000 10.000 0.000 10.000 0.000   /

\setcoordinatesystem units <15.000 pt,13 pt>

 \color{black}
\put {O} at 0 0 
\setplotarea x from 0 to 10.000000, y from 0 to 11.000000
\axis top /
\axis right /
\axis bottom 
 label {{\Large \hspace{2.3cm} $L$}} /
\axis bottom 
 ticks numbered from 0 to 10 by 2 /
\axis left 
 label {{\Large $f_L \hbox{$\,$}$}} /
\axis left 
 ticks numbered from 0 to 11 by 4 /

\normalcolor
\endpicture

%% file: Bargraphs/Bargraph5-2k1.tex
\beginpicture

 \setcoordinatesystem units <15 pt,13 pt>

 \color{red}

\setplotsymbol ({\LARGE $\cdot$})

\plot 0.000 0.000 0.000 11.000 1.000 11.000 1.000 0.000   /

 \color{blue}
\setshadegrid span <0.500000pt>
\hshade 0.000 0.000 1.000 11.000 0.000 1.000  /

 \color{red}

\plot 1.000 0.000 1.000 6.841 2.000 6.841 2.000 0.000   /

 \color{blue}
\setshadegrid span <0.793701pt>
\hshade 0.000 1.000 2.000 6.841 1.000 2.000  /

 \color{red}

\plot 2.000 0.000 2.000 3.539 3.000 3.539 3.000 0.000   /

 \color{red}
\setshadegrid span <1.040042pt>
\hshade 0.000 2.000 3.000 3.539 2.000 3.000  /

 \color{red}

\plot 3.000 0.000 3.000 0.803 4.000 0.803 4.000 0.000   /

 \color{red}
\setshadegrid span <1.259921pt>
\hshade 0.000 3.000 4.000 0.803 3.000 4.000  /

 \color{red}

\plot 4.000 0.000 4.000 0.038 5.000 0.038 5.000 0.000   /

 \color{red}
\setshadegrid span <1.462009pt>
\hshade 0.000 4.000 5.000 0.038 4.000 5.000  /

 \color{red}

\plot 5.000 0.000 5.000 0.000 6.000 0.000 6.000 0.000   /

\plot 6.000 0.000 6.000 0.000 7.000 0.000 7.000 0.000   /

\plot 7.000 0.000 7.000 0.000 8.000 0.000 8.000 0.000   /

\plot 8.000 0.000 8.000 0.000 9.000 0.000 9.000 0.000   /

\plot 9.000 0.000 9.000 0.000 10.000 0.000 10.000 0.000   /

 \setcoordinatesystem units <15 pt,13 pt>

 \color{black}
\put {O} at 0 0 
\setplotarea x from 0 to 10.000000, y from 0 to 11.000000
\axis top /
\axis right /
\axis bottom 
 label {{\Large \hspace{2.3cm} $L$}} /
\axis bottom 
 ticks numbered from 0 to 10 by 2 /
\axis left 
 /
\axis left 
 ticks numbered from 0 to 11 by 4 /

\normalcolor
\endpicture

%% file: Bargraphs/Bargraph6-1k1.tex
\beginpicture

 \setcoordinatesystem units <15.000 pt,4.000 pt>

 \color{red}

\setplotsymbol ({\LARGE $\cdot$})

\plot 0.000 0.000 0.000 25.000 1.000 25.000 1.000 0.000   /

 \color{blue}
\setshadegrid span <0.500000pt>
\hshade 0.000 0.000 1.000 25.000 0.000 1.000  /

 \color{red}

\plot 1.000 0.000 1.000 15.458 2.000 15.458 2.000 0.000   /

 \color{blue}
\setshadegrid span <0.793701pt>
\hshade 0.000 1.000 2.000 15.458 1.000 2.000  /

 \color{red}

\plot 2.000 0.000 2.000 -1.387  3.000 -1.387 3.000 0.000  /

 \color{blue}
\setshadegrid span <1.26pt>
\hshade -1.387 2.000 3.000 0.000 2.000 3.000  /

 \color{red}
\plot 3.000 0.000 3.000 0.943 4.000 0.943 4.000 0.000   /

 \color{red}
\setshadegrid span <1.259921pt>
\hshade 0.000 3.000 4.000 0.943 3.000 4.000  /

 \color{red}

\plot 4.000 0.000 4.000 0.024 5.000 0.024 5.000 0.000   /

 \color{red}
\setshadegrid span <1.462009pt>
\hshade 0.000 4.000 5.000 0.024 4.000 5.000  /

 \color{red}

\plot 5.000 0.000 5.000 0.000 6.000 0.000 6.000 0.000   /

\plot 6.000 0.000 6.000 0.000 7.000 0.000 7.000 0.000   /

\plot 7.000 0.000 7.000 0.000 8.000 0.000 8.000 0.000   /

\plot 8.000 0.000 8.000 0.000 9.000 0.000 9.000 0.000   /

\plot 9.000 0.000 9.000 0.000 10.000 0.000 10.000 0.000   /

 \setcoordinatesystem units <15.000 pt,4.000 pt>

 \color{black}
\put {O} at 0 0 
\setplotarea x from 0 to 10.000000, y from 0 to 25.000000
\axis top /
\axis right /
\axis bottom 
 label {{\Large \hspace{2.3cm} $L$}} /
\axis bottom 
 ticks numbered from 0 to 10 by 2 /
\axis left 
 label {{\Large $f_L \hbox{$\,$}$}} /
\axis left 
 ticks numbered from 0 to 25 by 8 /

\normalcolor
\endpicture

%% file: Bargraphs/Bargraph6-2k1.tex
\beginpicture

 \setcoordinatesystem units <15.000 pt,4.35 pt>

 \color{red}

\setplotsymbol ({\LARGE $\cdot$})

\plot 0.000 0.000 0.000 23.000 1.000 23.000 1.000 0.000   /

 \color{blue}
\setshadegrid span <0.500000pt>
\hshade 0.000 0.000 1.000 23.000 0.000 1.000  /

 \color{red}

\plot 1.000 0.000 1.000 13.879 2.000 13.879 2.000 0.000   /

 \color{blue}
\setshadegrid span <0.793701pt>
\hshade 0.000 1.000 2.000 13.879 1.000 2.000  /

 \color{red}

\plot 2.000 0.000 2.000 2.422 3.000 2.422 3.000 0.000   /

 \color{blue}
\setshadegrid span <1.040042pt>
\hshade 0.000 2.000 3.000 2.422 2.000 3.000  /

 \color{red}

\plot 3.000 0.000 3.000 1.590 4.000 1.590 4.000 0.000   /

 \color{red}
\setshadegrid span <1.259921pt>
\hshade 0.000 3.000 4.000 1.590 3.000 4.000  /

 \color{red}

\plot 4.000 0.000 4.000 0.115 5.000 0.115 5.000 0.000   /

 \color{red}
\setshadegrid span <1.462009pt>
\hshade 0.000 4.000 5.000 0.115 4.000 5.000  /

 \color{red}

\plot 5.000 0.000 5.000 0.000 6.000 0.000 6.000 0.000   /

\plot 6.000 0.000 6.000 0.000 7.000 0.000 7.000 0.000   /

\plot 7.000 0.000 7.000 0.000 8.000 0.000 8.000 0.000   /

\plot 8.000 0.000 8.000 0.000 9.000 0.000 9.000 0.000   /

\plot 9.000 0.000 9.000 0.000 10.000 0.000 10.000 0.000   /

 \setcoordinatesystem units <15.000 pt,4.35 pt>

 \color{black}
\put {O} at 0 0 
\setplotarea x from 0 to 10.000000, y from 0 to 23.000000
\axis top /
\axis right /
\axis bottom 
 label {{\Large \hspace{2.3cm} $L$}} /
\axis bottom 
 ticks numbered from 0 to 10 by 2 /
\axis left 
 label {{\Large $f_L \hbox{$\,$}$}} /
\axis left 
 ticks numbered from 0 to 23 by 8 /

\normalcolor
\endpicture

%% file: Bargraphs/Bargraph6-3k1.tex
\beginpicture

 \setcoordinatesystem units <15.000 pt,2.7 pt>

 \color{red}

\setplotsymbol ({\LARGE $\cdot$})

\plot 0.000 0.000 0.000 37.000 1.000 37.000 1.000 0.000   /

 \color{blue}
\setshadegrid span <0.500000pt>
\hshade 0.000 0.000 1.000 37.000 0.000 1.000  /

 \color{red}

\plot 1.000 0.000 1.000 22.798 2.000 22.798 2.000 0.000   /

 \color{blue}
\setshadegrid span <0.793701pt>
\hshade 0.000 1.000 2.000 22.798 1.000 2.000  /

 \color{red}

\plot 2.000 0.000 2.000 1.456 3.000 1.456 3.000 0.000   /

 \color{blue}
\setshadegrid span <1.040042pt>
\hshade 0.000 2.000 3.000 1.456 2.000 3.000  /

 \color{red}

\plot 3.000 0.000 3.000 1.750 4.000 1.750 4.000 0.000   /

 \color{red}
\setshadegrid span <1.259921pt>
\hshade 0.000 3.000 4.000 1.750 3.000 4.000  /

 \color{red}

\plot 4.000 0.000 4.000 0.041 5.000 0.041 5.000 0.000   /

 \color{red}
\setshadegrid span <1.462009pt>
\hshade 0.000 4.000 5.000 0.041 4.000 5.000  /

 \color{red}

\plot 5.000 0.000 5.000 0.000 6.000 0.000 6.000 0.000   /

\plot 6.000 0.000 6.000 0.000 7.000 0.000 7.000 0.000   /

\plot 7.000 0.000 7.000 0.000 8.000 0.000 8.000 0.000   /

\plot 8.000 0.000 8.000 0.000 9.000 0.000 9.000 0.000   /

\plot 9.000 0.000 9.000 0.000 10.000 0.000 10.000 0.000   /

 \setcoordinatesystem units <15.000 pt,2.7 pt>

 \color{black}
\put {O} at 0 0 
\setplotarea x from 0 to 10.000000, y from 0 to 37.000000
\axis top /
\axis right /
\axis bottom 
 label {{\Large \hspace{2.3cm} $L$}} /
\axis bottom 
 ticks numbered from 0 to 10 by 2 /
\axis left 
 label {{\Large $f_L \hbox{$\,$}$}} /
\axis left 
 ticks numbered from 0 to 37 by 15 /

\normalcolor
\endpicture

%% file: Bargraphs/Bargraph3-1p+3-1pk1.tex
\beginpicture

 \setcoordinatesystem units <15.000 pt,5.9 pt>

 \color{red}

\setplotsymbol ({\LARGE $\cdot$})

\plot 0.000 0.000 0.000 17.000 1.000 17.000 1.000 0.000   /

 \color{blue}
\setshadegrid span <0.500000pt>
\hshade 0.000 0.000 1.000 17.000 0.000 1.000  /

 \color{red}

\plot 1.000 0.000 1.000 10.284 2.000 10.284 2.000 0.000   /

 \color{blue}
\setshadegrid span <0.793701pt>
\hshade 0.000 1.000 2.000 10.284 1.000 2.000  /

 \color{red}

\plot 2.000 0.000 2.000 6.882 3.000 6.882 3.000 0.000   /

 \color{red}
\setshadegrid span <1.040042pt>
\hshade 0.000 2.000 3.000 6.882 2.000 3.000  /

 \color{red}

\plot 3.000 0.000 3.000 0.296 4.000 0.296 4.000 0.000   /

 \color{red}
\setshadegrid span <1.259921pt>
\hshade 0.000 3.000 4.000 0.296 3.000 4.000  /

 \color{red}

\plot 4.000 0.000 4.000 0.378 5.000 0.378 5.000 0.000   /

 \color{red}
\setshadegrid span <1.462009pt>
\hshade 0.000 4.000 5.000 0.378 4.000 5.000  /

 \color{red}

\plot 5.000 0.000 5.000 0.000 6.000 0.000 6.000 0.000   /

\plot 6.000 0.000 6.000 0.001 7.000 0.001 7.000 0.000   /

\plot 7.000 0.000 7.000 0.000 8.000 0.000 8.000 0.000   /

\plot 8.000 0.000 8.000 0.000 9.000 0.000 9.000 0.000   /

\plot 9.000 0.000 9.000 0.000 10.000 0.000 10.000 0.000   /

 \setcoordinatesystem units <15.000 pt,5.9 pt>

 \color{black}
\put {O} at 0 0 
\setplotarea x from 0 to 10.000000, y from 0 to 17.000000
\axis top /
\axis right /
\axis bottom 
 label {{\Large \hspace{2.3cm} $L$}} /
\axis bottom 
 ticks numbered from 0 to 10 by 2 /
\axis left 
 label {{\Large $f_L \hbox{$\,$}$}} /
\axis left 
 ticks numbered from 0 to 17 by 6 /

\normalcolor
\endpicture

%% file: Bargraphs/Bargraph3-1p+3-1mk1.tex
\beginpicture

 \setcoordinatesystem units <15.000 pt,11 pt>

 \color{red}

\setplotsymbol ({\LARGE $\cdot$})

\plot 0.000 0.000 0.000 9.000 1.000 9.000 1.000 0.000   /

 \color{blue}
\setshadegrid span <0.500000pt>
\hshade 0.000 0.000 1.000 9.000 0.000 1.000  /

 \color{red}

\plot 1.000 0.000 1.000 5.135 2.000 5.135 2.000 0.000   /

 \color{blue}
\setshadegrid span <0.793701pt>
\hshade 0.000 1.000 2.000 5.135 1.000 2.000  /

 \color{red}

\plot 2.000 0.000 2.000 4.839 3.000 4.839 3.000 0.000   /

 \color{red}
\setshadegrid span <1.040042pt>
\hshade 0.000 2.000 3.000 4.839 2.000 3.000  /

 \color{red}

\plot 3.000 0.000 3.000 0.328 4.000 0.328 4.000 0.000   /

 \color{red}
\setshadegrid span <1.259921pt>
\hshade 0.000 3.000 4.000 0.328 3.000 4.000  /

 \color{red}

\plot 4.000 0.000 4.000 0.331 5.000 0.331 5.000 0.000   /

 \color{red}
\setshadegrid span <1.462009pt>
\hshade 0.000 4.000 5.000 0.331 4.000 5.000  /

 \color{red}

\plot 5.000 0.000 5.000 0.000 6.000 0.000 6.000 0.000   /

\plot 6.000 0.000 6.000 0.000 7.000 0.000 7.000 0.000   /

\plot 7.000 0.000 7.000 0.000 8.000 0.000 8.000 0.000   /

\plot 8.000 0.000 8.000 0.000 9.000 0.000 9.000 0.000   /

\plot 9.000 0.000 9.000 0.000 10.000 0.000 10.000 0.000   /

 \setcoordinatesystem units <15.000 pt,11 pt>

 \color{black}
\put {O} at 0 0 
\setplotarea x from 0 to 10.000000, y from 0 to 9.000000
\axis top /
\axis right /
\axis bottom 
 label {{\Large \hspace{2.3cm} $L$}} /
\axis bottom 
 ticks numbered from 0 to 10 by 2 /
\axis left 
/
\axis left 
 ticks numbered from 0 to 9 by 4 /

\normalcolor
\endpicture

%% file: Bargraphs/Bargraph3-1p+3-1pk1L.tex
\beginpicture

 \setcoordinatesystem units <15.000 pt,35 pt>

 \color{red}

\setplotsymbol ({\LARGE $\cdot$})

\plot 0.000 0.000 0.000 2.890 1.000 2.890 1.000 0.000   /

 \color{blue}
\setshadegrid span <0.500000pt>
\hshade 0.000 0.000 1.000 2.890 0.000 1.000  /

 \color{red}

\plot 1.000 0.000 1.000 2.423 2.000 2.423 2.000 0.000   /

 \color{blue}
\setshadegrid span <0.793701pt>
\hshade 0.000 1.000 2.000 2.423 1.000 2.000  /

 \color{red}

\plot 2.000 0.000 2.000 2.065 3.000 2.065 3.000 0.000   /

 \color{red}
\setshadegrid span <1.040042pt>
\hshade 0.000 2.000 3.000 2.065 2.000 3.000  /

 \color{red}

\plot 3.000 0.000 3.000 0.260 4.000 0.260 4.000 0.000   /

 \color{red}
\setshadegrid span <1.259921pt>
\hshade 0.000 3.000 4.000 0.260 3.000 4.000  /

 \color{red}

\plot 4.000 0.000 4.000 0.320 5.000 0.320 5.000 0.000   /

 \color{red}
\setshadegrid span <1.462009pt>
\hshade 0.000 4.000 5.000 0.320 4.000 5.000  /

 \color{red}

\plot 5.000 0.000 5.000 0.000 6.000 0.000 6.000 0.000   /

\plot 6.000 0.000 6.000 0.001 7.000 0.001 7.000 0.000   /

\plot 7.000 0.000 7.000 0.000 8.000 0.000 8.000 0.000   /

\plot 8.000 0.000 8.000 0.000 9.000 0.000 9.000 0.000   /

\plot 9.000 0.000 9.000 0.000 10.000 0.000 10.000 0.000   /

 \setcoordinatesystem units <15.000 pt,35 pt>

 \color{black}
\put {O} at 0 0 
\setplotarea x from 0 to 10.000000, y from 0 to 2.890372
\axis top /
\axis right /
\axis bottom 
 label {{\Large \hspace{2.3cm} $L$}} /
\axis bottom 
 ticks numbered from 0 to 10 by 2 /
\axis left 
 label {{\Large $f_L \hbox{$\q$}$}} /
\axis left 
 ticks withvalues 0 3.2 17 / quantity 3 / 
\normalcolor
\endpicture

%% file: Bargraphs/Bargraph3-1p+3-1mk1L.tex
\beginpicture

 \setcoordinatesystem units <15.000 pt,43.9 pt>

 \color{red}

\setplotsymbol ({\LARGE $\cdot$})

\plot 0.000 0.000 0.000 2.303 1.000 2.303 1.000 0.000   /

 \color{blue}
\setshadegrid span <0.500000pt>
\hshade 0.000 0.000 1.000 2.303 0.000 1.000  /

 \color{red}

\plot 1.000 0.000 1.000 1.814 2.000 1.814 2.000 0.000   /

 \color{blue}
\setshadegrid span <0.793701pt>
\hshade 0.000 1.000 2.000 1.814 1.000 2.000  /

 \color{red}

\plot 2.000 0.000 2.000 1.765 3.000 1.765 3.000 0.000   /

 \color{red}
\setshadegrid span <1.040042pt>
\hshade 0.000 2.000 3.000 1.765 2.000 3.000  /

 \color{red}

\plot 3.000 0.000 3.000 0.284 4.000 0.284 4.000 0.000   /

 \color{red}
\setshadegrid span <1.259921pt>
\hshade 0.000 3.000 4.000 0.284 3.000 4.000  /

 \color{red}

\plot 4.000 0.000 4.000 0.286 5.000 0.286 5.000 0.000   /

 \color{red}
\setshadegrid span <1.462009pt>
\hshade 0.000 4.000 5.000 0.286 4.000 5.000  /

 \color{red}

\plot 5.000 0.000 5.000 0.000 6.000 0.000 6.000 0.000   /

\plot 6.000 0.000 6.000 0.000 7.000 0.000 7.000 0.000   /

\plot 7.000 0.000 7.000 0.000 8.000 0.000 8.000 0.000   /

\plot 8.000 0.000 8.000 0.000 9.000 0.000 9.000 0.000   /

\plot 9.000 0.000 9.000 0.000 10.000 0.000 10.000 0.000   /

 \setcoordinatesystem units <15.000 pt,43.9 pt>

 \color{black}
\put {O} at 0 0 
\setplotarea x from 0 to 10.000000, y from 0 to 2.302585
\axis top /
\axis right /
\axis bottom 
 label {{\Large \hspace{2.3cm} $L$}} /
\axis bottom 
 ticks numbered from 0 to 10 by 2 /
\axis left 
 /
\axis left 
 ticks withvalues 0 2.2 9 / quantity 3 / 
\normalcolor
\endpicture

%% file: Compressgraphs/Compress0-1.tex
\beginpicture

 \setcoordinatesystem units <13.000 pt,100 pt>

\put { } at -14 0
 \color{red}

\setplotsymbol ({\LARGE $\cdot$})

\plot -10.0000 0.0000  -9.9500 0.0000  -9.9000 0.0000  -9.8500 0.0000  -9.8000 0.0000  -9.7500 0.0000  -9.7000 0.0000  -9.6500 0.0000  -9.6000 0.0000  -9.5500 0.0000  -9.5000 0.0000  -9.4500 0.0000  -9.4000 0.0000  -9.3500 0.0000  -9.3000 0.0000  -9.2500 0.0000  -9.2000 0.0001  -9.1500 0.0001  -9.1000 0.0001  -9.0500 0.0001  -9.0000 0.0001  -8.9500 0.0001  -8.9000 0.0001  -8.8500 0.0001  -8.8000 0.0001  -8.7500 0.0001  -8.7000 0.0001  -8.6500 0.0001  -8.6000 0.0001  -8.5500 0.0001  -8.5000 0.0001  -8.4500 0.0001  -8.4000 0.0001  -8.3500 0.0001  -8.3000 0.0001  -8.2500 0.0001  -8.2000 0.0001  -8.1500 0.0001  -8.1000 0.0002  -8.0500 0.0002  -8.0000 0.0002  -7.9500 0.0002  -7.9000 0.0002  -7.8500 0.0002  -7.8000 0.0002  -7.7500 0.0002  -7.7000 0.0002  -7.6500 0.0002  -7.6000 0.0003  -7.5500 0.0003  -7.5000 0.0003  -7.4500 0.0003  -7.4000 0.0003  -7.3500 0.0003  -7.3000 0.0003  -7.2500 0.0004  -7.2000 0.0004  -7.1500 0.0004  -7.1000 0.0004  -7.0500 0.0004  -7.0000 0.0005  -6.9500 0.0005  -6.9000 0.0005  -6.8500 0.0005  -6.8000 0.0006  -6.7500 0.0006  -6.7000 0.0006  -6.6500 0.0006  -6.6000 0.0007  -6.5500 0.0007  -6.5000 0.0008  -6.4500 0.0008  -6.4000 0.0008  -6.3500 0.0009  -6.3000 0.0009  -6.2500 0.0010  -6.2000 0.0010  -6.1500 0.0011  -6.1000 0.0011  -6.0500 0.0012  -6.0000 0.0012  -5.9500 0.0013  -5.9000 0.0014  -5.8500 0.0014  -5.8000 0.0015  -5.7500 0.0016  -5.7000 0.0017  -5.6500 0.0018  -5.6000 0.0018  -5.5500 0.0019  -5.5000 0.0020  -5.4500 0.0021  -5.4000 0.0023  -5.3500 0.0024  -5.3000 0.0025  -5.2500 0.0026  -5.2000 0.0028  -5.1500 0.0029  -5.1000 0.0030  -5.0500 0.0032  -5.0000 0.0034  -4.9500 0.0035  -4.9000 0.0037  -4.8500 0.0039  -4.8000 0.0041  -4.7500 0.0043  -4.7000 0.0045  -4.6500 0.0048  -4.6000 0.0050  -4.5500 0.0053  -4.5000 0.0055  -4.4500 0.0058  -4.4000 0.0061  -4.3500 0.0064  -4.3000 0.0067  -4.2500 0.0071  -4.2000 0.0074  -4.1500 0.0078  -4.1000 0.0082  -4.0500 0.0086  -4.0000 0.0091  -3.9500 0.0095  -3.9000 0.0100  -3.8500 0.0105  -3.8000 0.0111  -3.7500 0.0116  -3.7000 0.0122  -3.6500 0.0128  -3.6000 0.0135  -3.5500 0.0142  -3.5000 0.0149  -3.4500 0.0156  -3.4000 0.0164  -3.3500 0.0172  -3.3000 0.0181  -3.2500 0.0190  -3.2000 0.0200  -3.1500 0.0210  -3.1000 0.0220  -3.0500 0.0231  -3.0000 0.0243  -2.9500 0.0255  -2.9000 0.0268  -2.8500 0.0281  -2.8000 0.0295  -2.7500 0.0310  -2.7000 0.0325  -2.6500 0.0341  -2.6000 0.0358  -2.5500 0.0376  -2.5000 0.0394  -2.4500 0.0414  -2.4000 0.0434  -2.3500 0.0455  -2.3000 0.0477  -2.2500 0.0501  -2.2000 0.0525  -2.1500 0.0550  -2.1000 0.0577  -2.0500 0.0605  -2.0000 0.0634  -1.9500 0.0664  -1.9000 0.0696  -1.8500 0.0729  -1.8000 0.0763  -1.7500 0.0799  -1.7000 0.0837  -1.6500 0.0876  -1.6000 0.0917  -1.5500 0.0959  -1.5000 0.1004  -1.4500 0.1050  -1.4000 0.1098  -1.3500 0.1147  -1.3000 0.1199  -1.2500 0.1253  -1.2000 0.1309  -1.1500 0.1367  -1.1000 0.1427  -1.0500 0.1489  -1.0000 0.1554  -0.9500 0.1620  -0.9000 0.1689  -0.8500 0.1761  -0.8000 0.1834  -0.7500 0.1911  -0.7000 0.1989  -0.6500 0.2070  -0.6000 0.2153  -0.5500 0.2239  -0.5000 0.2327  -0.4500 0.2417  -0.4000 0.2510  -0.3500 0.2605  -0.3000 0.2703  -0.2500 0.2803  -0.2000 0.2905  -0.1500 0.3009  -0.1000 0.3115  -0.0500 0.3223  0.0000 0.3333  0.0500 0.3445  0.1000 0.3559  0.1500 0.3675  0.2000 0.3792  0.2500 0.3910  0.3000 0.4030  0.3500 0.4150  0.4000 0.4272  0.4500 0.4395  0.5000 0.4519  0.5500 0.4643  0.6000 0.4767  0.6500 0.4892  0.7000 0.5017  0.7500 0.5142  0.8000 0.5267  0.8500 0.5391  0.9000 0.5515  0.9500 0.5639  1.0000 0.5761  1.0500 0.5883  1.1000 0.6003  1.1500 0.6123  1.2000 0.6241  1.2500 0.6357  1.3000 0.6472  1.3500 0.6586  1.4000 0.6697  1.4500 0.6807  1.5000 0.6914  1.5500 0.7020  1.6000 0.7124  1.6500 0.7225  1.7000 0.7324  1.7500 0.7421  1.8000 0.7515  1.8500 0.7608  1.9000 0.7697  1.9500 0.7785  2.0000 0.7870  2.0500 0.7952  2.1000 0.8033  2.1500 0.8111  2.2000 0.8186  2.2500 0.8259  2.3000 0.8330  2.3500 0.8398  2.4000 0.8464  2.4500 0.8528  2.5000 0.8590  2.5500 0.8649  2.6000 0.8707  2.6500 0.8762  2.7000 0.8815  2.7500 0.8866  2.8000 0.8916  2.8500 0.8963  2.9000 0.9009  2.9500 0.9052  3.0000 0.9094  3.0500 0.9135  3.1000 0.9173  3.1500 0.9211  3.2000 0.9246  3.2500 0.9280  3.3000 0.9313  3.3500 0.9344  3.4000 0.9374  3.4500 0.9403  3.5000 0.9430  3.5500 0.9457  3.6000 0.9482  3.6500 0.9506  3.7000 0.9529  3.7500 0.9551  3.8000 0.9572  3.8500 0.9592  3.9000 0.9611  3.9500 0.9629  4.0000 0.9647  4.0500 0.9663  4.1000 0.9679  4.1500 0.9694  4.2000 0.9709  4.2500 0.9723  4.3000 0.9736  4.3500 0.9748  4.4000 0.9760  4.4500 0.9772  4.5000 0.9783  4.5500 0.9793  4.6000 0.9803  4.6500 0.9812  4.7000 0.9821  4.7500 0.9830  4.8000 0.9838  4.8500 0.9846  4.9000 0.9853  4.9500 0.9860  5.0000 0.9867  5.0500 0.9873  5.1000 0.9880  5.1500 0.9885  5.2000 0.9891  5.2500 0.9896  5.3000 0.9901  5.3500 0.9906  5.4000 0.9910  5.4500 0.9915  5.5000 0.9919  5.5500 0.9923  5.6000 0.9927  5.6500 0.9930  5.7000 0.9934  5.7500 0.9937  5.8000 0.9940  5.8500 0.9943  5.9000 0.9946  5.9500 0.9948  6.0000 0.9951  6.0500 0.9953  6.1000 0.9955  6.1500 0.9958  6.2000 0.9960  6.2500 0.9962  6.3000 0.9963  6.3500 0.9965  6.4000 0.9967  6.4500 0.9968  6.5000 0.9970  6.5500 0.9971  6.6000 0.9973  6.6500 0.9974  6.7000 0.9975  6.7500 0.9977  6.8000 0.9978  6.8500 0.9979  6.9000 0.9980  6.9500 0.9981  7.0000 0.9982  7.0500 0.9983  7.1000 0.9984  7.1500 0.9984  7.2000 0.9985  7.2500 0.9986  7.3000 0.9987  7.3500 0.9987  7.4000 0.9988  7.4500 0.9988  7.5000 0.9989  7.5500 0.9989  7.6000 0.9990  7.6500 0.9990  7.7000 0.9991  7.7500 0.9991  7.8000 0.9992  7.8500 0.9992  7.9000 0.9993  7.9500 0.9993  8.0000 0.9993  8.0500 0.9994  8.1000 0.9994  8.1500 0.9994  8.2000 0.9995  8.2500 0.9995  8.3000 0.9995  8.3500 0.9995  8.4000 0.9996  8.4500 0.9996  8.5000 0.9996  8.5500 0.9996  8.6000 0.9996  8.6500 0.9996  8.7000 0.9997  8.7500 0.9997  8.8000 0.9997  8.8500 0.9997  8.9000 0.9997  8.9500 0.9997  9.0000 0.9998  9.0500 0.9998  9.1000 0.9998  9.1500 0.9998  9.2000 0.9998  9.2500 0.9998  9.3000 0.9998  9.3500 0.9998  9.4000 0.9998  9.4500 0.9998  9.5000 0.9999  9.5500 0.9999  9.6000 0.9999  9.6500 0.9999  9.7000 0.9999  9.7500 0.9999  9.8000 0.9999  9.8500 0.9999  9.9000 0.9999  9.9500 0.9999  10.0000 0.9999   /

 \color{blue}

\setplotsymbol ({\LARGE $\cdot$})

\plot -10.0000 1.0000  -9.9500 1.0000  -9.9000 1.0000  -9.8500 1.0000  -9.8000 1.0000  -9.7500 1.0000  -9.7000 1.0000  -9.6500 1.0000  -9.6000 1.0000  -9.5500 1.0000  -9.5000 1.0000  -9.4500 1.0000  -9.4000 1.0000  -9.3500 1.0000  -9.3000 1.0000  -9.2500 1.0000  -9.2000 0.9999  -9.1500 0.9999  -9.1000 0.9999  -9.0500 0.9999  -9.0000 0.9999  -8.9500 0.9999  -8.9000 0.9999  -8.8500 0.9999  -8.8000 0.9999  -8.7500 0.9999  -8.7000 0.9999  -8.6500 0.9999  -8.6000 0.9999  -8.5500 0.9999  -8.5000 0.9999  -8.4500 0.9999  -8.4000 0.9999  -8.3500 0.9999  -8.3000 0.9999  -8.2500 0.9999  -8.2000 0.9999  -8.1500 0.9999  -8.1000 0.9998  -8.0500 0.9998  -8.0000 0.9998  -7.9500 0.9998  -7.9000 0.9998  -7.8500 0.9998  -7.8000 0.9998  -7.7500 0.9998  -7.7000 0.9998  -7.6500 0.9998  -7.6000 0.9997  -7.5500 0.9997  -7.5000 0.9997  -7.4500 0.9997  -7.4000 0.9997  -7.3500 0.9997  -7.3000 0.9997  -7.2500 0.9996  -7.2000 0.9996  -7.1500 0.9996  -7.1000 0.9996  -7.0500 0.9996  -7.0000 0.9995  -6.9500 0.9995  -6.9000 0.9995  -6.8500 0.9995  -6.8000 0.9994  -6.7500 0.9994  -6.7000 0.9994  -6.6500 0.9994  -6.6000 0.9993  -6.5500 0.9993  -6.5000 0.9992  -6.4500 0.9992  -6.4000 0.9992  -6.3500 0.9991  -6.3000 0.9991  -6.2500 0.9990  -6.2000 0.9990  -6.1500 0.9989  -6.1000 0.9989  -6.0500 0.9988  -6.0000 0.9988  -5.9500 0.9987  -5.9000 0.9986  -5.8500 0.9986  -5.8000 0.9985  -5.7500 0.9984  -5.7000 0.9983  -5.6500 0.9982  -5.6000 0.9982  -5.5500 0.9981  -5.5000 0.9980  -5.4500 0.9979  -5.4000 0.9977  -5.3500 0.9976  -5.3000 0.9975  -5.2500 0.9974  -5.2000 0.9972  -5.1500 0.9971  -5.1000 0.9970  -5.0500 0.9968  -5.0000 0.9966  -4.9500 0.9965  -4.9000 0.9963  -4.8500 0.9961  -4.8000 0.9959  -4.7500 0.9957  -4.7000 0.9955  -4.6500 0.9952  -4.6000 0.9950  -4.5500 0.9947  -4.5000 0.9945  -4.4500 0.9942  -4.4000 0.9939  -4.3500 0.9936  -4.3000 0.9933  -4.2500 0.9929  -4.2000 0.9926  -4.1500 0.9922  -4.1000 0.9918  -4.0500 0.9914  -4.0000 0.9909  -3.9500 0.9905  -3.9000 0.9900  -3.8500 0.9895  -3.8000 0.9889  -3.7500 0.9884  -3.7000 0.9878  -3.6500 0.9872  -3.6000 0.9865  -3.5500 0.9858  -3.5000 0.9851  -3.4500 0.9844  -3.4000 0.9836  -3.3500 0.9828  -3.3000 0.9819  -3.2500 0.9810  -3.2000 0.9800  -3.1500 0.9790  -3.1000 0.9780  -3.0500 0.9769  -3.0000 0.9757  -2.9500 0.9745  -2.9000 0.9732  -2.8500 0.9719  -2.8000 0.9705  -2.7500 0.9690  -2.7000 0.9675  -2.6500 0.9659  -2.6000 0.9642  -2.5500 0.9624  -2.5000 0.9606  -2.4500 0.9586  -2.4000 0.9566  -2.3500 0.9545  -2.3000 0.9523  -2.2500 0.9499  -2.2000 0.9475  -2.1500 0.9450  -2.1000 0.9423  -2.0500 0.9395  -2.0000 0.9366  -1.9500 0.9336  -1.9000 0.9304  -1.8500 0.9271  -1.8000 0.9237  -1.7500 0.9201  -1.7000 0.9163  -1.6500 0.9124  -1.6000 0.9083  -1.5500 0.9041  -1.5000 0.8996  -1.4500 0.8950  -1.4000 0.8902  -1.3500 0.8853  -1.3000 0.8801  -1.2500 0.8747  -1.2000 0.8691  -1.1500 0.8633  -1.1000 0.8573  -1.0500 0.8511  -1.0000 0.8446  -0.9500 0.8380  -0.9000 0.8311  -0.8500 0.8239  -0.8000 0.8166  -0.7500 0.8089  -0.7000 0.8011  -0.6500 0.7930  -0.6000 0.7847  -0.5500 0.7761  -0.5000 0.7673  -0.4500 0.7583  -0.4000 0.7490  -0.3500 0.7395  -0.3000 0.7297  -0.2500 0.7197  -0.2000 0.7095  -0.1500 0.6991  -0.1000 0.6885  -0.0500 0.6777  0.0000 0.6667  0.0500 0.6555  0.1000 0.6441  0.1500 0.6325  0.2000 0.6208  0.2500 0.6090  0.3000 0.5970  0.3500 0.5850  0.4000 0.5728  0.4500 0.5605  0.5000 0.5481  0.5500 0.5357  0.6000 0.5233  0.6500 0.5108  0.7000 0.4983  0.7500 0.4858  0.8000 0.4733  0.8500 0.4609  0.9000 0.4485  0.9500 0.4361  1.0000 0.4239  1.0500 0.4117  1.1000 0.3997  1.1500 0.3877  1.2000 0.3759  1.2500 0.3643  1.3000 0.3528  1.3500 0.3414  1.4000 0.3303  1.4500 0.3193  1.5000 0.3086  1.5500 0.2980  1.6000 0.2876  1.6500 0.2775  1.7000 0.2676  1.7500 0.2579  1.8000 0.2485  1.8500 0.2392  1.9000 0.2303  1.9500 0.2215  2.0000 0.2130  2.0500 0.2048  2.1000 0.1967  2.1500 0.1889  2.2000 0.1814  2.2500 0.1741  2.3000 0.1670  2.3500 0.1602  2.4000 0.1536  2.4500 0.1472  2.5000 0.1410  2.5500 0.1351  2.6000 0.1293  2.6500 0.1238  2.7000 0.1185  2.7500 0.1134  2.8000 0.1084  2.8500 0.1037  2.9000 0.0991  2.9500 0.0948  3.0000 0.0906  3.0500 0.0865  3.1000 0.0827  3.1500 0.0789  3.2000 0.0754  3.2500 0.0720  3.3000 0.0687  3.3500 0.0656  3.4000 0.0626  3.4500 0.0597  3.5000 0.0570  3.5500 0.0543  3.6000 0.0518  3.6500 0.0494  3.7000 0.0471  3.7500 0.0449  3.8000 0.0428  3.8500 0.0408  3.9000 0.0389  3.9500 0.0371  4.0000 0.0353  4.0500 0.0337  4.1000 0.0321  4.1500 0.0306  4.2000 0.0291  4.2500 0.0277  4.3000 0.0264  4.3500 0.0252  4.4000 0.0240  4.4500 0.0228  4.5000 0.0217  4.5500 0.0207  4.6000 0.0197  4.6500 0.0188  4.7000 0.0179  4.7500 0.0170  4.8000 0.0162  4.8500 0.0154  4.9000 0.0147  4.9500 0.0140  5.0000 0.0133  5.0500 0.0127  5.1000 0.0120  5.1500 0.0115  5.2000 0.0109  5.2500 0.0104  5.3000 0.0099  5.3500 0.0094  5.4000 0.0090  5.4500 0.0085  5.5000 0.0081  5.5500 0.0077  5.6000 0.0073  5.6500 0.0070  5.7000 0.0066  5.7500 0.0063  5.8000 0.0060  5.8500 0.0057  5.9000 0.0054  5.9500 0.0052  6.0000 0.0049  6.0500 0.0047  6.1000 0.0045  6.1500 0.0042  6.2000 0.0040  6.2500 0.0038  6.3000 0.0037  6.3500 0.0035  6.4000 0.0033  6.4500 0.0032  6.5000 0.0030  6.5500 0.0029  6.6000 0.0027  6.6500 0.0026  6.7000 0.0025  6.7500 0.0023  6.8000 0.0022  6.8500 0.0021  6.9000 0.0020  6.9500 0.0019  7.0000 0.0018  7.0500 0.0017  7.1000 0.0016  7.1500 0.0016  7.2000 0.0015  7.2500 0.0014  7.3000 0.0013  7.3500 0.0013  7.4000 0.0012  7.4500 0.0012  7.5000 0.0011  7.5500 0.0011  7.6000 0.0010  7.6500 0.0010  7.7000 0.0009  7.7500 0.0009  7.8000 0.0008  7.8500 0.0008  7.9000 0.0007  7.9500 0.0007  8.0000 0.0007  8.0500 0.0006  8.1000 0.0006  8.1500 0.0006  8.2000 0.0005  8.2500 0.0005  8.3000 0.0005  8.3500 0.0005  8.4000 0.0004  8.4500 0.0004  8.5000 0.0004  8.5500 0.0004  8.6000 0.0004  8.6500 0.0004  8.7000 0.0003  8.7500 0.0003  8.8000 0.0003  8.8500 0.0003  8.9000 0.0003  8.9500 0.0003  9.0000 0.0002  9.0500 0.0002  9.1000 0.0002  9.1500 0.0002  9.2000 0.0002  9.2500 0.0002  9.3000 0.0002  9.3500 0.0002  9.4000 0.0002  9.4500 0.0002  9.5000 0.0001  9.5500 0.0001  9.6000 0.0001  9.6500 0.0001  9.7000 0.0001  9.7500 0.0001  9.8000 0.0001  9.8500 0.0001  9.9000 0.0001  9.9500 0.0001  10.0000 0.0001   /

 \color{black}
\put {O} at 0 0 
\setplotarea x from -10.000000 to 10.000000, y from 0.000000 to 1.200000
\axis top /
\axis right /
\axis left /
\axis bottom 
 label {{\Large \hspace{2cm} $f$}} /
\axis bottom 
 ticks out withvalues -10.00 0 10.00 / quantity 3 /
\axis left shiftedto x=0 
 label {{\Large $\hbox{$\qquad$}$}} /
\axis left shiftedto y=0 
 ticks out withvalues  { } { } 0.4 { } { } 1.0 { } / quantity 7 /

\put {\Large $\kappa$} at 5 0.9
\put {\Large $\LA h \RA$} at -5 0.9

\normalcolor
\endpicture

%% file: Compressgraphs/Compress3-1L.tex
\beginpicture

 \setcoordinatesystem units <7.500 pt,31.5 pt>

 \color{red}

\setplotsymbol ({\LARGE $\cdot$})

\plot -10.0000 3.9832  -9.9500 3.9823  -9.9000 3.9814  -9.8500 3.9805  -9.8000 3.9795  -9.7500 3.9785  -9.7000 3.9774  -9.6500 3.9763  -9.6000 3.9751  -9.5500 3.9739  -9.5000 3.9726  -9.4500 3.9712  -9.4000 3.9698  -9.3500 3.9683  -9.3000 3.9667  -9.2500 3.9650  -9.2000 3.9633  -9.1500 3.9615  -9.1000 3.9596  -9.0500 3.9576  -9.0000 3.9555  -8.9500 3.9534  -8.9000 3.9511  -8.8500 3.9487  -8.8000 3.9462  -8.7500 3.9436  -8.7000 3.9409  -8.6500 3.9381  -8.6000 3.9351  -8.5500 3.9320  -8.5000 3.9288  -8.4500 3.9254  -8.4000 3.9218  -8.3500 3.9182  -8.3000 3.9143  -8.2500 3.9103  -8.2000 3.9062  -8.1500 3.9018  -8.1000 3.8973  -8.0500 3.8926  -8.0000 3.8877  -7.9500 3.8826  -7.9000 3.8774  -7.8500 3.8719  -7.8000 3.8662  -7.7500 3.8603  -7.7000 3.8542  -7.6500 3.8478  -7.6000 3.8413  -7.5500 3.8345  -7.5000 3.8274  -7.4500 3.8202  -7.4000 3.8127  -7.3500 3.8050  -7.3000 3.7970  -7.2500 3.7888  -7.2000 3.7803  -7.1500 3.7716  -7.1000 3.7627  -7.0500 3.7535  -7.0000 3.7441  -6.9500 3.7345  -6.9000 3.7246  -6.8500 3.7145  -6.8000 3.7042  -6.7500 3.6937  -6.7000 3.6830  -6.6500 3.6720  -6.6000 3.6609  -6.5500 3.6496  -6.5000 3.6382  -6.4500 3.6265  -6.4000 3.6147  -6.3500 3.6028  -6.3000 3.5908  -6.2500 3.5786  -6.2000 3.5664  -6.1500 3.5541  -6.1000 3.5417  -6.0500 3.5292  -6.0000 3.5167  -5.9500 3.5042  -5.9000 3.4917  -5.8500 3.4792  -5.8000 3.4667  -5.7500 3.4543  -5.7000 3.4419  -5.6500 3.4296  -5.6000 3.4174  -5.5500 3.4053  -5.5000 3.3932  -5.4500 3.3814  -5.4000 3.3696  -5.3500 3.3580  -5.3000 3.3466  -5.2500 3.3353  -5.2000 3.3242  -5.1500 3.3133  -5.1000 3.3026  -5.0500 3.2921  -5.0000 3.2818  -4.9500 3.2718  -4.9000 3.2619  -4.8500 3.2523  -4.8000 3.2430  -4.7500 3.2338  -4.7000 3.2249  -4.6500 3.2162  -4.6000 3.2078  -4.5500 3.1996  -4.5000 3.1917  -4.4500 3.1840  -4.4000 3.1765  -4.3500 3.1692  -4.3000 3.1622  -4.2500 3.1554  -4.2000 3.1488  -4.1500 3.1425  -4.1000 3.1363  -4.0500 3.1304  -4.0000 3.1247  -3.9500 3.1192  -3.9000 3.1138  -3.8500 3.1087  -3.8000 3.1038  -3.7500 3.0990  -3.7000 3.0944  -3.6500 3.0900  -3.6000 3.0857  -3.5500 3.0817  -3.5000 3.0777  -3.4500 3.0739  -3.4000 3.0703  -3.3500 3.0668  -3.3000 3.0634  -3.2500 3.0601  -3.2000 3.0570  -3.1500 3.0540  -3.1000 3.0511  -3.0500 3.0483  -3.0000 3.0456  -2.9500 3.0430  -2.9000 3.0404  -2.8500 3.0380  -2.8000 3.0357  -2.7500 3.0334  -2.7000 3.0312  -2.6500 3.0290  -2.6000 3.0270  -2.5500 3.0249  -2.5000 3.0230  -2.4500 3.0211  -2.4000 3.0192  -2.3500 3.0174  -2.3000 3.0156  -2.2500 3.0139  -2.2000 3.0121  -2.1500 3.0104  -2.1000 3.0088  -2.0500 3.0071  -2.0000 3.0055  -1.9500 3.0038  -1.9000 3.0022  -1.8500 3.0006  -1.8000 2.9989  -1.7500 2.9973  -1.7000 2.9957  -1.6500 2.9940  -1.6000 2.9923  -1.5500 2.9906  -1.5000 2.9889  -1.4500 2.9872  -1.4000 2.9854  -1.3500 2.9835  -1.3000 2.9817  -1.2500 2.9797  -1.2000 2.9777  -1.1500 2.9757  -1.1000 2.9736  -1.0500 2.9714  -1.0000 2.9692  -0.9500 2.9668  -0.9000 2.9644  -0.8500 2.9619  -0.8000 2.9592  -0.7500 2.9565  -0.7000 2.9536  -0.6500 2.9506  -0.6000 2.9475  -0.5500 2.9443  -0.5000 2.9408  -0.4500 2.9373  -0.4000 2.9335  -0.3500 2.9296  -0.3000 2.9255  -0.2500 2.9211  -0.2000 2.9166  -0.1500 2.9118  -0.1000 2.9068  -0.0500 2.9015  0.0000 2.8960  0.0500 2.8901  0.1000 2.8840  0.1500 2.8775  0.2000 2.8707  0.2500 2.8636  0.3000 2.8561  0.3500 2.8482  0.4000 2.8398  0.4500 2.8311  0.5000 2.8218  0.5500 2.8121  0.6000 2.8019  0.6500 2.7912  0.7000 2.7799  0.7500 2.7681  0.8000 2.7556  0.8500 2.7425  0.9000 2.7288  0.9500 2.7144  1.0000 2.6994  1.0500 2.6836  1.1000 2.6671  1.1500 2.6498  1.2000 2.6317  1.2500 2.6129  1.3000 2.5932  1.3500 2.5728  1.4000 2.5515  1.4500 2.5294  1.5000 2.5064  1.5500 2.4826  1.6000 2.4580  1.6500 2.4326  1.7000 2.4064  1.7500 2.3795  1.8000 2.3518  1.8500 2.3233  1.9000 2.2943  1.9500 2.2646  2.0000 2.2343  2.0500 2.2035  2.1000 2.1723  2.1500 2.1406  2.2000 2.1087  2.2500 2.0764  2.3000 2.0440  2.3500 2.0116  2.4000 1.9790  2.4500 1.9466  2.5000 1.9142  2.5500 1.8821  2.6000 1.8502  2.6500 1.8187  2.7000 1.7876  2.7500 1.7570  2.8000 1.7269  2.8500 1.6974  2.9000 1.6685  2.9500 1.6403  3.0000 1.6128  3.0500 1.5861  3.1000 1.5602  3.1500 1.5350  3.2000 1.5107  3.2500 1.4872  3.3000 1.4645  3.3500 1.4427  3.4000 1.4217  3.4500 1.4015  3.5000 1.3821  3.5500 1.3636  3.6000 1.3458  3.6500 1.3288  3.7000 1.3125  3.7500 1.2970  3.8000 1.2822  3.8500 1.2681  3.9000 1.2546  3.9500 1.2418  4.0000 1.2296  4.0500 1.2180  4.1000 1.2069  4.1500 1.1964  4.2000 1.1865  4.2500 1.1770  4.3000 1.1680  4.3500 1.1594  4.4000 1.1513  4.4500 1.1436  4.5000 1.1363  4.5500 1.1294  4.6000 1.1228  4.6500 1.1166  4.7000 1.1106  4.7500 1.1050  4.8000 1.0997  4.8500 1.0946  4.9000 1.0898  4.9500 1.0853  5.0000 1.0810  5.0500 1.0769  5.1000 1.0730  5.1500 1.0693  5.2000 1.0658  5.2500 1.0625  5.3000 1.0594  5.3500 1.0564  5.4000 1.0536  5.4500 1.0509  5.5000 1.0483  5.5500 1.0459  5.6000 1.0436  5.6500 1.0415  5.7000 1.0394  5.7500 1.0374  5.8000 1.0356  5.8500 1.0338  5.9000 1.0321  5.9500 1.0305  6.0000 1.0290  6.0500 1.0276  6.1000 1.0262  6.1500 1.0249  6.2000 1.0237  6.2500 1.0225  6.3000 1.0214  6.3500 1.0203  6.4000 1.0193  6.4500 1.0184  6.5000 1.0175  6.5500 1.0166  6.6000 1.0158  6.6500 1.0150  6.7000 1.0143  6.7500 1.0136  6.8000 1.0129  6.8500 1.0123  6.9000 1.0117  6.9500 1.0111  7.0000 1.0105  7.0500 1.0100  7.1000 1.0095  7.1500 1.0091  7.2000 1.0086  7.2500 1.0082  7.3000 1.0078  7.3500 1.0074  7.4000 1.0071  7.4500 1.0067  7.5000 1.0064  7.5500 1.0061  7.6000 1.0058  7.6500 1.0055  7.7000 1.0052  7.7500 1.0050  7.8000 1.0047  7.8500 1.0045  7.9000 1.0043  7.9500 1.0041  8.0000 1.0039  8.0500 1.0037  8.1000 1.0035  8.1500 1.0033  8.2000 1.0032  8.2500 1.0030  8.3000 1.0029  8.3500 1.0027  8.4000 1.0026  8.4500 1.0025  8.5000 1.0023  8.5500 1.0022  8.6000 1.0021  8.6500 1.0020  8.7000 1.0019  8.7500 1.0018  8.8000 1.0017  8.8500 1.0016  8.9000 1.0016  8.9500 1.0015  9.0000 1.0014  9.0500 1.0013  9.1000 1.0013  9.1500 1.0012  9.2000 1.0012  9.2500 1.0011  9.3000 1.0011  9.3500 1.0010  9.4000 1.0010  9.4500 1.0009  9.5000 1.0009  9.5500 1.0008  9.6000 1.0008  9.6500 1.0007  9.7000 1.0007  9.7500 1.0007  9.8000 1.0006  9.8500 1.0006  9.9000 1.0006  9.9500 1.0005  10.0000 1.0005   /

 \color{black}
\put {O} at 0 0 
\setplotarea x from -10.000000 to 10.000000, y from 0.000000 to 4.200000
\axis top /
\axis right /
\axis left /
\axis bottom 
 label {{\Large \hspace{2cm} $f$}} /
\axis bottom 
 ticks out withvalues -10 0 10 / quantity 3 /
\axis left shiftedto x=0 
 label {{\Large $\LA h \RA$}} /
\axis left shiftedto x=0 
 ticks out withvalues 0 { } 1.4 { } {2.8} {} {} {} / quantity 7 /

\normalcolor
\endpicture

%% file: Compressgraphs/Compress3-1F.tex
\beginpicture

 \setcoordinatesystem units <7.500 pt,262.5 pt>

 \color{blue}

\setplotsymbol ({\LARGE $\cdot$})

\plot -10.0000 0.0042  -9.9500 0.0044  -9.9000 0.0046  -9.8500 0.0048  -9.8000 0.0050  -9.7500 0.0053  -9.7000 0.0055  -9.6500 0.0058  -9.6000 0.0061  -9.5500 0.0064  -9.5000 0.0067  -9.4500 0.0070  -9.4000 0.0074  -9.3500 0.0077  -9.3000 0.0081  -9.2500 0.0085  -9.2000 0.0089  -9.1500 0.0093  -9.1000 0.0098  -9.0500 0.0103  -9.0000 0.0107  -8.9500 0.0112  -8.9000 0.0118  -8.8500 0.0123  -8.8000 0.0129  -8.7500 0.0135  -8.7000 0.0141  -8.6500 0.0148  -8.6000 0.0154  -8.5500 0.0161  -8.5000 0.0168  -8.4500 0.0176  -8.4000 0.0184  -8.3500 0.0192  -8.3000 0.0200  -8.2500 0.0209  -8.2000 0.0218  -8.1500 0.0227  -8.1000 0.0236  -8.0500 0.0246  -8.0000 0.0256  -7.9500 0.0267  -7.9000 0.0278  -7.8500 0.0289  -7.8000 0.0300  -7.7500 0.0311  -7.7000 0.0323  -7.6500 0.0335  -7.6000 0.0348  -7.5500 0.0360  -7.5000 0.0373  -7.4500 0.0386  -7.4000 0.0399  -7.3500 0.0413  -7.3000 0.0426  -7.2500 0.0440  -7.2000 0.0454  -7.1500 0.0467  -7.1000 0.0481  -7.0500 0.0495  -7.0000 0.0509  -6.9500 0.0522  -6.9000 0.0536  -6.8500 0.0549  -6.8000 0.0562  -6.7500 0.0575  -6.7000 0.0588  -6.6500 0.0600  -6.6000 0.0612  -6.5500 0.0624  -6.5000 0.0635  -6.4500 0.0646  -6.4000 0.0656  -6.3500 0.0665  -6.3000 0.0674  -6.2500 0.0682  -6.2000 0.0689  -6.1500 0.0696  -6.1000 0.0702  -6.0500 0.0707  -6.0000 0.0711  -5.9500 0.0714  -5.9000 0.0717  -5.8500 0.0718  -5.8000 0.0719  -5.7500 0.0719  -5.7000 0.0718  -5.6500 0.0716  -5.6000 0.0713  -5.5500 0.0709  -5.5000 0.0705  -5.4500 0.0699  -5.4000 0.0693  -5.3500 0.0686  -5.3000 0.0679  -5.2500 0.0670  -5.2000 0.0661  -5.1500 0.0652  -5.1000 0.0642  -5.0500 0.0631  -5.0000 0.0620  -4.9500 0.0608  -4.9000 0.0596  -4.8500 0.0584  -4.8000 0.0571  -4.7500 0.0558  -4.7000 0.0545  -4.6500 0.0532  -4.6000 0.0518  -4.5500 0.0505  -4.5000 0.0491  -4.4500 0.0478  -4.4000 0.0464  -4.3500 0.0450  -4.3000 0.0437  -4.2500 0.0424  -4.2000 0.0410  -4.1500 0.0397  -4.1000 0.0385  -4.0500 0.0372  -4.0000 0.0360  -3.9500 0.0347  -3.9000 0.0336  -3.8500 0.0324  -3.8000 0.0313  -3.7500 0.0302  -3.7000 0.0291  -3.6500 0.0281  -3.6000 0.0270  -3.5500 0.0261  -3.5000 0.0251  -3.4500 0.0242  -3.4000 0.0233  -3.3500 0.0225  -3.3000 0.0217  -3.2500 0.0209  -3.2000 0.0201  -3.1500 0.0194  -3.1000 0.0187  -3.0500 0.0181  -3.0000 0.0174  -2.9500 0.0168  -2.9000 0.0163  -2.8500 0.0157  -2.8000 0.0152  -2.7500 0.0148  -2.7000 0.0143  -2.6500 0.0139  -2.6000 0.0135  -2.5500 0.0131  -2.5000 0.0128  -2.4500 0.0125  -2.4000 0.0122  -2.3500 0.0120  -2.3000 0.0117  -2.2500 0.0115  -2.2000 0.0113  -2.1500 0.0112  -2.1000 0.0111  -2.0500 0.0110  -2.0000 0.0109  -1.9500 0.0109  -1.9000 0.0108  -1.8500 0.0108  -1.8000 0.0109  -1.7500 0.0109  -1.7000 0.0110  -1.6500 0.0111  -1.6000 0.0113  -1.5500 0.0114  -1.5000 0.0116  -1.4500 0.0119  -1.4000 0.0121  -1.3500 0.0124  -1.3000 0.0128  -1.2500 0.0131  -1.2000 0.0135  -1.1500 0.0139  -1.1000 0.0144  -1.0500 0.0149  -1.0000 0.0155  -0.9500 0.0161  -0.9000 0.0167  -0.8500 0.0174  -0.8000 0.0181  -0.7500 0.0189  -0.7000 0.0198  -0.6500 0.0207  -0.6000 0.0216  -0.5500 0.0227  -0.5000 0.0238  -0.4500 0.0249  -0.4000 0.0262  -0.3500 0.0275  -0.3000 0.0289  -0.2500 0.0304  -0.2000 0.0319  -0.1500 0.0336  -0.1000 0.0354  -0.0500 0.0373  0.0000 0.0393  0.0500 0.0414  0.1000 0.0437  0.1500 0.0460  0.2000 0.0486  0.2500 0.0512  0.3000 0.0540  0.3500 0.0570  0.4000 0.0602  0.4500 0.0635  0.5000 0.0670  0.5500 0.0708  0.6000 0.0747  0.6500 0.0788  0.7000 0.0832  0.7500 0.0878  0.8000 0.0926  0.8500 0.0976  0.9000 0.1029  0.9500 0.1085  1.0000 0.1143  1.0500 0.1204  1.1000 0.1267  1.1500 0.1333  1.2000 0.1402  1.2500 0.1473  1.3000 0.1547  1.3500 0.1623  1.4000 0.1701  1.4500 0.1782  1.5000 0.1864  1.5500 0.1949  1.6000 0.2034  1.6500 0.2122  1.7000 0.2209  1.7500 0.2298  1.8000 0.2387  1.8500 0.2475  1.9000 0.2563  1.9500 0.2649  2.0000 0.2734  2.0500 0.2816  2.1000 0.2896  2.1500 0.2972  2.2000 0.3045  2.2500 0.3113  2.3000 0.3176  2.3500 0.3234  2.4000 0.3286  2.4500 0.3331  2.5000 0.3370  2.5500 0.3402  2.6000 0.3427  2.6500 0.3445  2.7000 0.3455  2.7500 0.3457  2.8000 0.3452  2.8500 0.3440  2.9000 0.3420  2.9500 0.3393  3.0000 0.3360  3.0500 0.3320  3.1000 0.3274  3.1500 0.3223  3.2000 0.3166  3.2500 0.3105  3.3000 0.3039  3.3500 0.2970  3.4000 0.2897  3.4500 0.2822  3.5000 0.2744  3.5500 0.2665  3.6000 0.2584  3.6500 0.2503  3.7000 0.2421  3.7500 0.2339  3.8000 0.2257  3.8500 0.2175  3.9000 0.2094  3.9500 0.2015  4.0000 0.1936  4.0500 0.1859  4.1000 0.1784  4.1500 0.1711  4.2000 0.1639  4.2500 0.1569  4.3000 0.1502  4.3500 0.1436  4.4000 0.1373  4.4500 0.1312  4.5000 0.1253  4.5500 0.1196  4.6000 0.1141  4.6500 0.1089  4.7000 0.1038  4.7500 0.0990  4.8000 0.0943  4.8500 0.0899  4.9000 0.0856  4.9500 0.0816  5.0000 0.0777  5.0500 0.0740  5.1000 0.0704  5.1500 0.0670  5.2000 0.0638  5.2500 0.0607  5.3000 0.0578  5.3500 0.0550  5.4000 0.0523  5.4500 0.0498  5.5000 0.0473  5.5500 0.0450  5.6000 0.0428  5.6500 0.0408  5.7000 0.0388  5.7500 0.0369  5.8000 0.0351  5.8500 0.0334  5.9000 0.0317  5.9500 0.0302  6.0000 0.0287  6.0500 0.0273  6.1000 0.0259  6.1500 0.0247  6.2000 0.0235  6.2500 0.0223  6.3000 0.0212  6.3500 0.0202  6.4000 0.0192  6.4500 0.0183  6.5000 0.0174  6.5500 0.0165  6.6000 0.0157  6.6500 0.0149  6.7000 0.0142  6.7500 0.0135  6.8000 0.0128  6.8500 0.0122  6.9000 0.0116  6.9500 0.0111  7.0000 0.0105  7.0500 0.0100  7.1000 0.0095  7.1500 0.0090  7.2000 0.0086  7.2500 0.0082  7.3000 0.0078  7.3500 0.0074  7.4000 0.0070  7.4500 0.0067  7.5000 0.0064  7.5500 0.0061  7.6000 0.0058  7.6500 0.0055  7.7000 0.0052  7.7500 0.0050  7.8000 0.0047  7.8500 0.0045  7.9000 0.0043  7.9500 0.0041  8.0000 0.0039  8.0500 0.0037  8.1000 0.0035  8.1500 0.0033  8.2000 0.0032  8.2500 0.0030  8.3000 0.0029  8.3500 0.0027  8.4000 0.0026  8.4500 0.0025  8.5000 0.0023  8.5500 0.0022  8.6000 0.0021  8.6500 0.0020  8.7000 0.0019  8.7500 0.0018  8.8000 0.0017  8.8500 0.0016  8.9000 0.0016  8.9500 0.0015  9.0000 0.0014  9.0500 0.0013  9.1000 0.0013  9.1500 0.0012  9.2000 0.0012  9.2500 0.0011  9.3000 0.0010  9.3500 0.0010  9.4000 0.0009  9.4500 0.0009  9.5000 0.0009  9.5500 0.0008  9.6000 0.0008  9.6500 0.0007  9.7000 0.0007  9.7500 0.0007  9.8000 0.0006  9.8500 0.0006  9.9000 0.0006  9.9500 0.0005  10.0000 0.0005   /

 \color{black}
\put {O} at 0 0 
\setplotarea x from -10.000000 to 10.000000, y from 0 to 0.500000
\axis top /
\axis right /
\axis left /
\axis bottom 
 label {{\Large \hspace{2cm} $f$}} /
\axis bottom 
 ticks out withvalues -10 0 10 / quantity 3 /
\axis left shiftedto x=0 
 label {{\Large $\kappa$}} /
\axis left shiftedto x=0 
 ticks out withvalues 0 {} 0.2 {} 0.4 {}  / quantity 6 /

\normalcolor
\endpicture

%% file: Compressgraphs/Compress3-1LA.tex
\beginpicture

 \setcoordinatesystem units <7.500 pt,31.5 pt>

 \color{red}

\setplotsymbol ({\LARGE $\cdot$})

\plot -10.0000 3.9832  -9.9500 3.9823  -9.9000 3.9814  -9.8500 3.9805  -9.8000 3.9795  -9.7500 3.9785  -9.7000 3.9774  -9.6500 3.9763  -9.6000 3.9751  -9.5500 3.9739  -9.5000 3.9726  -9.4500 3.9712  -9.4000 3.9698  -9.3500 3.9683  -9.3000 3.9667  -9.2500 3.9650  -9.2000 3.9633  -9.1500 3.9615  -9.1000 3.9596  -9.0500 3.9576  -9.0000 3.9555  -8.9500 3.9534  -8.9000 3.9511  -8.8500 3.9487  -8.8000 3.9462  -8.7500 3.9436  -8.7000 3.9409  -8.6500 3.9381  -8.6000 3.9351  -8.5500 3.9320  -8.5000 3.9288  -8.4500 3.9254  -8.4000 3.9218  -8.3500 3.9182  -8.3000 3.9143  -8.2500 3.9103  -8.2000 3.9062  -8.1500 3.9018  -8.1000 3.8973  -8.0500 3.8926  -8.0000 3.8877  -7.9500 3.8826  -7.9000 3.8774  -7.8500 3.8719  -7.8000 3.8662  -7.7500 3.8603  -7.7000 3.8542  -7.6500 3.8478  -7.6000 3.8413  -7.5500 3.8345  -7.5000 3.8274  -7.4500 3.8202  -7.4000 3.8127  -7.3500 3.8050  -7.3000 3.7970  -7.2500 3.7888  -7.2000 3.7803  -7.1500 3.7716  -7.1000 3.7627  -7.0500 3.7535  -7.0000 3.7441  -6.9500 3.7345  -6.9000 3.7246  -6.8500 3.7145  -6.8000 3.7042  -6.7500 3.6937  -6.7000 3.6830  -6.6500 3.6720  -6.6000 3.6609  -6.5500 3.6496  -6.5000 3.6382  -6.4500 3.6265  -6.4000 3.6147  -6.3500 3.6028  -6.3000 3.5908  -6.2500 3.5786  -6.2000 3.5664  -6.1500 3.5541  -6.1000 3.5417  -6.0500 3.5292  -6.0000 3.5167  -5.9500 3.5042  -5.9000 3.4917  -5.8500 3.4792  -5.8000 3.4667  -5.7500 3.4543  -5.7000 3.4419  -5.6500 3.4296  -5.6000 3.4174  -5.5500 3.4053  -5.5000 3.3932  -5.4500 3.3814  -5.4000 3.3696  -5.3500 3.3580  -5.3000 3.3466  -5.2500 3.3353  -5.2000 3.3242  -5.1500 3.3133  -5.1000 3.3026  -5.0500 3.2921  -5.0000 3.2818  -4.9500 3.2718  -4.9000 3.2620  -4.8500 3.2523  -4.8000 3.2430  -4.7500 3.2338  -4.7000 3.2249  -4.6500 3.2163  -4.6000 3.2078  -4.5500 3.1996  -4.5000 3.1917  -4.4500 3.1840  -4.4000 3.1765  -4.3500 3.1692  -4.3000 3.1622  -4.2500 3.1554  -4.2000 3.1488  -4.1500 3.1425  -4.1000 3.1363  -4.0500 3.1304  -4.0000 3.1247  -3.9500 3.1192  -3.9000 3.1138  -3.8500 3.1087  -3.8000 3.1038  -3.7500 3.0990  -3.7000 3.0944  -3.6500 3.0900  -3.6000 3.0858  -3.5500 3.0817  -3.5000 3.0777  -3.4500 3.0739  -3.4000 3.0703  -3.3500 3.0668  -3.3000 3.0634  -3.2500 3.0602  -3.2000 3.0570  -3.1500 3.0540  -3.1000 3.0511  -3.0500 3.0483  -3.0000 3.0456  -2.9500 3.0430  -2.9000 3.0405  -2.8500 3.0381  -2.8000 3.0357  -2.7500 3.0334  -2.7000 3.0312  -2.6500 3.0291  -2.6000 3.0271  -2.5500 3.0251  -2.5000 3.0231  -2.4500 3.0212  -2.4000 3.0194  -2.3500 3.0175  -2.3000 3.0158  -2.2500 3.0140  -2.2000 3.0123  -2.1500 3.0107  -2.1000 3.0090  -2.0500 3.0074  -2.0000 3.0058  -1.9500 3.0042  -1.9000 3.0026  -1.8500 3.0010  -1.8000 2.9994  -1.7500 2.9978  -1.7000 2.9962  -1.6500 2.9946  -1.6000 2.9930  -1.5500 2.9914  -1.5000 2.9897  -1.4500 2.9881  -1.4000 2.9864  -1.3500 2.9846  -1.3000 2.9829  -1.2500 2.9811  -1.2000 2.9792  -1.1500 2.9774  -1.1000 2.9754  -1.0500 2.9734  -1.0000 2.9714  -0.9500 2.9693  -0.9000 2.9671  -0.8500 2.9648  -0.8000 2.9625  -0.7500 2.9601  -0.7000 2.9576  -0.6500 2.9550  -0.6000 2.9523  -0.5500 2.9495  -0.5000 2.9466  -0.4500 2.9437  -0.4000 2.9405  -0.3500 2.9373  -0.3000 2.9340  -0.2500 2.9305  -0.2000 2.9268  -0.1500 2.9231  -0.1000 2.9192  -0.0500 2.9151  0.0000 2.9109  0.0500 2.9065  0.1000 2.9019  0.1500 2.8972  0.2000 2.8922  0.2500 2.8871  0.3000 2.8818  0.3500 2.8763  0.4000 2.8706  0.4500 2.8647  0.5000 2.8586  0.5500 2.8522  0.6000 2.8457  0.6500 2.8389  0.7000 2.8318  0.7500 2.8246  0.8000 2.8171  0.8500 2.8093  0.9000 2.8014  0.9500 2.7931  1.0000 2.7847  1.0500 2.7760  1.1000 2.7670  1.1500 2.7578  1.2000 2.7483  1.2500 2.7387  1.3000 2.7287  1.3500 2.7186  1.4000 2.7082  1.4500 2.6975  1.5000 2.6867  1.5500 2.6756  1.6000 2.6643  1.6500 2.6529  1.7000 2.6412  1.7500 2.6294  1.8000 2.6173  1.8500 2.6052  1.9000 2.5928  1.9500 2.5804  2.0000 2.5678  2.0500 2.5551  2.1000 2.5423  2.1500 2.5294  2.2000 2.5164  2.2500 2.5034  2.3000 2.4903  2.3500 2.4773  2.4000 2.4642  2.4500 2.4510  2.5000 2.4380  2.5500 2.4249  2.6000 2.4119  2.6500 2.3989  2.7000 2.3860  2.7500 2.3732  2.8000 2.3604  2.8500 2.3478  2.9000 2.3353  2.9500 2.3228  3.0000 2.3106  3.0500 2.2984  3.1000 2.2864  3.1500 2.2745  3.2000 2.2628  3.2500 2.2512  3.3000 2.2398  3.3500 2.2285  3.4000 2.2174  3.4500 2.2065  3.5000 2.1957  3.5500 2.1851  3.6000 2.1746  3.6500 2.1643  3.7000 2.1541  3.7500 2.1441  3.8000 2.1342  3.8500 2.1244  3.9000 2.1148  3.9500 2.1053  4.0000 2.0959  4.0500 2.0866  4.1000 2.0774  4.1500 2.0683  4.2000 2.0592  4.2500 2.0502  4.3000 2.0413  4.3500 2.0324  4.4000 2.0236  4.4500 2.0148  4.5000 2.0060  4.5500 1.9972  4.6000 1.9884  4.6500 1.9796  4.7000 1.9708  4.7500 1.9620  4.8000 1.9531  4.8500 1.9441  4.9000 1.9351  4.9500 1.9260  5.0000 1.9168  5.0500 1.9076  5.1000 1.8982  5.1500 1.8887  5.2000 1.8791  5.2500 1.8694  5.3000 1.8596  5.3500 1.8496  5.4000 1.8395  5.4500 1.8292  5.5000 1.8188  5.5500 1.8082  5.6000 1.7975  5.6500 1.7866  5.7000 1.7756  5.7500 1.7644  5.8000 1.7530  5.8500 1.7415  5.9000 1.7299  5.9500 1.7181  6.0000 1.7061  6.0500 1.6940  6.1000 1.6818  6.1500 1.6694  6.2000 1.6569  6.2500 1.6443  6.3000 1.6316  6.3500 1.6188  6.4000 1.6060  6.4500 1.5930  6.5000 1.5800  6.5500 1.5670  6.6000 1.5539  6.6500 1.5408  6.7000 1.5277  6.7500 1.5147  6.8000 1.5016  6.8500 1.4886  6.9000 1.4756  6.9500 1.4627  7.0000 1.4499  7.0500 1.4372  7.1000 1.4245  7.1500 1.4121  7.2000 1.3997  7.2500 1.3875  7.3000 1.3754  7.3500 1.3635  7.4000 1.3518  7.4500 1.3403  7.5000 1.3290  7.5500 1.3179  7.6000 1.3070  7.6500 1.2963  7.7000 1.2859  7.7500 1.2756  7.8000 1.2657  7.8500 1.2559  7.9000 1.2464  7.9500 1.2372  8.0000 1.2282  8.0500 1.2194  8.1000 1.2109  8.1500 1.2026  8.2000 1.1946  8.2500 1.1869  8.3000 1.1793  8.3500 1.1721  8.4000 1.1650  8.4500 1.1582  8.5000 1.1516  8.5500 1.1453  8.6000 1.1392  8.6500 1.1332  8.7000 1.1276  8.7500 1.1221  8.8000 1.1168  8.8500 1.1117  8.9000 1.1068  8.9500 1.1021  9.0000 1.0976  9.0500 1.0933  9.1000 1.0892  9.1500 1.0852  9.2000 1.0813  9.2500 1.0777  9.3000 1.0742  9.3500 1.0708  9.4000 1.0676  9.4500 1.0645  9.5000 1.0615  9.5500 1.0587  9.6000 1.0560  9.6500 1.0534  9.7000 1.0509  9.7500 1.0486  9.8000 1.0463  9.8500 1.0441  9.9000 1.0421  9.9500 1.0401  10.0000 1.0382   /

 \color{black}
\put {O} at 0 0 
\setplotarea x from -10.000000 to 10.000000, y from 0.000000 to 4.200000
\axis top /
\axis right /
\axis left /
\axis bottom 
 label {{\Large \hspace{2cm} $f$}} /
\axis bottom 
 ticks out withvalues -10 0 10 / quantity 3 /
\axis left shiftedto x=0 
 label {{\Large $\LA h \RA$}} /
\axis left shiftedto x=0 
 ticks out withvalues 0 { } 1.4 { } {2.8} {} {} {} / quantity 7 /

\normalcolor
\endpicture

%% file: Compressgraphs/Compress3-1FA.tex
\beginpicture

 \setcoordinatesystem units <7.500 pt,262.50 pt>

 \color{blue}

\setplotsymbol ({\LARGE $\cdot$})

\plot -10.0000 0.0042  -9.9500 0.0044  -9.9000 0.0046  -9.8500 0.0048  -9.8000 0.0050  -9.7500 0.0053  -9.7000 0.0055  -9.6500 0.0058  -9.6000 0.0061  -9.5500 0.0064  -9.5000 0.0067  -9.4500 0.0070  -9.4000 0.0074  -9.3500 0.0077  -9.3000 0.0081  -9.2500 0.0085  -9.2000 0.0089  -9.1500 0.0093  -9.1000 0.0098  -9.0500 0.0103  -9.0000 0.0107  -8.9500 0.0112  -8.9000 0.0118  -8.8500 0.0123  -8.8000 0.0129  -8.7500 0.0135  -8.7000 0.0141  -8.6500 0.0148  -8.6000 0.0154  -8.5500 0.0161  -8.5000 0.0168  -8.4500 0.0176  -8.4000 0.0184  -8.3500 0.0192  -8.3000 0.0200  -8.2500 0.0209  -8.2000 0.0218  -8.1500 0.0227  -8.1000 0.0236  -8.0500 0.0246  -8.0000 0.0256  -7.9500 0.0267  -7.9000 0.0278  -7.8500 0.0289  -7.8000 0.0300  -7.7500 0.0311  -7.7000 0.0323  -7.6500 0.0335  -7.6000 0.0348  -7.5500 0.0360  -7.5000 0.0373  -7.4500 0.0386  -7.4000 0.0399  -7.3500 0.0413  -7.3000 0.0426  -7.2500 0.0440  -7.2000 0.0454  -7.1500 0.0467  -7.1000 0.0481  -7.0500 0.0495  -7.0000 0.0509  -6.9500 0.0522  -6.9000 0.0536  -6.8500 0.0549  -6.8000 0.0562  -6.7500 0.0575  -6.7000 0.0588  -6.6500 0.0600  -6.6000 0.0612  -6.5500 0.0624  -6.5000 0.0635  -6.4500 0.0646  -6.4000 0.0656  -6.3500 0.0665  -6.3000 0.0674  -6.2500 0.0682  -6.2000 0.0689  -6.1500 0.0696  -6.1000 0.0702  -6.0500 0.0707  -6.0000 0.0711  -5.9500 0.0714  -5.9000 0.0717  -5.8500 0.0718  -5.8000 0.0719  -5.7500 0.0719  -5.7000 0.0718  -5.6500 0.0716  -5.6000 0.0713  -5.5500 0.0709  -5.5000 0.0705  -5.4500 0.0699  -5.4000 0.0693  -5.3500 0.0686  -5.3000 0.0679  -5.2500 0.0670  -5.2000 0.0661  -5.1500 0.0652  -5.1000 0.0642  -5.0500 0.0631  -5.0000 0.0620  -4.9500 0.0608  -4.9000 0.0596  -4.8500 0.0584  -4.8000 0.0571  -4.7500 0.0558  -4.7000 0.0545  -4.6500 0.0532  -4.6000 0.0518  -4.5500 0.0505  -4.5000 0.0491  -4.4500 0.0477  -4.4000 0.0464  -4.3500 0.0450  -4.3000 0.0437  -4.2500 0.0424  -4.2000 0.0410  -4.1500 0.0397  -4.1000 0.0385  -4.0500 0.0372  -4.0000 0.0360  -3.9500 0.0347  -3.9000 0.0336  -3.8500 0.0324  -3.8000 0.0313  -3.7500 0.0302  -3.7000 0.0291  -3.6500 0.0280  -3.6000 0.0270  -3.5500 0.0261  -3.5000 0.0251  -3.4500 0.0242  -3.4000 0.0233  -3.3500 0.0225  -3.3000 0.0216  -3.2500 0.0209  -3.2000 0.0201  -3.1500 0.0194  -3.1000 0.0187  -3.0500 0.0180  -3.0000 0.0174  -2.9500 0.0168  -2.9000 0.0162  -2.8500 0.0157  -2.8000 0.0152  -2.7500 0.0147  -2.7000 0.0143  -2.6500 0.0138  -2.6000 0.0134  -2.5500 0.0131  -2.5000 0.0127  -2.4500 0.0124  -2.4000 0.0121  -2.3500 0.0119  -2.3000 0.0116  -2.2500 0.0114  -2.2000 0.0112  -2.1500 0.0110  -2.1000 0.0109  -2.0500 0.0108  -2.0000 0.0107  -1.9500 0.0106  -1.9000 0.0106  -1.8500 0.0106  -1.8000 0.0106  -1.7500 0.0106  -1.7000 0.0106  -1.6500 0.0107  -1.6000 0.0108  -1.5500 0.0109  -1.5000 0.0111  -1.4500 0.0113  -1.4000 0.0115  -1.3500 0.0117  -1.3000 0.0119  -1.2500 0.0122  -1.2000 0.0125  -1.1500 0.0129  -1.1000 0.0132  -1.0500 0.0136  -1.0000 0.0140  -0.9500 0.0145  -0.9000 0.0149  -0.8500 0.0155  -0.8000 0.0160  -0.7500 0.0166  -0.7000 0.0172  -0.6500 0.0178  -0.6000 0.0185  -0.5500 0.0192  -0.5000 0.0200  -0.4500 0.0207  -0.4000 0.0216  -0.3500 0.0224  -0.3000 0.0233  -0.2500 0.0243  -0.2000 0.0253  -0.1500 0.0263  -0.1000 0.0273  -0.0500 0.0285  0.0000 0.0296  0.0500 0.0308  0.1000 0.0321  0.1500 0.0334  0.2000 0.0347  0.2500 0.0361  0.3000 0.0375  0.3500 0.0390  0.4000 0.0405  0.4500 0.0420  0.5000 0.0436  0.5500 0.0453  0.6000 0.0470  0.6500 0.0487  0.7000 0.0504  0.7500 0.0522  0.8000 0.0541  0.8500 0.0559  0.9000 0.0578  0.9500 0.0598  1.0000 0.0617  1.0500 0.0637  1.1000 0.0657  1.1500 0.0676  1.2000 0.0696  1.2500 0.0717  1.3000 0.0737  1.3500 0.0757  1.4000 0.0776  1.4500 0.0796  1.5000 0.0816  1.5500 0.0835  1.6000 0.0854  1.6500 0.0872  1.7000 0.0890  1.7500 0.0908  1.8000 0.0925  1.8500 0.0941  1.9000 0.0956  1.9500 0.0971  2.0000 0.0985  2.0500 0.0998  2.1000 0.1011  2.1500 0.1022  2.2000 0.1032  2.2500 0.1042  2.3000 0.1050  2.3500 0.1058  2.4000 0.1064  2.4500 0.1069  2.5000 0.1073  2.5500 0.1076  2.6000 0.1078  2.6500 0.1079  2.7000 0.1079  2.7500 0.1077  2.8000 0.1075  2.8500 0.1072  2.9000 0.1068  2.9500 0.1064  3.0000 0.1058  3.0500 0.1052  3.1000 0.1045  3.1500 0.1037  3.2000 0.1030  3.2500 0.1021  3.3000 0.1012  3.3500 0.1003  3.4000 0.0994  3.4500 0.0984  3.5000 0.0975  3.5500 0.0965  3.6000 0.0956  3.6500 0.0947  3.7000 0.0938  3.7500 0.0929  3.8000 0.0921  3.8500 0.0913  3.9000 0.0906  3.9500 0.0899  4.0000 0.0893  4.0500 0.0887  4.1000 0.0882  4.1500 0.0878  4.2000 0.0875  4.2500 0.0873  4.3000 0.0871  4.3500 0.0871  4.4000 0.0872  4.4500 0.0873  4.5000 0.0876  4.5500 0.0879  4.6000 0.0884  4.6500 0.0890  4.7000 0.0897  4.7500 0.0905  4.8000 0.0914  4.8500 0.0924  4.9000 0.0936  4.9500 0.0948  5.0000 0.0962  5.0500 0.0977  5.1000 0.0992  5.1500 0.1009  5.2000 0.1027  5.2500 0.1046  5.3000 0.1066  5.3500 0.1086  5.4000 0.1108  5.4500 0.1130  5.5000 0.1153  5.5500 0.1177  5.6000 0.1202  5.6500 0.1227  5.7000 0.1252  5.7500 0.1278  5.8000 0.1304  5.8500 0.1331  5.9000 0.1357  5.9500 0.1384  6.0000 0.1410  6.0500 0.1437  6.1000 0.1463  6.1500 0.1489  6.2000 0.1514  6.2500 0.1538  6.3000 0.1562  6.3500 0.1585  6.4000 0.1607  6.4500 0.1628  6.5000 0.1648  6.5500 0.1667  6.6000 0.1684  6.6500 0.1700  6.7000 0.1714  6.7500 0.1726  6.8000 0.1737  6.8500 0.1746  6.9000 0.1753  6.9500 0.1759  7.0000 0.1762  7.0500 0.1764  7.1000 0.1763  7.1500 0.1760  7.2000 0.1756  7.2500 0.1749  7.3000 0.1741  7.3500 0.1730  7.4000 0.1718  7.4500 0.1704  7.5000 0.1688  7.5500 0.1670  7.6000 0.1651  7.6500 0.1630  7.7000 0.1607  7.7500 0.1584  7.8000 0.1558  7.8500 0.1532  7.9000 0.1504  7.9500 0.1476  8.0000 0.1446  8.0500 0.1416  8.1000 0.1385  8.1500 0.1353  8.2000 0.1321  8.2500 0.1288  8.3000 0.1255  8.3500 0.1222  8.4000 0.1189  8.4500 0.1156  8.5000 0.1122  8.5500 0.1089  8.6000 0.1056  8.6500 0.1023  8.7000 0.0991  8.7500 0.0959  8.8000 0.0927  8.8500 0.0896  8.9000 0.0865  8.9500 0.0835  9.0000 0.0805  9.0500 0.0776  9.1000 0.0747  9.1500 0.0720  9.2000 0.0693  9.2500 0.0666  9.3000 0.0640  9.3500 0.0615  9.4000 0.0591  9.4500 0.0568  9.5000 0.0545  9.5500 0.0523  9.6000 0.0501  9.6500 0.0481  9.7000 0.0461  9.7500 0.0441  9.8000 0.0423  9.8500 0.0405  9.9000 0.0387  9.9500 0.0370  10.0000 0.0354   /

 \color{black}
\put {O} at 0 0 
\setplotarea x from -10.000000 to 10.000000, y from 0 to 0.500000
\axis top /
\axis right /
\axis left /
\axis bottom 
 label {{\Large \hspace{2cm} $f$}} /
\axis bottom 
 ticks out withvalues -10 0 10 / quantity 3 /
\axis left shiftedto x=0 
 label {{\Large $\kappa$}} /
\axis left shiftedto x=0 
 ticks out withvalues 0 {} 0.2 {} 0.4 {}  / quantity 6 /

\normalcolor
\endpicture

%% file: Compressgraphs/Compress4-1L.tex
\beginpicture

 \setcoordinatesystem units <5.000 pt,28.5 pt>

 \color{red}

\setplotsymbol ({\LARGE $\cdot$})

\plot -15.0000 4.0000  -14.9250 4.0000  -14.8500 4.0000  -14.7750 4.0000  -14.7000 4.0000  -14.6250 4.0000  -14.5500 4.0000  -14.4750 4.0000  -14.4000 4.0000  -14.3250 4.0000  -14.2500 4.0000  -14.1750 4.0000  -14.1000 4.0000  -14.0250 4.0000  -13.9500 4.0000  -13.8750 4.0000  -13.8000 4.0000  -13.7250 4.0000  -13.6500 4.0000  -13.5750 4.0000  -13.5000 4.0000  -13.4250 4.0000  -13.3500 4.0000  -13.2750 4.0000  -13.2000 4.0000  -13.1250 4.0000  -13.0500 4.0000  -12.9750 4.0000  -12.9000 4.0000  -12.8250 4.0000  -12.7500 4.0000  -12.6750 4.0000  -12.6000 4.0000  -12.5250 4.0000  -12.4500 4.0000  -12.3750 4.0000  -12.3000 4.0000  -12.2250 4.0000  -12.1500 4.0000  -12.0750 4.0000  -12.0000 4.0000  -11.9250 4.0000  -11.8500 4.0000  -11.7750 4.0000  -11.7000 4.0000  -11.6250 4.0000  -11.5500 4.0000  -11.4750 4.0000  -11.4000 4.0000  -11.3250 4.0000  -11.2500 4.0000  -11.1750 4.0000  -11.1000 4.0000  -11.0250 4.0000  -10.9500 3.9999  -10.8750 3.9999  -10.8000 3.9999  -10.7250 3.9999  -10.6500 3.9999  -10.5750 3.9999  -10.5000 3.9999  -10.4250 3.9999  -10.3500 3.9999  -10.2750 3.9999  -10.2000 3.9999  -10.1250 3.9999  -10.0500 3.9999  -9.9750 3.9999  -9.9000 3.9999  -9.8250 3.9998  -9.7500 3.9998  -9.6750 3.9998  -9.6000 3.9998  -9.5250 3.9998  -9.4500 3.9998  -9.3750 3.9998  -9.3000 3.9997  -9.2250 3.9997  -9.1500 3.9997  -9.0750 3.9997  -9.0000 3.9996  -8.9250 3.9996  -8.8500 3.9996  -8.7750 3.9995  -8.7000 3.9995  -8.6250 3.9995  -8.5500 3.9994  -8.4750 3.9994  -8.4000 3.9993  -8.3250 3.9993  -8.2500 3.9992  -8.1750 3.9992  -8.1000 3.9991  -8.0250 3.9990  -7.9500 3.9990  -7.8750 3.9989  -7.8000 3.9988  -7.7250 3.9987  -7.6500 3.9986  -7.5750 3.9985  -7.5000 3.9984  -7.4250 3.9983  -7.3500 3.9981  -7.2750 3.9980  -7.2000 3.9978  -7.1250 3.9976  -7.0500 3.9975  -6.9750 3.9973  -6.9000 3.9970  -6.8250 3.9968  -6.7500 3.9966  -6.6750 3.9963  -6.6000 3.9960  -6.5250 3.9957  -6.4500 3.9954  -6.3750 3.9950  -6.3000 3.9946  -6.2250 3.9942  -6.1500 3.9938  -6.0750 3.9933  -6.0000 3.9928  -5.9250 3.9922  -5.8500 3.9916  -5.7750 3.9909  -5.7000 3.9902  -5.6250 3.9895  -5.5500 3.9887  -5.4750 3.9878  -5.4000 3.9868  -5.3250 3.9858  -5.2500 3.9847  -5.1750 3.9835  -5.1000 3.9823  -5.0250 3.9809  -4.9500 3.9794  -4.8750 3.9778  -4.8000 3.9761  -4.7250 3.9743  -4.6500 3.9723  -4.5750 3.9702  -4.5000 3.9679  -4.4250 3.9654  -4.3500 3.9628  -4.2750 3.9599  -4.2000 3.9569  -4.1250 3.9536  -4.0500 3.9501  -3.9750 3.9463  -3.9000 3.9422  -3.8250 3.9378  -3.7500 3.9331  -3.6750 3.9281  -3.6000 3.9227  -3.5250 3.9169  -3.4500 3.9106  -3.3750 3.9040  -3.3000 3.8968  -3.2250 3.8892  -3.1500 3.8810  -3.0750 3.8723  -3.0000 3.8630  -2.9250 3.8530  -2.8500 3.8424  -2.7750 3.8310  -2.7000 3.8189  -2.6250 3.8061  -2.5500 3.7924  -2.4750 3.7778  -2.4000 3.7624  -2.3250 3.7460  -2.2500 3.7287  -2.1750 3.7104  -2.1000 3.6910  -2.0250 3.6706  -1.9500 3.6491  -1.8750 3.6265  -1.8000 3.6027  -1.7250 3.5778  -1.6500 3.5518  -1.5750 3.5246  -1.5000 3.4963  -1.4250 3.4669  -1.3500 3.4364  -1.2750 3.4048  -1.2000 3.3722  -1.1250 3.3386  -1.0500 3.3041  -0.9750 3.2688  -0.9000 3.2328  -0.8250 3.1960  -0.7500 3.1587  -0.6750 3.1209  -0.6000 3.0828  -0.5250 3.0444  -0.4500 3.0059  -0.3750 2.9673  -0.3000 2.9289  -0.2250 2.8906  -0.1500 2.8527  -0.0750 2.8153  0.0000 2.7783  0.0750 2.7420  0.1500 2.7065  0.2250 2.6717  0.3000 2.6379  0.3750 2.6050  0.4500 2.5731  0.5250 2.5422  0.6000 2.5124  0.6750 2.4838  0.7500 2.4562  0.8250 2.4298  0.9000 2.4046  0.9750 2.3805  1.0500 2.3575  1.1250 2.3356  1.2000 2.3149  1.2750 2.2951  1.3500 2.2765  1.4250 2.2588  1.5000 2.2421  1.5750 2.2264  1.6500 2.2115  1.7250 2.1976  1.8000 2.1844  1.8750 2.1721  1.9500 2.1605  2.0250 2.1496  2.1000 2.1393  2.1750 2.1298  2.2500 2.1208  2.3250 2.1124  2.4000 2.1045  2.4750 2.0972  2.5500 2.0903  2.6250 2.0839  2.7000 2.0779  2.7750 2.0722  2.8500 2.0670  2.9250 2.0621  3.0000 2.0575  3.0750 2.0532  3.1500 2.0492  3.2250 2.0454  3.3000 2.0419  3.3750 2.0386  3.4500 2.0355  3.5250 2.0326  3.6000 2.0298  3.6750 2.0273  3.7500 2.0248  3.8250 2.0226  3.9000 2.0204  3.9750 2.0183  4.0500 2.0164  4.1250 2.0145  4.2000 2.0127  4.2750 2.0110  4.3500 2.0093  4.4250 2.0077  4.5000 2.0061  4.5750 2.0046  4.6500 2.0031  4.7250 2.0016  4.8000 2.0001  4.8750 1.9987  4.9500 1.9972  5.0250 1.9957  5.1000 1.9942  5.1750 1.9926  5.2500 1.9910  5.3250 1.9894  5.4000 1.9877  5.4750 1.9859  5.5500 1.9841  5.6250 1.9822  5.7000 1.9802  5.7750 1.9781  5.8500 1.9759  5.9250 1.9735  6.0000 1.9710  6.0750 1.9684  6.1500 1.9656  6.2250 1.9627  6.3000 1.9595  6.3750 1.9562  6.4500 1.9527  6.5250 1.9489  6.6000 1.9449  6.6750 1.9406  6.7500 1.9361  6.8250 1.9312  6.9000 1.9261  6.9750 1.9206  7.0500 1.9148  7.1250 1.9086  7.2000 1.9021  7.2750 1.8951  7.3500 1.8878  7.4250 1.8800  7.5000 1.8717  7.5750 1.8630  7.6500 1.8538  7.7250 1.8441  7.8000 1.8339  7.8750 1.8232  7.9500 1.8120  8.0250 1.8002  8.1000 1.7879  8.1750 1.7751  8.2500 1.7617  8.3250 1.7478  8.4000 1.7333  8.4750 1.7184  8.5500 1.7029  8.6250 1.6870  8.7000 1.6707  8.7750 1.6539  8.8500 1.6367  8.9250 1.6191  9.0000 1.6013  9.0750 1.5832  9.1500 1.5648  9.2250 1.5463  9.3000 1.5276  9.3750 1.5089  9.4500 1.4902  9.5250 1.4714  9.6000 1.4528  9.6750 1.4343  9.7500 1.4159  9.8250 1.3978  9.9000 1.3800  9.9750 1.3625  10.0500 1.3454  10.1250 1.3286  10.2000 1.3123  10.2750 1.2964  10.3500 1.2810  10.4250 1.2661  10.5000 1.2517  10.5750 1.2378  10.6500 1.2245  10.7250 1.2117  10.8000 1.1995  10.8750 1.1878  10.9500 1.1766  11.0250 1.1659  11.1000 1.1558  11.1750 1.1462  11.2500 1.1371  11.3250 1.1285  11.4000 1.1203  11.4750 1.1126  11.5500 1.1053  11.6250 1.0984  11.7000 1.0920  11.7750 1.0859  11.8500 1.0802  11.9250 1.0748  12.0000 1.0698  12.0750 1.0651  12.1500 1.0607  12.2250 1.0565  12.3000 1.0527  12.3750 1.0490  12.4500 1.0457  12.5250 1.0425  12.6000 1.0396  12.6750 1.0368  12.7500 1.0342  12.8250 1.0318  12.9000 1.0296  12.9750 1.0275  13.0500 1.0256  13.1250 1.0238  13.2000 1.0221  13.2750 1.0205  13.3500 1.0191  13.4250 1.0177  13.5000 1.0165  13.5750 1.0153  13.6500 1.0142  13.7250 1.0132  13.8000 1.0123  13.8750 1.0114  13.9500 1.0106  14.0250 1.0098  14.1000 1.0091  14.1750 1.0085  14.2500 1.0078  14.3250 1.0073  14.4000 1.0068  14.4750 1.0063  14.5500 1.0058  14.6250 1.0054  14.7000 1.0050  14.7750 1.0047  14.8500 1.0043  14.9250 1.0040  15.0000 1.0037   /

 \color{black}
\put {O} at 0 0 
\setplotarea x from -15.000000 to 15.000000, y from 0.000000 to 4.200000
\axis top /
\axis right /
\axis left /
\axis bottom 
 label {{\Large \hspace{2cm} $f$}} /
\axis bottom 
 ticks out withvalues -15 0 15 / quantity 3 /
\axis left shiftedto x=0 
 label {{\Large $\LA h \RA$}} /
\axis left shiftedto x=0 
  ticks out withvalues 0 { } 1.4 { } {2.8} {} {} {} / quantity 7 /

\normalcolor
\endpicture

%% file: Compressgraphs/Compress4-1F.tex
\beginpicture

 \setcoordinatesystem units <5.000 pt,600.000 pt>

 \color{blue}

\setplotsymbol ({\LARGE $\cdot$})

\plot -15.0000 0.0000  -14.9250 0.0000  -14.8500 0.0000  -14.7750 0.0000  -14.7000 0.0000  -14.6250 0.0000  -14.5500 0.0000  -14.4750 0.0000  -14.4000 0.0000  -14.3250 0.0000  -14.2500 0.0000  -14.1750 0.0000  -14.1000 0.0000  -14.0250 0.0000  -13.9500 0.0000  -13.8750 0.0000  -13.8000 0.0000  -13.7250 0.0000  -13.6500 0.0000  -13.5750 0.0000  -13.5000 0.0000  -13.4250 0.0000  -13.3500 0.0000  -13.2750 0.0000  -13.2000 0.0000  -13.1250 0.0000  -13.0500 0.0000  -12.9750 0.0000  -12.9000 0.0000  -12.8250 0.0000  -12.7500 0.0000  -12.6750 0.0000  -12.6000 0.0000  -12.5250 0.0000  -12.4500 0.0000  -12.3750 0.0000  -12.3000 0.0000  -12.2250 0.0000  -12.1500 0.0000  -12.0750 0.0000  -12.0000 0.0000  -11.9250 0.0000  -11.8500 0.0000  -11.7750 0.0000  -11.7000 0.0000  -11.6250 0.0000  -11.5500 0.0000  -11.4750 0.0000  -11.4000 0.0000  -11.3250 0.0000  -11.2500 0.0000  -11.1750 0.0000  -11.1000 0.0000  -11.0250 0.0000  -10.9500 0.0000  -10.8750 0.0000  -10.8000 0.0000  -10.7250 0.0000  -10.6500 0.0000  -10.5750 0.0000  -10.5000 0.0000  -10.4250 0.0000  -10.3500 0.0000  -10.2750 0.0000  -10.2000 0.0000  -10.1250 0.0000  -10.0500 0.0000  -9.9750 0.0000  -9.9000 0.0000  -9.8250 0.0000  -9.7500 0.0000  -9.6750 0.0000  -9.6000 0.0000  -9.5250 0.0001  -9.4500 0.0001  -9.3750 0.0001  -9.3000 0.0001  -9.2250 0.0001  -9.1500 0.0001  -9.0750 0.0001  -9.0000 0.0001  -8.9250 0.0001  -8.8500 0.0001  -8.7750 0.0001  -8.7000 0.0001  -8.6250 0.0001  -8.5500 0.0001  -8.4750 0.0002  -8.4000 0.0002  -8.3250 0.0002  -8.2500 0.0002  -8.1750 0.0002  -8.1000 0.0002  -8.0250 0.0002  -7.9500 0.0003  -7.8750 0.0003  -7.8000 0.0003  -7.7250 0.0003  -7.6500 0.0003  -7.5750 0.0004  -7.5000 0.0004  -7.4250 0.0004  -7.3500 0.0005  -7.2750 0.0005  -7.2000 0.0005  -7.1250 0.0006  -7.0500 0.0006  -6.9750 0.0007  -6.9000 0.0007  -6.8250 0.0008  -6.7500 0.0009  -6.6750 0.0009  -6.6000 0.0010  -6.5250 0.0011  -6.4500 0.0012  -6.3750 0.0012  -6.3000 0.0013  -6.2250 0.0014  -6.1500 0.0016  -6.0750 0.0017  -6.0000 0.0018  -5.9250 0.0019  -5.8500 0.0021  -5.7750 0.0023  -5.7000 0.0024  -5.6250 0.0026  -5.5500 0.0028  -5.4750 0.0030  -5.4000 0.0033  -5.3250 0.0035  -5.2500 0.0038  -5.1750 0.0041  -5.1000 0.0044  -5.0250 0.0048  -4.9500 0.0051  -4.8750 0.0055  -4.8000 0.0059  -4.7250 0.0064  -4.6500 0.0069  -4.5750 0.0074  -4.5000 0.0080  -4.4250 0.0086  -4.3500 0.0092  -4.2750 0.0099  -4.2000 0.0107  -4.1250 0.0115  -4.0500 0.0124  -3.9750 0.0133  -3.9000 0.0143  -3.8250 0.0154  -3.7500 0.0165  -3.6750 0.0177  -3.6000 0.0190  -3.5250 0.0204  -3.4500 0.0219  -3.3750 0.0235  -3.3000 0.0253  -3.2250 0.0271  -3.1500 0.0290  -3.0750 0.0311  -3.0000 0.0333  -2.9250 0.0356  -2.8500 0.0381  -2.7750 0.0407  -2.7000 0.0435  -2.6250 0.0465  -2.5500 0.0496  -2.4750 0.0529  -2.4000 0.0563  -2.3250 0.0599  -2.2500 0.0637  -2.1750 0.0677  -2.1000 0.0718  -2.0250 0.0762  -1.9500 0.0806  -1.8750 0.0852  -1.8000 0.0900  -1.7250 0.0949  -1.6500 0.0999  -1.5750 0.1049  -1.5000 0.1101  -1.4250 0.1153  -1.3500 0.1205  -1.2750 0.1257  -1.2000 0.1309  -1.1250 0.1359  -1.0500 0.1409  -0.9750 0.1456  -0.9000 0.1502  -0.8250 0.1545  -0.7500 0.1586  -0.6750 0.1623  -0.6000 0.1656  -0.5250 0.1685  -0.4500 0.1710  -0.3750 0.1731  -0.3000 0.1746  -0.2250 0.1757  -0.1500 0.1762  -0.0750 0.1762  0.0000 0.1758  0.0750 0.1747  0.1500 0.1732  0.2250 0.1712  0.3000 0.1688  0.3750 0.1659  0.4500 0.1627  0.5250 0.1590  0.6000 0.1551  0.6750 0.1509  0.7500 0.1464  0.8250 0.1417  0.9000 0.1368  0.9750 0.1319  1.0500 0.1268  1.1250 0.1217  1.2000 0.1166  1.2750 0.1114  1.3500 0.1063  1.4250 0.1013  1.5000 0.0964  1.5750 0.0915  1.6500 0.0868  1.7250 0.0822  1.8000 0.0778  1.8750 0.0735  1.9500 0.0694  2.0250 0.0654  2.1000 0.0616  2.1750 0.0580  2.2500 0.0546  2.3250 0.0513  2.4000 0.0482  2.4750 0.0452  2.5500 0.0424  2.6250 0.0398  2.7000 0.0373  2.7750 0.0350  2.8500 0.0328  2.9250 0.0307  3.0000 0.0288  3.0750 0.0270  3.1500 0.0253  3.2250 0.0237  3.3000 0.0223  3.3750 0.0209  3.4500 0.0196  3.5250 0.0185  3.6000 0.0174  3.6750 0.0164  3.7500 0.0155  3.8250 0.0147  3.9000 0.0139  3.9750 0.0133  4.0500 0.0127  4.1250 0.0121  4.2000 0.0116  4.2750 0.0112  4.3500 0.0108  4.4250 0.0105  4.5000 0.0103  4.5750 0.0101  4.6500 0.0100  4.7250 0.0099  4.8000 0.0099  4.8750 0.0099  4.9500 0.0099  5.0250 0.0101  5.1000 0.0103  5.1750 0.0105  5.2500 0.0108  5.3250 0.0111  5.4000 0.0116  5.4750 0.0120  5.5500 0.0126  5.6250 0.0132  5.7000 0.0138  5.7750 0.0146  5.8500 0.0154  5.9250 0.0163  6.0000 0.0172  6.0750 0.0183  6.1500 0.0194  6.2250 0.0207  6.3000 0.0220  6.3750 0.0234  6.4500 0.0250  6.5250 0.0266  6.6000 0.0284  6.6750 0.0303  6.7500 0.0323  6.8250 0.0344  6.9000 0.0367  6.9750 0.0391  7.0500 0.0417  7.1250 0.0444  7.2000 0.0473  7.2750 0.0503  7.3500 0.0535  7.4250 0.0569  7.5000 0.0604  7.5750 0.0640  7.6500 0.0679  7.7250 0.0718  7.8000 0.0760  7.8750 0.0802  7.9500 0.0846  8.0250 0.0892  8.1000 0.0938  8.1750 0.0985  8.2500 0.1033  8.3250 0.1082  8.4000 0.1131  8.4750 0.1179  8.5500 0.1228  8.6250 0.1276  8.7000 0.1324  8.7750 0.1370  8.8500 0.1415  8.9250 0.1458  9.0000 0.1498  9.0750 0.1536  9.1500 0.1572  9.2250 0.1604  9.3000 0.1632  9.3750 0.1657  9.4500 0.1678  9.5250 0.1694  9.6000 0.1706  9.6750 0.1713  9.7500 0.1716  9.8250 0.1714  9.9000 0.1708  9.9750 0.1696  10.0500 0.1681  10.1250 0.1661  10.2000 0.1637  10.2750 0.1609  10.3500 0.1577  10.4250 0.1543  10.5000 0.1505  10.5750 0.1465  10.6500 0.1422  10.7250 0.1377  10.8000 0.1331  10.8750 0.1284  10.9500 0.1236  11.0250 0.1187  11.1000 0.1138  11.1750 0.1089  11.2500 0.1040  11.3250 0.0992  11.4000 0.0945  11.4750 0.0898  11.5500 0.0852  11.6250 0.0808  11.7000 0.0765  11.7750 0.0723  11.8500 0.0683  11.9250 0.0644  12.0000 0.0607  12.0750 0.0571  12.1500 0.0537  12.2250 0.0505  12.3000 0.0474  12.3750 0.0445  12.4500 0.0417  12.5250 0.0390  12.6000 0.0365  12.6750 0.0342  12.7500 0.0320  12.8250 0.0299  12.9000 0.0279  12.9750 0.0261  13.0500 0.0243  13.1250 0.0227  13.2000 0.0211  13.2750 0.0197  13.3500 0.0184  13.4250 0.0171  13.5000 0.0159  13.5750 0.0148  13.6500 0.0138  13.7250 0.0129  13.8000 0.0120  13.8750 0.0111  13.9500 0.0103  14.0250 0.0096  14.1000 0.0089  14.1750 0.0083  14.2500 0.0077  14.3250 0.0072  14.4000 0.0067  14.4750 0.0062  14.5500 0.0058  14.6250 0.0053  14.7000 0.0050  14.7750 0.0046  14.8500 0.0043  14.9250 0.0040  15.0000 0.0037   /

 \color{black}
\put {O} at 0 0 
\setplotarea x from -15.000000 to 15.000000, y from 0 to 0.200000
\axis top /
\axis right /
\axis left /
\axis bottom 
 label {{\Large \hspace{2cm} $f$}} /
\axis bottom 
 ticks out withvalues -15 0 15 / quantity 3 /
\axis left shiftedto x=0 
 label {{\Large $\kappa$}} /
\axis left shiftedto x=0 
 ticks out withvalues 0 0.05 {} 0.15 {}  / quantity 5 /

\normalcolor
\endpicture

%% file: Compressgraphs/Compress5-12L.tex
\beginpicture

 \setcoordinatesystem units <5.000 pt,25 pt>

 \color{red}

\setplotsymbol ({\LARGE $\cdot$})

\plot -15.0000 5.0000  -14.9250 5.0000  -14.8500 5.0000  -14.7750 5.0000  -14.7000 5.0000  -14.6250 5.0000  -14.5500 5.0000  -14.4750 5.0000  -14.4000 5.0000  -14.3250 5.0000  -14.2500 5.0000  -14.1750 5.0000  -14.1000 5.0000  -14.0250 5.0000  -13.9500 5.0000  -13.8750 5.0000  -13.8000 5.0000  -13.7250 4.9999  -13.6500 4.9999  -13.5750 4.9999  -13.5000 4.9999  -13.4250 4.9999  -13.3500 4.9999  -13.2750 4.9999  -13.2000 4.9999  -13.1250 4.9999  -13.0500 4.9999  -12.9750 4.9999  -12.9000 4.9999  -12.8250 4.9999  -12.7500 4.9999  -12.6750 4.9998  -12.6000 4.9998  -12.5250 4.9998  -12.4500 4.9998  -12.3750 4.9998  -12.3000 4.9998  -12.2250 4.9998  -12.1500 4.9997  -12.0750 4.9997  -12.0000 4.9997  -11.9250 4.9997  -11.8500 4.9997  -11.7750 4.9996  -11.7000 4.9996  -11.6250 4.9996  -11.5500 4.9995  -11.4750 4.9995  -11.4000 4.9995  -11.3250 4.9994  -11.2500 4.9994  -11.1750 4.9993  -11.1000 4.9993  -11.0250 4.9992  -10.9500 4.9992  -10.8750 4.9991  -10.8000 4.9990  -10.7250 4.9989  -10.6500 4.9989  -10.5750 4.9988  -10.5000 4.9987  -10.4250 4.9986  -10.3500 4.9985  -10.2750 4.9983  -10.2000 4.9982  -10.1250 4.9981  -10.0500 4.9979  -9.9750 4.9978  -9.9000 4.9976  -9.8250 4.9974  -9.7500 4.9972  -9.6750 4.9970  -9.6000 4.9967  -9.5250 4.9965  -9.4500 4.9962  -9.3750 4.9959  -9.3000 4.9956  -9.2250 4.9953  -9.1500 4.9949  -9.0750 4.9945  -9.0000 4.9941  -8.9250 4.9936  -8.8500 4.9931  -8.7750 4.9926  -8.7000 4.9920  -8.6250 4.9914  -8.5500 4.9907  -8.4750 4.9900  -8.4000 4.9892  -8.3250 4.9884  -8.2500 4.9875  -8.1750 4.9866  -8.1000 4.9855  -8.0250 4.9844  -7.9500 4.9832  -7.8750 4.9819  -7.8000 4.9805  -7.7250 4.9791  -7.6500 4.9775  -7.5750 4.9758  -7.5000 4.9739  -7.4250 4.9719  -7.3500 4.9698  -7.2750 4.9675  -7.2000 4.9651  -7.1250 4.9625  -7.0500 4.9597  -6.9750 4.9567  -6.9000 4.9535  -6.8250 4.9500  -6.7500 4.9463  -6.6750 4.9424  -6.6000 4.9382  -6.5250 4.9337  -6.4500 4.9289  -6.3750 4.9237  -6.3000 4.9183  -6.2250 4.9125  -6.1500 4.9063  -6.0750 4.8997  -6.0000 4.8927  -5.9250 4.8853  -5.8500 4.8775  -5.7750 4.8691  -5.7000 4.8604  -5.6250 4.8511  -5.5500 4.8413  -5.4750 4.8310  -5.4000 4.8202  -5.3250 4.8088  -5.2500 4.7969  -5.1750 4.7844  -5.1000 4.7714  -5.0250 4.7578  -4.9500 4.7437  -4.8750 4.7291  -4.8000 4.7139  -4.7250 4.6982  -4.6500 4.6820  -4.5750 4.6654  -4.5000 4.6483  -4.4250 4.6308  -4.3500 4.6130  -4.2750 4.5948  -4.2000 4.5763  -4.1250 4.5575  -4.0500 4.5386  -3.9750 4.5195  -3.9000 4.5003  -3.8250 4.4810  -3.7500 4.4617  -3.6750 4.4425  -3.6000 4.4233  -3.5250 4.4043  -3.4500 4.3854  -3.3750 4.3668  -3.3000 4.3484  -3.2250 4.3302  -3.1500 4.3124  -3.0750 4.2949  -3.0000 4.2778  -2.9250 4.2610  -2.8500 4.2446  -2.7750 4.2285  -2.7000 4.2128  -2.6250 4.1975  -2.5500 4.1825  -2.4750 4.1678  -2.4000 4.1534  -2.3250 4.1393  -2.2500 4.1254  -2.1750 4.1117  -2.1000 4.0981  -2.0250 4.0846  -1.9500 4.0712  -1.8750 4.0577  -1.8000 4.0441  -1.7250 4.0303  -1.6500 4.0162  -1.5750 4.0017  -1.5000 3.9867  -1.4250 3.9712  -1.3500 3.9550  -1.2750 3.9379  -1.2000 3.9198  -1.1250 3.9006  -1.0500 3.8802  -0.9750 3.8583  -0.9000 3.8348  -0.8250 3.8096  -0.7500 3.7824  -0.6750 3.7531  -0.6000 3.7215  -0.5250 3.6876  -0.4500 3.6511  -0.3750 3.6119  -0.3000 3.5700  -0.2250 3.5254  -0.1500 3.4781  -0.0750 3.4281  0.0000 3.3756  0.0750 3.3207  0.1500 3.2638  0.2250 3.2051  0.3000 3.1451  0.3750 3.0840  0.4500 3.0225  0.5250 2.9609  0.6000 2.8998  0.6750 2.8396  0.7500 2.7806  0.8250 2.7235  0.9000 2.6683  0.9750 2.6155  1.0500 2.5653  1.1250 2.5179  1.2000 2.4732  1.2750 2.4315  1.3500 2.3926  1.4250 2.3565  1.5000 2.3232  1.5750 2.2925  1.6500 2.2642  1.7250 2.2384  1.8000 2.2147  1.8750 2.1930  1.9500 2.1732  2.0250 2.1551  2.1000 2.1385  2.1750 2.1234  2.2500 2.1095  2.3250 2.0968  2.4000 2.0850  2.4750 2.0742  2.5500 2.0642  2.6250 2.0548  2.7000 2.0461  2.7750 2.0379  2.8500 2.0302  2.9250 2.0229  3.0000 2.0158  3.0750 2.0091  3.1500 2.0025  3.2250 1.9961  3.3000 1.9898  3.3750 1.9836  3.4500 1.9774  3.5250 1.9712  3.6000 1.9649  3.6750 1.9586  3.7500 1.9521  3.8250 1.9455  3.9000 1.9387  3.9750 1.9316  4.0500 1.9243  4.1250 1.9168  4.2000 1.9089  4.2750 1.9008  4.3500 1.8922  4.4250 1.8833  4.5000 1.8740  4.5750 1.8643  4.6500 1.8541  4.7250 1.8435  4.8000 1.8324  4.8750 1.8209  4.9500 1.8088  5.0250 1.7962  5.1000 1.7832  5.1750 1.7696  5.2500 1.7555  5.3250 1.7409  5.4000 1.7258  5.4750 1.7103  5.5500 1.6942  5.6250 1.6778  5.7000 1.6609  5.7750 1.6436  5.8500 1.6260  5.9250 1.6081  6.0000 1.5899  6.0750 1.5714  6.1500 1.5528  6.2250 1.5341  6.3000 1.5153  6.3750 1.4964  6.4500 1.4776  6.5250 1.4588  6.6000 1.4402  6.6750 1.4217  6.7500 1.4035  6.8250 1.3855  6.9000 1.3679  6.9750 1.3506  7.0500 1.3337  7.1250 1.3172  7.2000 1.3011  7.2750 1.2856  7.3500 1.2705  7.4250 1.2559  7.5000 1.2419  7.5750 1.2284  7.6500 1.2154  7.7250 1.2030  7.8000 1.1911  7.8750 1.1798  7.9500 1.1690  8.0250 1.1587  8.1000 1.1490  8.1750 1.1397  8.2500 1.1309  8.3250 1.1226  8.4000 1.1148  8.4750 1.1074  8.5500 1.1004  8.6250 1.0938  8.7000 1.0876  8.7750 1.0818  8.8500 1.0763  8.9250 1.0712  9.0000 1.0664  9.0750 1.0619  9.1500 1.0577  9.2250 1.0537  9.3000 1.0501  9.3750 1.0466  9.4500 1.0434  9.5250 1.0404  9.6000 1.0376  9.6750 1.0350  9.7500 1.0325  9.8250 1.0302  9.9000 1.0281  9.9750 1.0261  10.0500 1.0243  10.1250 1.0226  10.2000 1.0210  10.2750 1.0195  10.3500 1.0181  10.4250 1.0168  10.5000 1.0156  10.5750 1.0145  10.6500 1.0135  10.7250 1.0125  10.8000 1.0116  10.8750 1.0108  10.9500 1.0100  11.0250 1.0093  11.1000 1.0086  11.1750 1.0080  11.2500 1.0074  11.3250 1.0069  11.4000 1.0064  11.4750 1.0060  11.5500 1.0055  11.6250 1.0051  11.7000 1.0048  11.7750 1.0044  11.8500 1.0041  11.9250 1.0038  12.0000 1.0035  12.0750 1.0033  12.1500 1.0030  12.2250 1.0028  12.3000 1.0026  12.3750 1.0024  12.4500 1.0023  12.5250 1.0021  12.6000 1.0019  12.6750 1.0018  12.7500 1.0017  12.8250 1.0015  12.9000 1.0014  12.9750 1.0013  13.0500 1.0012  13.1250 1.0011  13.2000 1.0011  13.2750 1.0010  13.3500 1.0009  13.4250 1.0009  13.5000 1.0008  13.5750 1.0007  13.6500 1.0007  13.7250 1.0006  13.8000 1.0006  13.8750 1.0005  13.9500 1.0005  14.0250 1.0005  14.1000 1.0004  14.1750 1.0004  14.2500 1.0004  14.3250 1.0003  14.4000 1.0003  14.4750 1.0003  14.5500 1.0003  14.6250 1.0003  14.7000 1.0002  14.7750 1.0002  14.8500 1.0002  14.9250 1.0002  15.0000 1.0002   /

 \color{blue}

\setplotsymbol ({\LARGE $\cdot$})

\plot -15.0000 5.0000  -14.9250 5.0000  -14.8500 5.0000  -14.7750 5.0000  -14.7000 5.0000  -14.6250 5.0000  -14.5500 5.0000  -14.4750 5.0000  -14.4000 5.0000  -14.3250 5.0000  -14.2500 5.0000  -14.1750 5.0000  -14.1000 5.0000  -14.0250 5.0000  -13.9500 5.0000  -13.8750 5.0000  -13.8000 5.0000  -13.7250 5.0000  -13.6500 5.0000  -13.5750 5.0000  -13.5000 5.0000  -13.4250 5.0000  -13.3500 5.0000  -13.2750 5.0000  -13.2000 5.0000  -13.1250 5.0000  -13.0500 5.0000  -12.9750 5.0000  -12.9000 5.0000  -12.8250 5.0000  -12.7500 5.0000  -12.6750 5.0000  -12.6000 5.0000  -12.5250 4.9999  -12.4500 4.9999  -12.3750 4.9999  -12.3000 4.9999  -12.2250 4.9999  -12.1500 4.9999  -12.0750 4.9999  -12.0000 4.9999  -11.9250 4.9999  -11.8500 4.9999  -11.7750 4.9999  -11.7000 4.9999  -11.6250 4.9999  -11.5500 4.9999  -11.4750 4.9999  -11.4000 4.9998  -11.3250 4.9998  -11.2500 4.9998  -11.1750 4.9998  -11.1000 4.9998  -11.0250 4.9998  -10.9500 4.9997  -10.8750 4.9997  -10.8000 4.9997  -10.7250 4.9997  -10.6500 4.9997  -10.5750 4.9996  -10.5000 4.9996  -10.4250 4.9996  -10.3500 4.9995  -10.2750 4.9995  -10.2000 4.9995  -10.1250 4.9994  -10.0500 4.9994  -9.9750 4.9993  -9.9000 4.9993  -9.8250 4.9992  -9.7500 4.9992  -9.6750 4.9991  -9.6000 4.9990  -9.5250 4.9989  -9.4500 4.9989  -9.3750 4.9988  -9.3000 4.9987  -9.2250 4.9986  -9.1500 4.9985  -9.0750 4.9984  -9.0000 4.9982  -8.9250 4.9981  -8.8500 4.9979  -8.7750 4.9978  -8.7000 4.9976  -8.6250 4.9974  -8.5500 4.9972  -8.4750 4.9970  -8.4000 4.9968  -8.3250 4.9965  -8.2500 4.9962  -8.1750 4.9960  -8.1000 4.9956  -8.0250 4.9953  -7.9500 4.9949  -7.8750 4.9945  -7.8000 4.9941  -7.7250 4.9937  -7.6500 4.9932  -7.5750 4.9927  -7.5000 4.9921  -7.4250 4.9915  -7.3500 4.9908  -7.2750 4.9901  -7.2000 4.9893  -7.1250 4.9885  -7.0500 4.9876  -6.9750 4.9867  -6.9000 4.9857  -6.8250 4.9846  -6.7500 4.9834  -6.6750 4.9821  -6.6000 4.9807  -6.5250 4.9793  -6.4500 4.9777  -6.3750 4.9760  -6.3000 4.9742  -6.2250 4.9722  -6.1500 4.9701  -6.0750 4.9678  -6.0000 4.9654  -5.9250 4.9628  -5.8500 4.9600  -5.7750 4.9570  -5.7000 4.9538  -5.6250 4.9503  -5.5500 4.9466  -5.4750 4.9427  -5.4000 4.9385  -5.3250 4.9340  -5.2500 4.9291  -5.1750 4.9240  -5.1000 4.9185  -5.0250 4.9127  -4.9500 4.9064  -4.8750 4.8998  -4.8000 4.8927  -4.7250 4.8852  -4.6500 4.8772  -4.5750 4.8687  -4.5000 4.8598  -4.4250 4.8503  -4.3500 4.8403  -4.2750 4.8297  -4.2000 4.8185  -4.1250 4.8068  -4.0500 4.7944  -3.9750 4.7814  -3.9000 4.7678  -3.8250 4.7536  -3.7500 4.7388  -3.6750 4.7233  -3.6000 4.7071  -3.5250 4.6904  -3.4500 4.6730  -3.3750 4.6550  -3.3000 4.6365  -3.2250 4.6173  -3.1500 4.5976  -3.0750 4.5774  -3.0000 4.5567  -2.9250 4.5356  -2.8500 4.5140  -2.7750 4.4920  -2.7000 4.4696  -2.6250 4.4470  -2.5500 4.4240  -2.4750 4.4008  -2.4000 4.3774  -2.3250 4.3538  -2.2500 4.3301  -2.1750 4.3062  -2.1000 4.2823  -2.0250 4.2582  -1.9500 4.2342  -1.8750 4.2101  -1.8000 4.1859  -1.7250 4.1618  -1.6500 4.1376  -1.5750 4.1135  -1.5000 4.0893  -1.4250 4.0651  -1.3500 4.0409  -1.2750 4.0167  -1.2000 3.9924  -1.1250 3.9680  -1.0500 3.9437  -0.9750 3.9192  -0.9000 3.8946  -0.8250 3.8700  -0.7500 3.8453  -0.6750 3.8204  -0.6000 3.7955  -0.5250 3.7705  -0.4500 3.7453  -0.3750 3.7201  -0.3000 3.6947  -0.2250 3.6693  -0.1500 3.6439  -0.0750 3.6184  0.0000 3.5928  0.0750 3.5673  0.1500 3.5418  0.2250 3.5164  0.3000 3.4910  0.3750 3.4658  0.4500 3.4406  0.5250 3.4156  0.6000 3.3908  0.6750 3.3662  0.7500 3.3418  0.8250 3.3176  0.9000 3.2936  0.9750 3.2699  1.0500 3.2464  1.1250 3.2231  1.2000 3.2001  1.2750 3.1773  1.3500 3.1547  1.4250 3.1323  1.5000 3.1101  1.5750 3.0880  1.6500 3.0661  1.7250 3.0442  1.8000 3.0224  1.8750 3.0007  1.9500 2.9790  2.0250 2.9573  2.1000 2.9355  2.1750 2.9137  2.2500 2.8918  2.3250 2.8697  2.4000 2.8476  2.4750 2.8253  2.5500 2.8028  2.6250 2.7802  2.7000 2.7573  2.7750 2.7344  2.8500 2.7112  2.9250 2.6878  3.0000 2.6643  3.0750 2.6407  3.1500 2.6169  3.2250 2.5930  3.3000 2.5690  3.3750 2.5449  3.4500 2.5208  3.5250 2.4967  3.6000 2.4726  3.6750 2.4486  3.7500 2.4247  3.8250 2.4008  3.9000 2.3771  3.9750 2.3535  4.0500 2.3301  4.1250 2.3069  4.2000 2.2839  4.2750 2.2611  4.3500 2.2385  4.4250 2.2161  4.5000 2.1939  4.5750 2.1720  4.6500 2.1502  4.7250 2.1286  4.8000 2.1072  4.8750 2.0859  4.9500 2.0648  5.0250 2.0438  5.1000 2.0229  5.1750 2.0020  5.2500 1.9811  5.3250 1.9603  5.4000 1.9395  5.4750 1.9186  5.5500 1.8977  5.6250 1.8766  5.7000 1.8556  5.7750 1.8344  5.8500 1.8131  5.9250 1.7917  6.0000 1.7702  6.0750 1.7486  6.1500 1.7269  6.2250 1.7051  6.3000 1.6832  6.3750 1.6613  6.4500 1.6394  6.5250 1.6175  6.6000 1.5957  6.6750 1.5739  6.7500 1.5523  6.8250 1.5308  6.9000 1.5095  6.9750 1.4885  7.0500 1.4678  7.1250 1.4473  7.2000 1.4272  7.2750 1.4076  7.3500 1.3883  7.4250 1.3695  7.5000 1.3513  7.5750 1.3335  7.6500 1.3163  7.7250 1.2996  7.8000 1.2835  7.8750 1.2680  7.9500 1.2531  8.0250 1.2388  8.1000 1.2250  8.1750 1.2119  8.2500 1.1994  8.3250 1.1875  8.4000 1.1761  8.4750 1.1653  8.5500 1.1551  8.6250 1.1454  8.7000 1.1362  8.7750 1.1275  8.8500 1.1193  8.9250 1.1115  9.0000 1.1043  9.0750 1.0974  9.1500 1.0910  9.2250 1.0849  9.3000 1.0792  9.3750 1.0739  9.4500 1.0689  9.5250 1.0642  9.6000 1.0598  9.6750 1.0557  9.7500 1.0519  9.8250 1.0483  9.9000 1.0450  9.9750 1.0418  10.0500 1.0389  10.1250 1.0362  10.2000 1.0337  10.2750 1.0313  10.3500 1.0291  10.4250 1.0271  10.5000 1.0251  10.5750 1.0234  10.6500 1.0217  10.7250 1.0202  10.8000 1.0187  10.8750 1.0174  10.9500 1.0162  11.0250 1.0150  11.1000 1.0139  11.1750 1.0130  11.2500 1.0120  11.3250 1.0112  11.4000 1.0104  11.4750 1.0096  11.5500 1.0089  11.6250 1.0083  11.7000 1.0077  11.7750 1.0071  11.8500 1.0066  11.9250 1.0062  12.0000 1.0057  12.0750 1.0053  12.1500 1.0049  12.2250 1.0046  12.3000 1.0042  12.3750 1.0039  12.4500 1.0037  12.5250 1.0034  12.6000 1.0031  12.6750 1.0029  12.7500 1.0027  12.8250 1.0025  12.9000 1.0023  12.9750 1.0022  13.0500 1.0020  13.1250 1.0019  13.2000 1.0017  13.2750 1.0016  13.3500 1.0015  13.4250 1.0014  13.5000 1.0013  13.5750 1.0012  13.6500 1.0011  13.7250 1.0010  13.8000 1.0009  13.8750 1.0009  13.9500 1.0008  14.0250 1.0008  14.1000 1.0007  14.1750 1.0007  14.2500 1.0006  14.3250 1.0006  14.4000 1.0005  14.4750 1.0005  14.5500 1.0004  14.6250 1.0004  14.7000 1.0004  14.7750 1.0004  14.8500 1.0003  14.9250 1.0003  15.0000 1.0003   /

 \color{black}
\put {O} at 0 0 
\setplotarea x from -15.000000 to 15.000000, y from 0.000000 to 6.00000
\axis top /
\axis right /
\axis left /
\axis bottom 
 label {{\Large \hspace{2cm} $f$}} /
\axis bottom 
 ticks out withvalues -15 0 15 / quantity 3 /
\axis left shiftedto x=0 
 label {{\Large $\LA h \RA \q$}} /
\axis left shiftedto x=0 
 ticks out withvalues {} 1 {} 3 {} 5 {} / quantity 7 /

\normalcolor
\endpicture

%% file: Compressgraphs/Compress5-12F.tex
\beginpicture

 \setcoordinatesystem units <5.000 pt,500. pt>

 \color{red}

\setplotsymbol ({\LARGE $\cdot$})

\plot -15.0000 0.0000  -14.9250 0.0000  -14.8500 0.0000  -14.7750 0.0000  -14.7000 0.0000  -14.6250 0.0000  -14.5500 0.0000  -14.4750 0.0000  -14.4000 0.0000  -14.3250 0.0000  -14.2500 0.0000  -14.1750 0.0000  -14.1000 0.0000  -14.0250 0.0000  -13.9500 0.0000  -13.8750 0.0000  -13.8000 0.0000  -13.7250 0.0000  -13.6500 0.0000  -13.5750 0.0000  -13.5000 0.0000  -13.4250 0.0000  -13.3500 0.0000  -13.2750 0.0000  -13.2000 0.0000  -13.1250 0.0000  -13.0500 0.0000  -12.9750 0.0000  -12.9000 0.0000  -12.8250 0.0000  -12.7500 0.0000  -12.6750 0.0000  -12.6000 0.0000  -12.5250 0.0000  -12.4500 0.0000  -12.3750 0.0000  -12.3000 0.0000  -12.2250 0.0000  -12.1500 0.0001  -12.0750 0.0001  -12.0000 0.0001  -11.9250 0.0001  -11.8500 0.0001  -11.7750 0.0001  -11.7000 0.0001  -11.6250 0.0001  -11.5500 0.0001  -11.4750 0.0001  -11.4000 0.0001  -11.3250 0.0001  -11.2500 0.0001  -11.1750 0.0001  -11.1000 0.0001  -11.0250 0.0002  -10.9500 0.0002  -10.8750 0.0002  -10.8000 0.0002  -10.7250 0.0002  -10.6500 0.0002  -10.5750 0.0002  -10.5000 0.0003  -10.4250 0.0003  -10.3500 0.0003  -10.2750 0.0003  -10.2000 0.0004  -10.1250 0.0004  -10.0500 0.0004  -9.9750 0.0004  -9.9000 0.0005  -9.8250 0.0005  -9.7500 0.0006  -9.6750 0.0006  -9.6000 0.0007  -9.5250 0.0007  -9.4500 0.0008  -9.3750 0.0008  -9.3000 0.0009  -9.2250 0.0009  -9.1500 0.0010  -9.0750 0.0011  -9.0000 0.0012  -8.9250 0.0013  -8.8500 0.0014  -8.7750 0.0015  -8.7000 0.0016  -8.6250 0.0017  -8.5500 0.0018  -8.4750 0.0020  -8.4000 0.0021  -8.3250 0.0023  -8.2500 0.0025  -8.1750 0.0027  -8.1000 0.0029  -8.0250 0.0031  -7.9500 0.0033  -7.8750 0.0036  -7.8000 0.0038  -7.7250 0.0041  -7.6500 0.0044  -7.5750 0.0048  -7.5000 0.0051  -7.4250 0.0055  -7.3500 0.0059  -7.2750 0.0063  -7.2000 0.0068  -7.1250 0.0073  -7.0500 0.0078  -6.9750 0.0084  -6.9000 0.0090  -6.8250 0.0096  -6.7500 0.0103  -6.6750 0.0110  -6.6000 0.0118  -6.5250 0.0126  -6.4500 0.0134  -6.3750 0.0143  -6.3000 0.0153  -6.2250 0.0163  -6.1500 0.0173  -6.0750 0.0184  -6.0000 0.0196  -5.9250 0.0208  -5.8500 0.0221  -5.7750 0.0234  -5.7000 0.0248  -5.6250 0.0262  -5.5500 0.0277  -5.4750 0.0292  -5.4000 0.0307  -5.3250 0.0323  -5.2500 0.0339  -5.1750 0.0355  -5.1000 0.0371  -5.0250 0.0388  -4.9500 0.0404  -4.8750 0.0420  -4.8000 0.0436  -4.7250 0.0452  -4.6500 0.0467  -4.5750 0.0482  -4.5000 0.0496  -4.4250 0.0509  -4.3500 0.0521  -4.2750 0.0532  -4.2000 0.0543  -4.1250 0.0551  -4.0500 0.0559  -3.9750 0.0565  -3.9000 0.0570  -3.8250 0.0574  -3.7500 0.0576  -3.6750 0.0577  -3.6000 0.0576  -3.5250 0.0574  -3.4500 0.0570  -3.3750 0.0566  -3.3000 0.0560  -3.2250 0.0554  -3.1500 0.0546  -3.0750 0.0538  -3.0000 0.0529  -2.9250 0.0520  -2.8500 0.0510  -2.7750 0.0501  -2.7000 0.0491  -2.6250 0.0482  -2.5500 0.0473  -2.4750 0.0465  -2.4000 0.0457  -2.3250 0.0451  -2.2500 0.0446  -2.1750 0.0442  -2.1000 0.0440  -2.0250 0.0439  -1.9500 0.0441  -1.8750 0.0445  -1.8000 0.0452  -1.7250 0.0461  -1.6500 0.0474  -1.5750 0.0490  -1.5000 0.0509  -1.4250 0.0533  -1.3500 0.0561  -1.2750 0.0594  -1.2000 0.0632  -1.1250 0.0676  -1.0500 0.0726  -0.9750 0.0783  -0.9000 0.0846  -0.8250 0.0917  -0.7500 0.0994  -0.6750 0.1080  -0.6000 0.1173  -0.5250 0.1273  -0.4500 0.1381  -0.3750 0.1495  -0.3000 0.1615  -0.2250 0.1739  -0.1500 0.1866  -0.0750 0.1995  0.0000 0.2122  0.0750 0.2246  0.1500 0.2364  0.2250 0.2472  0.3000 0.2570  0.3750 0.2652  0.4500 0.2719  0.5250 0.2766  0.6000 0.2794  0.6750 0.2801  0.7500 0.2787  0.8250 0.2752  0.9000 0.2698  0.9750 0.2627  1.0500 0.2540  1.1250 0.2440  1.2000 0.2329  1.2750 0.2211  1.3500 0.2088  1.4250 0.1963  1.5000 0.1837  1.5750 0.1712  1.6500 0.1591  1.7250 0.1475  1.8000 0.1364  1.8750 0.1259  1.9500 0.1161  2.0250 0.1070  2.1000 0.0987  2.1750 0.0910  2.2500 0.0840  2.3250 0.0777  2.4000 0.0720  2.4750 0.0669  2.5500 0.0624  2.6250 0.0585  2.7000 0.0550  2.7750 0.0520  2.8500 0.0495  2.9250 0.0473  3.0000 0.0455  3.0750 0.0441  3.1500 0.0431  3.2250 0.0423  3.3000 0.0419  3.3750 0.0417  3.4500 0.0418  3.5250 0.0422  3.6000 0.0428  3.6750 0.0436  3.7500 0.0447  3.8250 0.0460  3.9000 0.0476  3.9750 0.0494  4.0500 0.0513  4.1250 0.0535  4.2000 0.0559  4.2750 0.0586  4.3500 0.0614  4.4250 0.0644  4.5000 0.0676  4.5750 0.0711  4.6500 0.0747  4.7250 0.0785  4.8000 0.0824  4.8750 0.0865  4.9500 0.0908  5.0250 0.0952  5.1000 0.0997  5.1750 0.1043  5.2500 0.1089  5.3250 0.1137  5.4000 0.1184  5.4750 0.1232  5.5500 0.1279  5.6250 0.1325  5.7000 0.1371  5.7750 0.1415  5.8500 0.1458  5.9250 0.1498  6.0000 0.1536  6.0750 0.1572  6.1500 0.1604  6.2250 0.1633  6.3000 0.1658  6.3750 0.1680  6.4500 0.1697  6.5250 0.1709  6.6000 0.1718  6.6750 0.1721  6.7500 0.1720  6.8250 0.1715  6.9000 0.1704  6.9750 0.1690  7.0500 0.1671  7.1250 0.1647  7.2000 0.1620  7.2750 0.1589  7.3500 0.1555  7.4250 0.1518  7.5000 0.1478  7.5750 0.1436  7.6500 0.1392  7.7250 0.1346  7.8000 0.1299  7.8750 0.1251  7.9500 0.1202  8.0250 0.1153  8.1000 0.1104  8.1750 0.1055  8.2500 0.1007  8.3250 0.0959  8.4000 0.0912  8.4750 0.0866  8.5500 0.0821  8.6250 0.0777  8.7000 0.0735  8.7750 0.0695  8.8500 0.0655  8.9250 0.0618  9.0000 0.0582  9.0750 0.0547  9.1500 0.0514  9.2250 0.0483  9.3000 0.0453  9.3750 0.0425  9.4500 0.0398  9.5250 0.0372  9.6000 0.0349  9.6750 0.0326  9.7500 0.0305  9.8250 0.0285  9.9000 0.0266  9.9750 0.0248  10.0500 0.0231  10.1250 0.0216  10.2000 0.0201  10.2750 0.0187  10.3500 0.0175  10.4250 0.0163  10.5000 0.0151  10.5750 0.0141  10.6500 0.0131  10.7250 0.0122  10.8000 0.0114  10.8750 0.0106  10.9500 0.0098  11.0250 0.0091  11.1000 0.0085  11.1750 0.0079  11.2500 0.0073  11.3250 0.0068  11.4000 0.0063  11.4750 0.0059  11.5500 0.0055  11.6250 0.0051  11.7000 0.0047  11.7750 0.0044  11.8500 0.0041  11.9250 0.0038  12.0000 0.0035  12.0750 0.0033  12.1500 0.0030  12.2250 0.0028  12.3000 0.0026  12.3750 0.0024  12.4500 0.0022  12.5250 0.0021  12.6000 0.0019  12.6750 0.0018  12.7500 0.0017  12.8250 0.0015  12.9000 0.0014  12.9750 0.0013  13.0500 0.0012  13.1250 0.0011  13.2000 0.0011  13.2750 0.0010  13.3500 0.0009  13.4250 0.0008  13.5000 0.0008  13.5750 0.0007  13.6500 0.0007  13.7250 0.0006  13.8000 0.0006  13.8750 0.0005  13.9500 0.0005  14.0250 0.0005  14.1000 0.0004  14.1750 0.0004  14.2500 0.0004  14.3250 0.0003  14.4000 0.0003  14.4750 0.0003  14.5500 0.0003  14.6250 0.0003  14.7000 0.0002  14.7750 0.0002  14.8500 0.0002  14.9250 0.0002  15.0000 0.0002   /

\color{blue}

\setplotsymbol ({\LARGE $\cdot$})

\plot -15.0000 0.0000  -14.9250 0.0000  -14.8500 0.0000  -14.7750 0.0000  -14.7000 0.0000  -14.6250 0.0000  -14.5500 0.0000  -14.4750 0.0000  -14.4000 0.0000  -14.3250 0.0000  -14.2500 0.0000  -14.1750 0.0000  -14.1000 0.0000  -14.0250 0.0000  -13.9500 0.0000  -13.8750 0.0000  -13.8000 0.0000  -13.7250 0.0000  -13.6500 0.0000  -13.5750 0.0000  -13.5000 0.0000  -13.4250 0.0000  -13.3500 0.0000  -13.2750 0.0000  -13.2000 0.0000  -13.1250 0.0000  -13.0500 0.0000  -12.9750 0.0000  -12.9000 0.0000  -12.8250 0.0000  -12.7500 0.0000  -12.6750 0.0000  -12.6000 0.0000  -12.5250 0.0000  -12.4500 0.0000  -12.3750 0.0000  -12.3000 0.0000  -12.2250 0.0000  -12.1500 0.0000  -12.0750 0.0000  -12.0000 0.0000  -11.9250 0.0000  -11.8500 0.0000  -11.7750 0.0000  -11.7000 0.0000  -11.6250 0.0000  -11.5500 0.0000  -11.4750 0.0000  -11.4000 0.0000  -11.3250 0.0000  -11.2500 0.0000  -11.1750 0.0000  -11.1000 0.0000  -11.0250 0.0000  -10.9500 0.0001  -10.8750 0.0001  -10.8000 0.0001  -10.7250 0.0001  -10.6500 0.0001  -10.5750 0.0001  -10.5000 0.0001  -10.4250 0.0001  -10.3500 0.0001  -10.2750 0.0001  -10.2000 0.0001  -10.1250 0.0001  -10.0500 0.0001  -9.9750 0.0001  -9.9000 0.0001  -9.8250 0.0002  -9.7500 0.0002  -9.6750 0.0002  -9.6000 0.0002  -9.5250 0.0002  -9.4500 0.0002  -9.3750 0.0002  -9.3000 0.0003  -9.2250 0.0003  -9.1500 0.0003  -9.0750 0.0003  -9.0000 0.0004  -8.9250 0.0004  -8.8500 0.0004  -8.7750 0.0004  -8.7000 0.0005  -8.6250 0.0005  -8.5500 0.0006  -8.4750 0.0006  -8.4000 0.0006  -8.3250 0.0007  -8.2500 0.0007  -8.1750 0.0008  -8.1000 0.0009  -8.0250 0.0009  -7.9500 0.0010  -7.8750 0.0011  -7.8000 0.0012  -7.7250 0.0013  -7.6500 0.0014  -7.5750 0.0015  -7.5000 0.0016  -7.4250 0.0017  -7.3500 0.0018  -7.2750 0.0020  -7.2000 0.0021  -7.1250 0.0023  -7.0500 0.0025  -6.9750 0.0026  -6.9000 0.0028  -6.8250 0.0031  -6.7500 0.0033  -6.6750 0.0035  -6.6000 0.0038  -6.5250 0.0041  -6.4500 0.0044  -6.3750 0.0047  -6.3000 0.0051  -6.2250 0.0055  -6.1500 0.0059  -6.0750 0.0063  -6.0000 0.0068  -5.9250 0.0073  -5.8500 0.0078  -5.7750 0.0083  -5.7000 0.0089  -5.6250 0.0096  -5.5500 0.0103  -5.4750 0.0110  -5.4000 0.0118  -5.3250 0.0126  -5.2500 0.0135  -5.1750 0.0144  -5.1000 0.0154  -5.0250 0.0164  -4.9500 0.0175  -4.8750 0.0186  -4.8000 0.0199  -4.7250 0.0211  -4.6500 0.0225  -4.5750 0.0239  -4.5000 0.0253  -4.4250 0.0268  -4.3500 0.0284  -4.2750 0.0300  -4.2000 0.0317  -4.1250 0.0334  -4.0500 0.0352  -3.9750 0.0370  -3.9000 0.0389  -3.8250 0.0408  -3.7500 0.0427  -3.6750 0.0446  -3.6000 0.0466  -3.5250 0.0485  -3.4500 0.0504  -3.3750 0.0523  -3.3000 0.0542  -3.2250 0.0561  -3.1500 0.0579  -3.0750 0.0596  -3.0000 0.0613  -2.9250 0.0629  -2.8500 0.0644  -2.7750 0.0658  -2.7000 0.0672  -2.6250 0.0684  -2.5500 0.0696  -2.4750 0.0706  -2.4000 0.0716  -2.3250 0.0725  -2.2500 0.0733  -2.1750 0.0740  -2.1000 0.0747  -2.0250 0.0753  -1.9500 0.0759  -1.8750 0.0764  -1.8000 0.0769  -1.7250 0.0774  -1.6500 0.0779  -1.5750 0.0783  -1.5000 0.0788  -1.4250 0.0794  -1.3500 0.0799  -1.2750 0.0805  -1.2000 0.0812  -1.1250 0.0818  -1.0500 0.0826  -0.9750 0.0834  -0.9000 0.0842  -0.8250 0.0850  -0.7500 0.0859  -0.6750 0.0869  -0.6000 0.0878  -0.5250 0.0888  -0.4500 0.0897  -0.3750 0.0906  -0.3000 0.0916  -0.2250 0.0924  -0.1500 0.0933  -0.0750 0.0940  0.0000 0.0947  0.0750 0.0953  0.1500 0.0959  0.2250 0.0963  0.3000 0.0967  0.3750 0.0970  0.4500 0.0971  0.5250 0.0972  0.6000 0.0972  0.6750 0.0971  0.7500 0.0970  0.8250 0.0968  0.9000 0.0966  0.9750 0.0963  1.0500 0.0960  1.1250 0.0957  1.2000 0.0955  1.2750 0.0953  1.3500 0.0951  1.4250 0.0950  1.5000 0.0950  1.5750 0.0950  1.6500 0.0952  1.7250 0.0955  1.8000 0.0959  1.8750 0.0965  1.9500 0.0972  2.0250 0.0980  2.1000 0.0990  2.1750 0.1001  2.2500 0.1013  2.3250 0.1026  2.4000 0.1041  2.4750 0.1056  2.5500 0.1073  2.6250 0.1090  2.7000 0.1108  2.7750 0.1126  2.8500 0.1144  2.9250 0.1162  3.0000 0.1180  3.0750 0.1198  3.1500 0.1215  3.2250 0.1232  3.3000 0.1247  3.3750 0.1262  3.4500 0.1275  3.5250 0.1287  3.6000 0.1297  3.6750 0.1307  3.7500 0.1315  3.8250 0.1321  3.9000 0.1327  3.9750 0.1331  4.0500 0.1334  4.1250 0.1336  4.2000 0.1338  4.2750 0.1339  4.3500 0.1340  4.4250 0.1340  4.5000 0.1341  4.5750 0.1342  4.6500 0.1344  4.7250 0.1346  4.8000 0.1350  4.8750 0.1354  4.9500 0.1361  5.0250 0.1368  5.1000 0.1378  5.1750 0.1389  5.2500 0.1402  5.3250 0.1417  5.4000 0.1434  5.4750 0.1453  5.5500 0.1473  5.6250 0.1495  5.7000 0.1519  5.7750 0.1544  5.8500 0.1570  5.9250 0.1597  6.0000 0.1624  6.0750 0.1652  6.1500 0.1679  6.2250 0.1706  6.3000 0.1733  6.3750 0.1758  6.4500 0.1781  6.5250 0.1803  6.6000 0.1822  6.6750 0.1839  6.7500 0.1852  6.8250 0.1863  6.9000 0.1869  6.9750 0.1872  7.0500 0.1871  7.1250 0.1866  7.2000 0.1857  7.2750 0.1844  7.3500 0.1826  7.4250 0.1805  7.5000 0.1779  7.5750 0.1750  7.6500 0.1717  7.7250 0.1681  7.8000 0.1642  7.8750 0.1600  7.9500 0.1555  8.0250 0.1509  8.1000 0.1460  8.1750 0.1410  8.2500 0.1360  8.3250 0.1308  8.4000 0.1256  8.4750 0.1204  8.5500 0.1151  8.6250 0.1100  8.7000 0.1048  8.7750 0.0998  8.8500 0.0949  8.9250 0.0900  9.0000 0.0853  9.0750 0.0808  9.1500 0.0764  9.2250 0.0721  9.3000 0.0680  9.3750 0.0641  9.4500 0.0603  9.5250 0.0567  9.6000 0.0533  9.6750 0.0501  9.7500 0.0470  9.8250 0.0440  9.9000 0.0412  9.9750 0.0386  10.0500 0.0361  10.1250 0.0338  10.2000 0.0316  10.2750 0.0295  10.3500 0.0275  10.4250 0.0257  10.5000 0.0240  10.5750 0.0223  10.6500 0.0208  10.7250 0.0194  10.8000 0.0181  10.8750 0.0168  10.9500 0.0157  11.0250 0.0146  11.1000 0.0136  11.1750 0.0126  11.2500 0.0118  11.3250 0.0109  11.4000 0.0102  11.4750 0.0094  11.5500 0.0088  11.6250 0.0082  11.7000 0.0076  11.7750 0.0070  11.8500 0.0065  11.9250 0.0061  12.0000 0.0057  12.0750 0.0052  12.1500 0.0049  12.2250 0.0045  12.3000 0.0042  12.3750 0.0039  12.4500 0.0036  12.5250 0.0034  12.6000 0.0031  12.6750 0.0029  12.7500 0.0027  12.8250 0.0025  12.9000 0.0023  12.9750 0.0022  13.0500 0.0020  13.1250 0.0019  13.2000 0.0017  13.2750 0.0016  13.3500 0.0015  13.4250 0.0014  13.5000 0.0013  13.5750 0.0012  13.6500 0.0011  13.7250 0.0010  13.8000 0.0009  13.8750 0.0009  13.9500 0.0008  14.0250 0.0008  14.1000 0.0007  14.1750 0.0007  14.2500 0.0006  14.3250 0.0006  14.4000 0.0005  14.4750 0.0005  14.5500 0.0004  14.6250 0.0004  14.7000 0.0004  14.7750 0.0004  14.8500 0.0003  14.9250 0.0003  15.0000 0.0003   /

 \color{black}
\put {O} at 0 0 
\setplotarea x from -15.000000 to 15.000000, y from 0 to 0.300000
\axis top /
\axis right /
\axis left /
\axis bottom 
 label {{\large \hspace{2cm} $f$}} /
\axis bottom 
 ticks out withvalues -15 0 15 / quantity 3 /
\axis left shiftedto x=0 
 label {{\Large $\kappa \,$}} /
\axis left shiftedto x=0 
 ticks out withvalues 0 0.1 0.2  / quantity 4 /

\normalcolor
\endpicture

%% file: Compressgraphs/Compress6-123L.tex
\beginpicture

 \setcoordinatesystem units <25.000 pt,25 pt>

 \color{red}

\setplotsymbol ({\LARGE $\cdot$})

\plot -3.0000 5.0000  -2.9850 5.0000  -2.9700 5.0000  -2.9550 5.0000  -2.9400 5.0000  -2.9250 5.0000  -2.9100 5.0000  -2.8950 5.0000  -2.8800 5.0000  -2.8650 5.0000  -2.8500 5.0000  -2.8350 5.0000  -2.8200 5.0000  -2.8050 5.0000  -2.7900 5.0000  -2.7750 5.0000  -2.7600 5.0000  -2.7450 5.0000  -2.7300 5.0000  -2.7150 5.0000  -2.7000 5.0000  -2.6850 5.0000  -2.6700 5.0000  -2.6550 5.0000  -2.6400 5.0000  -2.6250 5.0000  -2.6100 5.0000  -2.5950 5.0000  -2.5800 5.0000  -2.5650 5.0000  -2.5500 5.0000  -2.5350 5.0000  -2.5200 5.0000  -2.5050 5.0000  -2.4900 5.0000  -2.4750 5.0000  -2.4600 5.0000  -2.4450 5.0000  -2.4300 5.0000  -2.4150 5.0000  -2.4000 5.0000  -2.3850 5.0000  -2.3700 5.0000  -2.3550 5.0000  -2.3400 5.0000  -2.3250 5.0000  -2.3100 5.0000  -2.2950 5.0000  -2.2800 5.0000  -2.2650 5.0000  -2.2500 5.0000  -2.2350 5.0000  -2.2200 5.0000  -2.2050 5.0000  -2.1900 5.0000  -2.1750 5.0000  -2.1600 5.0000  -2.1450 5.0000  -2.1300 5.0000  -2.1150 5.0000  -2.1000 5.0000  -2.0850 5.0000  -2.0700 5.0000  -2.0550 5.0000  -2.0400 5.0000  -2.0250 5.0000  -2.0100 5.0000  -1.9950 5.0000  -1.9800 5.0000  -1.9650 5.0000  -1.9500 5.0000  -1.9350 5.0000  -1.9200 5.0000  -1.9050 5.0000  -1.8900 5.0000  -1.8750 5.0000  -1.8600 5.0000  -1.8450 5.0000  -1.8300 5.0000  -1.8150 5.0000  -1.8000 5.0000  -1.7850 5.0000  -1.7700 5.0000  -1.7550 5.0000  -1.7400 5.0000  -1.7250 5.0000  -1.7100 5.0000  -1.6950 5.0000  -1.6800 5.0000  -1.6650 5.0000  -1.6500 5.0000  -1.6350 5.0000  -1.6200 5.0000  -1.6050 5.0000  -1.5900 5.0000  -1.5750 5.0000  -1.5600 5.0000  -1.5450 5.0000  -1.5300 5.0000  -1.5150 5.0000  -1.5000 5.0000  -1.4850 5.0000  -1.4700 5.0000  -1.4550 5.0000  -1.4400 5.0000  -1.4250 5.0000  -1.4100 5.0000  -1.3950 5.0000  -1.3800 5.0000  -1.3650 5.0000  -1.3500 5.0000  -1.3350 5.0000  -1.3200 5.0000  -1.3050 4.9999  -1.2900 4.9999  -1.2750 4.9999  -1.2600 4.9999  -1.2450 4.9999  -1.2300 4.9999  -1.2150 4.9999  -1.2000 4.9998  -1.1850 4.9998  -1.1700 4.9998  -1.1550 4.9998  -1.1400 4.9997  -1.1250 4.9997  -1.1100 4.9996  -1.0950 4.9996  -1.0800 4.9995  -1.0650 4.9994  -1.0500 4.9993  -1.0350 4.9992  -1.0200 4.9991  -1.0050 4.9989  -0.9900 4.9987  -0.9750 4.9985  -0.9600 4.9983  -0.9450 4.9980  -0.9300 4.9977  -0.9150 4.9973  -0.9000 4.9969  -0.8850 4.9964  -0.8700 4.9958  -0.8550 4.9951  -0.8400 4.9943  -0.8250 4.9934  -0.8100 4.9923  -0.7950 4.9911  -0.7800 4.9897  -0.7650 4.9880  -0.7500 4.9861  -0.7350 4.9839  -0.7200 4.9813  -0.7050 4.9784  -0.6900 4.9750  -0.6750 4.9710  -0.6600 4.9665  -0.6450 4.9613  -0.6300 4.9553  -0.6150 4.9484  -0.6000 4.9405  -0.5850 4.9315  -0.5700 4.9212  -0.5550 4.9096  -0.5400 4.8963  -0.5250 4.8814  -0.5100 4.8647  -0.4950 4.8460  -0.4800 4.8251  -0.4650 4.8020  -0.4500 4.7767  -0.4350 4.7489  -0.4200 4.7189  -0.4050 4.6865  -0.3900 4.6519  -0.3750 4.6152  -0.3600 4.5766  -0.3450 4.5364  -0.3300 4.4947  -0.3150 4.4517  -0.3000 4.4077  -0.2850 4.3626  -0.2700 4.3166  -0.2550 4.2693  -0.2400 4.2205  -0.2250 4.1696  -0.2100 4.1157  -0.1950 4.0580  -0.1800 3.9950  -0.1650 3.9254  -0.1500 3.8477  -0.1350 3.7604  -0.1200 3.6625  -0.1050 3.5536  -0.0900 3.4341  -0.0750 3.3057  -0.0600 3.1712  -0.0450 3.0344  -0.0300 2.8995  -0.0150 2.7708  0.0000 2.6517  0.0150 2.5447  0.0300 2.4509  0.0450 2.3705  0.0600 2.3029  0.0750 2.2467  0.0900 2.2007  0.1050 2.1632  0.1200 2.1329  0.1350 2.1084  0.1500 2.0886  0.1650 2.0727  0.1800 2.0598  0.1950 2.0494  0.2100 2.0409  0.2250 2.0341  0.2400 2.0284  0.2550 2.0238  0.2700 2.0200  0.2850 2.0168  0.3000 2.0142  0.3150 2.0120  0.3300 2.0102  0.3450 2.0086  0.3600 2.0074  0.3750 2.0063  0.3900 2.0053  0.4050 2.0046  0.4200 2.0039  0.4350 2.0033  0.4500 2.0028  0.4650 2.0024  0.4800 2.0021  0.4950 2.0018  0.5100 2.0015  0.5250 2.0013  0.5400 2.0011  0.5550 2.0009  0.5700 2.0008  0.5850 2.0007  0.6000 2.0005  0.6150 2.0004  0.6300 2.0004  0.6450 2.0003  0.6600 2.0002  0.6750 2.0001  0.6900 2.0001  0.7050 2.0000  0.7200 1.9999  0.7350 1.9999  0.7500 1.9998  0.7650 1.9997  0.7800 1.9996  0.7950 1.9995  0.8100 1.9994  0.8250 1.9993  0.8400 1.9992  0.8550 1.9990  0.8700 1.9989  0.8850 1.9987  0.9000 1.9985  0.9150 1.9982  0.9300 1.9979  0.9450 1.9976  0.9600 1.9972  0.9750 1.9967  0.9900 1.9962  1.0050 1.9956  1.0200 1.9948  1.0350 1.9940  1.0500 1.9930  1.0650 1.9919  1.0800 1.9906  1.0950 1.9891  1.1100 1.9874  1.1250 1.9853  1.1400 1.9830  1.1550 1.9803  1.1700 1.9772  1.1850 1.9736  1.2000 1.9695  1.2150 1.9647  1.2300 1.9592  1.2450 1.9529  1.2600 1.9457  1.2750 1.9375  1.2900 1.9281  1.3050 1.9174  1.3200 1.9053  1.3350 1.8917  1.3500 1.8763  1.3650 1.8591  1.3800 1.8400  1.3950 1.8188  1.4100 1.7954  1.4250 1.7699  1.4400 1.7423  1.4550 1.7126  1.4700 1.6809  1.4850 1.6475  1.5000 1.6125  1.5150 1.5764  1.5300 1.5394  1.5450 1.5020  1.5600 1.4646  1.5750 1.4275  1.5900 1.3913  1.6050 1.3562  1.6200 1.3226  1.6350 1.2907  1.6500 1.2608  1.6650 1.2329  1.6800 1.2072  1.6950 1.1836  1.7100 1.1622  1.7250 1.1428  1.7400 1.1254  1.7550 1.1099  1.7700 1.0960  1.7850 1.0838  1.8000 1.0730  1.8150 1.0634  1.8300 1.0551  1.8450 1.0478  1.8600 1.0414  1.8750 1.0358  1.8900 1.0310  1.9050 1.0268  1.9200 1.0232  1.9350 1.0200  1.9500 1.0173  1.9650 1.0149  1.9800 1.0128  1.9950 1.0111  2.0100 1.0095  2.0250 1.0082  2.0400 1.0071  2.0550 1.0061  2.0700 1.0053  2.0850 1.0045  2.1000 1.0039  2.1150 1.0034  2.1300 1.0029  2.1450 1.0025  2.1600 1.0021  2.1750 1.0018  2.1900 1.0016  2.2050 1.0014  2.2200 1.0012  2.2350 1.0010  2.2500 1.0009  2.2650 1.0008  2.2800 1.0006  2.2950 1.0006  2.3100 1.0005  2.3250 1.0004  2.3400 1.0004  2.3550 1.0003  2.3700 1.0003  2.3850 1.0002  2.4000 1.0002  2.4150 1.0002  2.4300 1.0001  2.4450 1.0001  2.4600 1.0001  2.4750 1.0001  2.4900 1.0001  2.5050 1.0001  2.5200 1.0001  2.5350 1.0001  2.5500 1.0000  2.5650 1.0000  2.5800 1.0000  2.5950 1.0000  2.6100 1.0000  2.6250 1.0000  2.6400 1.0000  2.6550 1.0000  2.6700 1.0000  2.6850 1.0000  2.7000 1.0000  2.7150 1.0000  2.7300 1.0000  2.7450 1.0000  2.7600 1.0000  2.7750 1.0000  2.7900 1.0000  2.8050 1.0000  2.8200 1.0000  2.8350 1.0000  2.8500 1.0000  2.8650 1.0000  2.8800 1.0000  2.8950 1.0000  2.9100 1.0000  2.9250 1.0000  2.9400 1.0000  2.9550 1.0000  2.9700 1.0000  2.9850 1.0000  3.0000 1.0000   /

 \color{blue}

\setplotsymbol ({\LARGE $\cdot$})

\plot -3.0000 5.0000  -2.9850 5.0000  -2.9700 5.0000  -2.9550 5.0000  -2.9400 5.0000  -2.9250 5.0000  -2.9100 5.0000  -2.8950 5.0000  -2.8800 5.0000  -2.8650 5.0000  -2.8500 5.0000  -2.8350 5.0000  -2.8200 5.0000  -2.8050 5.0000  -2.7900 5.0000  -2.7750 5.0000  -2.7600 5.0000  -2.7450 5.0000  -2.7300 5.0000  -2.7150 5.0000  -2.7000 5.0000  -2.6850 5.0000  -2.6700 5.0000  -2.6550 5.0000  -2.6400 5.0000  -2.6250 5.0000  -2.6100 5.0000  -2.5950 5.0000  -2.5800 5.0000  -2.5650 5.0000  -2.5500 5.0000  -2.5350 5.0000  -2.5200 5.0000  -2.5050 5.0000  -2.4900 5.0000  -2.4750 5.0000  -2.4600 5.0000  -2.4450 5.0000  -2.4300 5.0000  -2.4150 5.0000  -2.4000 5.0000  -2.3850 5.0000  -2.3700 5.0000  -2.3550 5.0000  -2.3400 5.0000  -2.3250 5.0000  -2.3100 5.0000  -2.2950 5.0000  -2.2800 5.0000  -2.2650 5.0000  -2.2500 5.0000  -2.2350 5.0000  -2.2200 5.0000  -2.2050 5.0000  -2.1900 5.0000  -2.1750 5.0000  -2.1600 5.0000  -2.1450 5.0000  -2.1300 5.0000  -2.1150 5.0000  -2.1000 5.0000  -2.0850 5.0000  -2.0700 5.0000  -2.0550 5.0000  -2.0400 5.0000  -2.0250 5.0000  -2.0100 5.0000  -1.9950 5.0000  -1.9800 5.0000  -1.9650 5.0000  -1.9500 5.0000  -1.9350 5.0000  -1.9200 5.0000  -1.9050 5.0000  -1.8900 5.0000  -1.8750 5.0000  -1.8600 5.0000  -1.8450 5.0000  -1.8300 5.0000  -1.8150 5.0000  -1.8000 5.0000  -1.7850 5.0000  -1.7700 5.0000  -1.7550 5.0000  -1.7400 5.0000  -1.7250 5.0000  -1.7100 5.0000  -1.6950 5.0000  -1.6800 5.0000  -1.6650 5.0000  -1.6500 5.0000  -1.6350 5.0000  -1.6200 5.0000  -1.6050 5.0000  -1.5900 5.0000  -1.5750 5.0000  -1.5600 5.0000  -1.5450 5.0000  -1.5300 5.0000  -1.5150 5.0000  -1.5000 5.0000  -1.4850 5.0000  -1.4700 5.0000  -1.4550 5.0000  -1.4400 5.0000  -1.4250 5.0000  -1.4100 5.0000  -1.3950 5.0000  -1.3800 5.0000  -1.3650 5.0000  -1.3500 5.0000  -1.3350 5.0000  -1.3200 5.0000  -1.3050 5.0000  -1.2900 5.0000  -1.2750 5.0000  -1.2600 5.0000  -1.2450 5.0000  -1.2300 5.0000  -1.2150 5.0000  -1.2000 5.0000  -1.1850 5.0000  -1.1700 4.9999  -1.1550 4.9999  -1.1400 4.9999  -1.1250 4.9999  -1.1100 4.9999  -1.0950 4.9999  -1.0800 4.9999  -1.0650 4.9998  -1.0500 4.9998  -1.0350 4.9998  -1.0200 4.9998  -1.0050 4.9997  -0.9900 4.9997  -0.9750 4.9996  -0.9600 4.9996  -0.9450 4.9995  -0.9300 4.9994  -0.9150 4.9993  -0.9000 4.9992  -0.8850 4.9991  -0.8700 4.9989  -0.8550 4.9987  -0.8400 4.9985  -0.8250 4.9983  -0.8100 4.9980  -0.7950 4.9977  -0.7800 4.9973  -0.7650 4.9969  -0.7500 4.9964  -0.7350 4.9958  -0.7200 4.9951  -0.7050 4.9944  -0.6900 4.9935  -0.6750 4.9924  -0.6600 4.9912  -0.6450 4.9898  -0.6300 4.9881  -0.6150 4.9862  -0.6000 4.9841  -0.5850 4.9815  -0.5700 4.9786  -0.5550 4.9752  -0.5400 4.9713  -0.5250 4.9668  -0.5100 4.9616  -0.4950 4.9556  -0.4800 4.9488  -0.4650 4.9410  -0.4500 4.9320  -0.4350 4.9218  -0.4200 4.9102  -0.4050 4.8970  -0.3900 4.8821  -0.3750 4.8654  -0.3600 4.8466  -0.3450 4.8258  -0.3300 4.8026  -0.3150 4.7771  -0.3000 4.7492  -0.2850 4.7188  -0.2700 4.6861  -0.2550 4.6511  -0.2400 4.6139  -0.2250 4.5748  -0.2100 4.5339  -0.1950 4.4917  -0.1800 4.4483  -0.1650 4.4041  -0.1500 4.3594  -0.1350 4.3143  -0.1200 4.2691  -0.1050 4.2238  -0.0900 4.1785  -0.0750 4.1331  -0.0600 4.0875  -0.0450 4.0413  -0.0300 3.9943  -0.0150 3.9461  0.0000 3.8963  0.0150 3.8444  0.0300 3.7900  0.0450 3.7326  0.0600 3.6718  0.0750 3.6075  0.0900 3.5393  0.1050 3.4674  0.1200 3.3919  0.1350 3.3132  0.1500 3.2318  0.1650 3.1484  0.1800 3.0641  0.1950 2.9798  0.2100 2.8965  0.2250 2.8153  0.2400 2.7371  0.2550 2.6627  0.2700 2.5927  0.2850 2.5277  0.3000 2.4677  0.3150 2.4129  0.3300 2.3633  0.3450 2.3187  0.3600 2.2788  0.3750 2.2434  0.3900 2.2120  0.4050 2.1844  0.4200 2.1601  0.4350 2.1388  0.4500 2.1202  0.4650 2.1040  0.4800 2.0900  0.4950 2.0777  0.5100 2.0671  0.5250 2.0579  0.5400 2.0499  0.5550 2.0430  0.5700 2.0370  0.5850 2.0318  0.6000 2.0273  0.6150 2.0234  0.6300 2.0200  0.6450 2.0171  0.6600 2.0146  0.6750 2.0123  0.6900 2.0104  0.7050 2.0087  0.7200 2.0071  0.7350 2.0058  0.7500 2.0045  0.7650 2.0034  0.7800 2.0023  0.7950 2.0013  0.8100 2.0003  0.8250 1.9994  0.8400 1.9984  0.8550 1.9974  0.8700 1.9963  0.8850 1.9951  0.9000 1.9938  0.9150 1.9924  0.9300 1.9909  0.9450 1.9891  0.9600 1.9871  0.9750 1.9848  0.9900 1.9822  1.0050 1.9792  1.0200 1.9758  1.0350 1.9719  1.0500 1.9674  1.0650 1.9622  1.0800 1.9563  1.0950 1.9495  1.1100 1.9417  1.1250 1.9329  1.1400 1.9228  1.1550 1.9114  1.1700 1.8985  1.1850 1.8839  1.2000 1.8676  1.2150 1.8494  1.2300 1.8292  1.2450 1.8069  1.2600 1.7824  1.2750 1.7558  1.2900 1.7270  1.3050 1.6963  1.3200 1.6636  1.3350 1.6294  1.3500 1.5937  1.3650 1.5571  1.3800 1.5199  1.3950 1.4824  1.4100 1.4451  1.4250 1.4084  1.4400 1.3727  1.4550 1.3384  1.4700 1.3056  1.4850 1.2748  1.5000 1.2459  1.5150 1.2192  1.5300 1.1946  1.5450 1.1721  1.5600 1.1518  1.5750 1.1335  1.5900 1.1171  1.6050 1.1024  1.6200 1.0894  1.6350 1.0779  1.6500 1.0678  1.6650 1.0589  1.6800 1.0511  1.6950 1.0443  1.7100 1.0384  1.7250 1.0332  1.7400 1.0287  1.7550 1.0248  1.7700 1.0214  1.7850 1.0185  1.8000 1.0160  1.8150 1.0138  1.8300 1.0119  1.8450 1.0102  1.8600 1.0088  1.8750 1.0076  1.8900 1.0066  1.9050 1.0056  1.9200 1.0049  1.9350 1.0042  1.9500 1.0036  1.9650 1.0031  1.9800 1.0027  1.9950 1.0023  2.0100 1.0020  2.0250 1.0017  2.0400 1.0015  2.0550 1.0013  2.0700 1.0011  2.0850 1.0009  2.1000 1.0008  2.1150 1.0007  2.1300 1.0006  2.1450 1.0005  2.1600 1.0004  2.1750 1.0004  2.1900 1.0003  2.2050 1.0003  2.2200 1.0002  2.2350 1.0002  2.2500 1.0002  2.2650 1.0002  2.2800 1.0001  2.2950 1.0001  2.3100 1.0001  2.3250 1.0001  2.3400 1.0001  2.3550 1.0001  2.3700 1.0001  2.3850 1.0000  2.4000 1.0000  2.4150 1.0000  2.4300 1.0000  2.4450 1.0000  2.4600 1.0000  2.4750 1.0000  2.4900 1.0000  2.5050 1.0000  2.5200 1.0000  2.5350 1.0000  2.5500 1.0000  2.5650 1.0000  2.5800 1.0000  2.5950 1.0000  2.6100 1.0000  2.6250 1.0000  2.6400 1.0000  2.6550 1.0000  2.6700 1.0000  2.6850 1.0000  2.7000 1.0000  2.7150 1.0000  2.7300 1.0000  2.7450 1.0000  2.7600 1.0000  2.7750 1.0000  2.7900 1.0000  2.8050 1.0000  2.8200 1.0000  2.8350 1.0000  2.8500 1.0000  2.8650 1.0000  2.8800 1.0000  2.8950 1.0000  2.9100 1.0000  2.9250 1.0000  2.9400 1.0000  2.9550 1.0000  2.9700 1.0000  2.9850 1.0000  3.0000 1.0000   /

 \color{green}

\setplotsymbol ({\LARGE $\cdot$})

\plot -3.0000 5.0000  -2.9850 5.0000  -2.9700 5.0000  -2.9550 5.0000  -2.9400 5.0000  -2.9250 5.0000  -2.9100 5.0000  -2.8950 5.0000  -2.8800 5.0000  -2.8650 5.0000  -2.8500 5.0000  -2.8350 5.0000  -2.8200 5.0000  -2.8050 5.0000  -2.7900 5.0000  -2.7750 5.0000  -2.7600 5.0000  -2.7450 5.0000  -2.7300 5.0000  -2.7150 5.0000  -2.7000 5.0000  -2.6850 5.0000  -2.6700 5.0000  -2.6550 5.0000  -2.6400 5.0000  -2.6250 5.0000  -2.6100 5.0000  -2.5950 5.0000  -2.5800 5.0000  -2.5650 5.0000  -2.5500 5.0000  -2.5350 5.0000  -2.5200 5.0000  -2.5050 5.0000  -2.4900 5.0000  -2.4750 5.0000  -2.4600 5.0000  -2.4450 5.0000  -2.4300 5.0000  -2.4150 5.0000  -2.4000 5.0000  -2.3850 5.0000  -2.3700 5.0000  -2.3550 5.0000  -2.3400 5.0000  -2.3250 5.0000  -2.3100 5.0000  -2.2950 5.0000  -2.2800 5.0000  -2.2650 5.0000  -2.2500 5.0000  -2.2350 5.0000  -2.2200 5.0000  -2.2050 5.0000  -2.1900 5.0000  -2.1750 5.0000  -2.1600 5.0000  -2.1450 5.0000  -2.1300 5.0000  -2.1150 5.0000  -2.1000 5.0000  -2.0850 5.0000  -2.0700 5.0000  -2.0550 5.0000  -2.0400 5.0000  -2.0250 5.0000  -2.0100 5.0000  -1.9950 5.0000  -1.9800 5.0000  -1.9650 5.0000  -1.9500 5.0000  -1.9350 5.0000  -1.9200 5.0000  -1.9050 5.0000  -1.8900 5.0000  -1.8750 5.0000  -1.8600 5.0000  -1.8450 5.0000  -1.8300 5.0000  -1.8150 5.0000  -1.8000 5.0000  -1.7850 5.0000  -1.7700 5.0000  -1.7550 5.0000  -1.7400 5.0000  -1.7250 5.0000  -1.7100 5.0000  -1.6950 5.0000  -1.6800 5.0000  -1.6650 5.0000  -1.6500 5.0000  -1.6350 5.0000  -1.6200 5.0000  -1.6050 5.0000  -1.5900 5.0000  -1.5750 5.0000  -1.5600 5.0000  -1.5450 5.0000  -1.5300 5.0000  -1.5150 5.0000  -1.5000 5.0000  -1.4850 5.0000  -1.4700 5.0000  -1.4550 5.0000  -1.4400 5.0000  -1.4250 5.0000  -1.4100 5.0000  -1.3950 5.0000  -1.3800 5.0000  -1.3650 5.0000  -1.3500 5.0000  -1.3350 5.0000  -1.3200 5.0000  -1.3050 5.0000  -1.2900 5.0000  -1.2750 4.9999  -1.2600 4.9999  -1.2450 4.9999  -1.2300 4.9999  -1.2150 4.9999  -1.2000 4.9999  -1.1850 4.9999  -1.1700 4.9998  -1.1550 4.9998  -1.1400 4.9998  -1.1250 4.9997  -1.1100 4.9997  -1.0950 4.9997  -1.0800 4.9996  -1.0650 4.9995  -1.0500 4.9995  -1.0350 4.9994  -1.0200 4.9993  -1.0050 4.9992  -0.9900 4.9990  -0.9750 4.9989  -0.9600 4.9987  -0.9450 4.9985  -0.9300 4.9982  -0.9150 4.9979  -0.9000 4.9976  -0.8850 4.9972  -0.8700 4.9968  -0.8550 4.9962  -0.8400 4.9956  -0.8250 4.9949  -0.8100 4.9941  -0.7950 4.9931  -0.7800 4.9920  -0.7650 4.9908  -0.7500 4.9893  -0.7350 4.9876  -0.7200 4.9856  -0.7050 4.9833  -0.6900 4.9807  -0.6750 4.9776  -0.6600 4.9741  -0.6450 4.9700  -0.6300 4.9653  -0.6150 4.9599  -0.6000 4.9537  -0.5850 4.9466  -0.5700 4.9385  -0.5550 4.9292  -0.5400 4.9187  -0.5250 4.9067  -0.5100 4.8932  -0.4950 4.8780  -0.4800 4.8609  -0.4650 4.8419  -0.4500 4.8208  -0.4350 4.7975  -0.4200 4.7720  -0.4050 4.7442  -0.3900 4.7143  -0.3750 4.6823  -0.3600 4.6484  -0.3450 4.6128  -0.3300 4.5758  -0.3150 4.5377  -0.3000 4.4988  -0.2850 4.4596  -0.2700 4.4204  -0.2550 4.3816  -0.2400 4.3435  -0.2250 4.3063  -0.2100 4.2703  -0.1950 4.2356  -0.1800 4.2021  -0.1650 4.1699  -0.1500 4.1387  -0.1350 4.1085  -0.1200 4.0788  -0.1050 4.0494  -0.0900 4.0198  -0.0750 3.9894  -0.0600 3.9577  -0.0450 3.9241  -0.0300 3.8877  -0.0150 3.8479  0.0000 3.8038  0.0150 3.7545  0.0300 3.6991  0.0450 3.6370  0.0600 3.5676  0.0750 3.4906  0.0900 3.4062  0.1050 3.3148  0.1200 3.2176  0.1350 3.1164  0.1500 3.0130  0.1650 2.9099  0.1800 2.8093  0.1950 2.7132  0.2100 2.6235  0.2250 2.5411  0.2400 2.4668  0.2550 2.4009  0.2700 2.3429  0.2850 2.2926  0.3000 2.2492  0.3150 2.2120  0.3300 2.1803  0.3450 2.1533  0.3600 2.1304  0.3750 2.1110  0.3900 2.0946  0.4050 2.0806  0.4200 2.0688  0.4350 2.0587  0.4500 2.0502  0.4650 2.0429  0.4800 2.0367  0.4950 2.0314  0.5100 2.0269  0.5250 2.0231  0.5400 2.0198  0.5550 2.0170  0.5700 2.0146  0.5850 2.0125  0.6000 2.0108  0.6150 2.0093  0.6300 2.0080  0.6450 2.0068  0.6600 2.0059  0.6750 2.0051  0.6900 2.0043  0.7050 2.0037  0.7200 2.0032  0.7350 2.0028  0.7500 2.0024  0.7650 2.0020  0.7800 2.0018  0.7950 2.0015  0.8100 2.0013  0.8250 2.0011  0.8400 2.0010  0.8550 2.0008  0.8700 2.0007  0.8850 2.0006  0.9000 2.0005  0.9150 2.0005  0.9300 2.0004  0.9450 2.0003  0.9600 2.0003  0.9750 2.0002  0.9900 2.0002  1.0050 2.0002  1.0200 2.0002  1.0350 2.0001  1.0500 2.0001  1.0650 2.0001  1.0800 2.0001  1.0950 2.0001  1.1100 2.0001  1.1250 2.0000  1.1400 2.0000  1.1550 2.0000  1.1700 2.0000  1.1850 2.0000  1.2000 2.0000  1.2150 2.0000  1.2300 2.0000  1.2450 2.0000  1.2600 2.0000  1.2750 2.0000  1.2900 2.0000  1.3050 2.0000  1.3200 1.9999  1.3350 1.9999  1.3500 1.9999  1.3650 1.9999  1.3800 1.9999  1.3950 1.9999  1.4100 1.9998  1.4250 1.9998  1.4400 1.9998  1.4550 1.9997  1.4700 1.9997  1.4850 1.9996  1.5000 1.9996  1.5150 1.9995  1.5300 1.9994  1.5450 1.9994  1.5600 1.9993  1.5750 1.9991  1.5900 1.9990  1.6050 1.9988  1.6200 1.9986  1.6350 1.9984  1.6500 1.9982  1.6650 1.9979  1.6800 1.9975  1.6950 1.9971  1.7100 1.9967  1.7250 1.9961  1.7400 1.9955  1.7550 1.9948  1.7700 1.9939  1.7850 1.9930  1.8000 1.9918  1.8150 1.9905  1.8300 1.9890  1.8450 1.9872  1.8600 1.9852  1.8750 1.9828  1.8900 1.9801  1.9050 1.9770  1.9200 1.9734  1.9350 1.9692  1.9500 1.9644  1.9650 1.9588  1.9800 1.9525  1.9950 1.9452  2.0100 1.9369  2.0250 1.9274  2.0400 1.9167  2.0550 1.9045  2.0700 1.8907  2.0850 1.8752  2.1000 1.8579  2.1150 1.8386  2.1300 1.8173  2.1450 1.7938  2.1600 1.7682  2.1750 1.7404  2.1900 1.7106  2.2050 1.6788  2.2200 1.6452  2.2350 1.6102  2.2500 1.5740  2.2650 1.5370  2.2800 1.4995  2.2950 1.4621  2.3100 1.4251  2.3250 1.3889  2.3400 1.3539  2.3550 1.3204  2.3700 1.2887  2.3850 1.2589  2.4000 1.2311  2.4150 1.2056  2.4300 1.1821  2.4450 1.1609  2.4600 1.1416  2.4750 1.1244  2.4900 1.1089  2.5050 1.0952  2.5200 1.0830  2.5350 1.0723  2.5500 1.0629  2.5650 1.0546  2.5800 1.0473  2.5950 1.0410  2.6100 1.0355  2.6250 1.0307  2.6400 1.0265  2.6550 1.0229  2.6700 1.0198  2.6850 1.0171  2.7000 1.0147  2.7150 1.0127  2.7300 1.0110  2.7450 1.0095  2.7600 1.0081  2.7750 1.0070  2.7900 1.0060  2.8050 1.0052  2.8200 1.0045  2.8350 1.0039  2.8500 1.0033  2.8650 1.0029  2.8800 1.0025  2.8950 1.0021  2.9100 1.0018  2.9250 1.0016  2.9400 1.0014  2.9550 1.0012  2.9700 1.0010  2.9850 1.0009  3.0000 1.0007   /

 \color{black}
\put {O} at 0 0 
\setplotarea x from -3.000000 to 3.000000, y from 0.000000 to 6.000000
\axis top /
\axis right /
\axis left /
\axis bottom 
 label {{\Large \hspace{2cm} $f$}} /
\axis bottom 
 ticks out withvalues -30 0 30 / quantity 3 /
\axis left shiftedto x=0 
 label {{\Large $\LA h \RA \,$}} /
\axis left shiftedto x=0 
 ticks out withvalues {} 1 2 {} 4 {} {} / quantity 7 /

\normalcolor
\endpicture

%% file: Compressgraphs/Compress6-123F.tex
\beginpicture

 \setcoordinatesystem units <25.000 pt,375.000 pt>

 \color{red}

\setplotsymbol ({\LARGE $\cdot$})

\plot -3.0000 0.0000  -2.9850 0.0000  -2.9700 0.0000  -2.9550 0.0000  -2.9400 0.0000  -2.9250 0.0000  -2.9100 0.0000  -2.8950 0.0000  -2.8800 0.0000  -2.8650 0.0000  -2.8500 0.0000  -2.8350 0.0000  -2.8200 0.0000  -2.8050 0.0000  -2.7900 0.0000  -2.7750 0.0000  -2.7600 0.0000  -2.7450 0.0000  -2.7300 0.0000  -2.7150 0.0000  -2.7000 0.0000  -2.6850 0.0000  -2.6700 0.0000  -2.6550 0.0000  -2.6400 0.0000  -2.6250 0.0000  -2.6100 0.0000  -2.5950 0.0000  -2.5800 0.0000  -2.5650 0.0000  -2.5500 0.0000  -2.5350 0.0000  -2.5200 0.0000  -2.5050 0.0000  -2.4900 0.0000  -2.4750 0.0000  -2.4600 0.0000  -2.4450 0.0000  -2.4300 0.0000  -2.4150 0.0000  -2.4000 0.0000  -2.3850 0.0000  -2.3700 0.0000  -2.3550 0.0000  -2.3400 0.0000  -2.3250 0.0000  -2.3100 0.0000  -2.2950 0.0000  -2.2800 0.0000  -2.2650 0.0000  -2.2500 0.0000  -2.2350 0.0000  -2.2200 0.0000  -2.2050 0.0000  -2.1900 0.0000  -2.1750 0.0000  -2.1600 0.0000  -2.1450 0.0000  -2.1300 0.0000  -2.1150 0.0000  -2.1000 0.0000  -2.0850 0.0000  -2.0700 0.0000  -2.0550 0.0000  -2.0400 0.0000  -2.0250 0.0000  -2.0100 0.0000  -1.9950 0.0000  -1.9800 0.0000  -1.9650 0.0000  -1.9500 0.0000  -1.9350 0.0000  -1.9200 0.0000  -1.9050 0.0000  -1.8900 0.0000  -1.8750 0.0000  -1.8600 0.0000  -1.8450 0.0000  -1.8300 0.0000  -1.8150 0.0000  -1.8000 0.0000  -1.7850 0.0000  -1.7700 0.0000  -1.7550 0.0000  -1.7400 0.0000  -1.7250 0.0000  -1.7100 0.0000  -1.6950 0.0000  -1.6800 0.0000  -1.6650 0.0000  -1.6500 0.0000  -1.6350 0.0000  -1.6200 0.0000  -1.6050 0.0000  -1.5900 0.0000  -1.5750 0.0000  -1.5600 0.0000  -1.5450 0.0000  -1.5300 0.0000  -1.5150 0.0000  -1.5000 0.0000  -1.4850 0.0000  -1.4700 0.0000  -1.4550 0.0000  -1.4400 0.0000  -1.4250 0.0000  -1.4100 0.0000  -1.3950 0.0000  -1.3800 0.0000  -1.3650 0.0000  -1.3500 0.0000  -1.3350 0.0000  -1.3200 0.0000  -1.3050 0.0000  -1.2900 0.0000  -1.2750 0.0000  -1.2600 0.0000  -1.2450 0.0000  -1.2300 0.0000  -1.2150 0.0000  -1.2000 0.0000  -1.1850 0.0000  -1.1700 0.0000  -1.1550 0.0000  -1.1400 0.0001  -1.1250 0.0001  -1.1100 0.0001  -1.0950 0.0001  -1.0800 0.0001  -1.0650 0.0001  -1.0500 0.0001  -1.0350 0.0002  -1.0200 0.0002  -1.0050 0.0002  -0.9900 0.0003  -0.9750 0.0003  -0.9600 0.0003  -0.9450 0.0004  -0.9300 0.0005  -0.9150 0.0005  -0.9000 0.0006  -0.8850 0.0007  -0.8700 0.0008  -0.8550 0.0010  -0.8400 0.0011  -0.8250 0.0013  -0.8100 0.0015  -0.7950 0.0018  -0.7800 0.0020  -0.7650 0.0024  -0.7500 0.0027  -0.7350 0.0032  -0.7200 0.0037  -0.7050 0.0043  -0.6900 0.0049  -0.6750 0.0057  -0.6600 0.0065  -0.6450 0.0075  -0.6300 0.0086  -0.6150 0.0099  -0.6000 0.0114  -0.5850 0.0130  -0.5700 0.0148  -0.5550 0.0169  -0.5400 0.0191  -0.5250 0.0216  -0.5100 0.0243  -0.4950 0.0272  -0.4800 0.0303  -0.4650 0.0336  -0.4500 0.0371  -0.4350 0.0406  -0.4200 0.0441  -0.4050 0.0477  -0.3900 0.0511  -0.3750 0.0544  -0.3600 0.0575  -0.3450 0.0603  -0.3300 0.0628  -0.3150 0.0652  -0.3000 0.0674  -0.2850 0.0696  -0.2700 0.0720  -0.2550 0.0749  -0.2400 0.0786  -0.2250 0.0835  -0.2100 0.0901  -0.1950 0.0988  -0.1800 0.1102  -0.1650 0.1247  -0.1500 0.1426  -0.1350 0.1639  -0.1200 0.1882  -0.1050 0.2146  -0.0900 0.2414  -0.0750 0.2663  -0.0600 0.2867  -0.0450 0.3000  -0.0300 0.3045  -0.0150 0.2993  0.0000 0.2850  0.0150 0.2634  0.0300 0.2367  0.0450 0.2077  0.0600 0.1785  0.0750 0.1508  0.0900 0.1258  0.1050 0.1038  0.1200 0.0851  0.1350 0.0694  0.1500 0.0565  0.1650 0.0459  0.1800 0.0374  0.1950 0.0305  0.2100 0.0249  0.2250 0.0204  0.2400 0.0167  0.2550 0.0138  0.2700 0.0114  0.2850 0.0095  0.3000 0.0079  0.3150 0.0066  0.3300 0.0055  0.3450 0.0047  0.3600 0.0039  0.3750 0.0033  0.3900 0.0028  0.4050 0.0024  0.4200 0.0020  0.4350 0.0017  0.4500 0.0015  0.4650 0.0013  0.4800 0.0011  0.4950 0.0009  0.5100 0.0008  0.5250 0.0007  0.5400 0.0006  0.5550 0.0005  0.5700 0.0005  0.5850 0.0004  0.6000 0.0004  0.6150 0.0003  0.6300 0.0003  0.6450 0.0003  0.6600 0.0002  0.6750 0.0002  0.6900 0.0002  0.7050 0.0002  0.7200 0.0002  0.7350 0.0002  0.7500 0.0002  0.7650 0.0003  0.7800 0.0003  0.7950 0.0003  0.8100 0.0004  0.8250 0.0004  0.8400 0.0005  0.8550 0.0005  0.8700 0.0006  0.8850 0.0007  0.9000 0.0008  0.9150 0.0009  0.9300 0.0011  0.9450 0.0012  0.9600 0.0014  0.9750 0.0017  0.9900 0.0019  1.0050 0.0022  1.0200 0.0026  1.0350 0.0030  1.0500 0.0035  1.0650 0.0040  1.0800 0.0047  1.0950 0.0054  1.1100 0.0063  1.1250 0.0073  1.1400 0.0084  1.1550 0.0097  1.1700 0.0113  1.1850 0.0130  1.2000 0.0150  1.2150 0.0173  1.2300 0.0200  1.2450 0.0230  1.2600 0.0264  1.2750 0.0302  1.2900 0.0346  1.3050 0.0395  1.3200 0.0450  1.3350 0.0511  1.3500 0.0578  1.3650 0.0651  1.3800 0.0731  1.3950 0.0816  1.4100 0.0906  1.4250 0.1001  1.4400 0.1098  1.4550 0.1196  1.4700 0.1293  1.4850 0.1385  1.5000 0.1472  1.5150 0.1549  1.5300 0.1614  1.5450 0.1664  1.5600 0.1698  1.5750 0.1714  1.5900 0.1712  1.6050 0.1691  1.6200 0.1652  1.6350 0.1598  1.6500 0.1529  1.6650 0.1449  1.6800 0.1361  1.6950 0.1266  1.7100 0.1169  1.7250 0.1071  1.7400 0.0975  1.7550 0.0881  1.7700 0.0792  1.7850 0.0708  1.8000 0.0630  1.8150 0.0559  1.8300 0.0493  1.8450 0.0434  1.8600 0.0381  1.8750 0.0334  1.8900 0.0291  1.9050 0.0254  1.9200 0.0221  1.9350 0.0192  1.9500 0.0167  1.9650 0.0145  1.9800 0.0125  1.9950 0.0108  2.0100 0.0094  2.0250 0.0081  2.0400 0.0070  2.0550 0.0060  2.0700 0.0052  2.0850 0.0045  2.1000 0.0039  2.1150 0.0033  2.1300 0.0029  2.1450 0.0025  2.1600 0.0021  2.1750 0.0018  2.1900 0.0016  2.2050 0.0014  2.2200 0.0012  2.2350 0.0010  2.2500 0.0009  2.2650 0.0008  2.2800 0.0006  2.2950 0.0006  2.3100 0.0005  2.3250 0.0004  2.3400 0.0004  2.3550 0.0003  2.3700 0.0003  2.3850 0.0002  2.4000 0.0002  2.4150 0.0002  2.4300 0.0001  2.4450 0.0001  2.4600 0.0001  2.4750 0.0001  2.4900 0.0001  2.5050 0.0001  2.5200 0.0001  2.5350 0.0001  2.5500 0.0000  2.5650 0.0000  2.5800 0.0000  2.5950 0.0000  2.6100 0.0000  2.6250 0.0000  2.6400 0.0000  2.6550 0.0000  2.6700 0.0000  2.6850 0.0000  2.7000 0.0000  2.7150 0.0000  2.7300 0.0000  2.7450 0.0000  2.7600 0.0000  2.7750 0.0000  2.7900 0.0000  2.8050 0.0000  2.8200 0.0000  2.8350 0.0000  2.8500 0.0000  2.8650 0.0000  2.8800 0.0000  2.8950 0.0000  2.9100 0.0000  2.9250 0.0000  2.9400 0.0000  2.9550 0.0000  2.9700 0.0000  2.9850 0.0000  3.0000 0.0000   /

 \color{blue}

\setplotsymbol ({\LARGE $\cdot$})

\plot -3.0000 0.0000  -2.9850 0.0000  -2.9700 0.0000  -2.9550 0.0000  -2.9400 0.0000  -2.9250 0.0000  -2.9100 0.0000  -2.8950 0.0000  -2.8800 0.0000  -2.8650 0.0000  -2.8500 0.0000  -2.8350 0.0000  -2.8200 0.0000  -2.8050 0.0000  -2.7900 0.0000  -2.7750 0.0000  -2.7600 0.0000  -2.7450 0.0000  -2.7300 0.0000  -2.7150 0.0000  -2.7000 0.0000  -2.6850 0.0000  -2.6700 0.0000  -2.6550 0.0000  -2.6400 0.0000  -2.6250 0.0000  -2.6100 0.0000  -2.5950 0.0000  -2.5800 0.0000  -2.5650 0.0000  -2.5500 0.0000  -2.5350 0.0000  -2.5200 0.0000  -2.5050 0.0000  -2.4900 0.0000  -2.4750 0.0000  -2.4600 0.0000  -2.4450 0.0000  -2.4300 0.0000  -2.4150 0.0000  -2.4000 0.0000  -2.3850 0.0000  -2.3700 0.0000  -2.3550 0.0000  -2.3400 0.0000  -2.3250 0.0000  -2.3100 0.0000  -2.2950 0.0000  -2.2800 0.0000  -2.2650 0.0000  -2.2500 0.0000  -2.2350 0.0000  -2.2200 0.0000  -2.2050 0.0000  -2.1900 0.0000  -2.1750 0.0000  -2.1600 0.0000  -2.1450 0.0000  -2.1300 0.0000  -2.1150 0.0000  -2.1000 0.0000  -2.0850 0.0000  -2.0700 0.0000  -2.0550 0.0000  -2.0400 0.0000  -2.0250 0.0000  -2.0100 0.0000  -1.9950 0.0000  -1.9800 0.0000  -1.9650 0.0000  -1.9500 0.0000  -1.9350 0.0000  -1.9200 0.0000  -1.9050 0.0000  -1.8900 0.0000  -1.8750 0.0000  -1.8600 0.0000  -1.8450 0.0000  -1.8300 0.0000  -1.8150 0.0000  -1.8000 0.0000  -1.7850 0.0000  -1.7700 0.0000  -1.7550 0.0000  -1.7400 0.0000  -1.7250 0.0000  -1.7100 0.0000  -1.6950 0.0000  -1.6800 0.0000  -1.6650 0.0000  -1.6500 0.0000  -1.6350 0.0000  -1.6200 0.0000  -1.6050 0.0000  -1.5900 0.0000  -1.5750 0.0000  -1.5600 0.0000  -1.5450 0.0000  -1.5300 0.0000  -1.5150 0.0000  -1.5000 0.0000  -1.4850 0.0000  -1.4700 0.0000  -1.4550 0.0000  -1.4400 0.0000  -1.4250 0.0000  -1.4100 0.0000  -1.3950 0.0000  -1.3800 0.0000  -1.3650 0.0000  -1.3500 0.0000  -1.3350 0.0000  -1.3200 0.0000  -1.3050 0.0000  -1.2900 0.0000  -1.2750 0.0000  -1.2600 0.0000  -1.2450 0.0000  -1.2300 0.0000  -1.2150 0.0000  -1.2000 0.0000  -1.1850 0.0000  -1.1700 0.0000  -1.1550 0.0000  -1.1400 0.0000  -1.1250 0.0000  -1.1100 0.0000  -1.0950 0.0000  -1.0800 0.0000  -1.0650 0.0000  -1.0500 0.0000  -1.0350 0.0000  -1.0200 0.0000  -1.0050 0.0001  -0.9900 0.0001  -0.9750 0.0001  -0.9600 0.0001  -0.9450 0.0001  -0.9300 0.0001  -0.9150 0.0001  -0.9000 0.0002  -0.8850 0.0002  -0.8700 0.0002  -0.8550 0.0003  -0.8400 0.0003  -0.8250 0.0003  -0.8100 0.0004  -0.7950 0.0005  -0.7800 0.0005  -0.7650 0.0006  -0.7500 0.0007  -0.7350 0.0008  -0.7200 0.0010  -0.7050 0.0011  -0.6900 0.0013  -0.6750 0.0015  -0.6600 0.0018  -0.6450 0.0020  -0.6300 0.0024  -0.6150 0.0027  -0.6000 0.0032  -0.5850 0.0036  -0.5700 0.0042  -0.5550 0.0049  -0.5400 0.0056  -0.5250 0.0065  -0.5100 0.0075  -0.4950 0.0086  -0.4800 0.0099  -0.4650 0.0113  -0.4500 0.0129  -0.4350 0.0148  -0.4200 0.0168  -0.4050 0.0191  -0.3900 0.0215  -0.3750 0.0243  -0.3600 0.0272  -0.3450 0.0304  -0.3300 0.0338  -0.3150 0.0373  -0.3000 0.0409  -0.2850 0.0446  -0.2700 0.0482  -0.2550 0.0518  -0.2400 0.0552  -0.2250 0.0583  -0.2100 0.0611  -0.1950 0.0636  -0.1800 0.0657  -0.1650 0.0674  -0.1500 0.0687  -0.1350 0.0698  -0.1200 0.0707  -0.1050 0.0715  -0.0900 0.0723  -0.0750 0.0734  -0.0600 0.0748  -0.0450 0.0767  -0.0300 0.0793  -0.0150 0.0827  0.0000 0.0869  0.0150 0.0921  0.0300 0.0982  0.0450 0.1054  0.0600 0.1135  0.0750 0.1224  0.0900 0.1319  0.1050 0.1418  0.1200 0.1518  0.1350 0.1614  0.1500 0.1702  0.1650 0.1778  0.1800 0.1839  0.1950 0.1879  0.2100 0.1897  0.2250 0.1891  0.2400 0.1861  0.2550 0.1809  0.2700 0.1738  0.2850 0.1650  0.3000 0.1550  0.3150 0.1441  0.3300 0.1328  0.3450 0.1213  0.3600 0.1100  0.3750 0.0991  0.3900 0.0888  0.4050 0.0791  0.4200 0.0701  0.4350 0.0619  0.4500 0.0545  0.4650 0.0478  0.4800 0.0419  0.4950 0.0366  0.5100 0.0319  0.5250 0.0278  0.5400 0.0241  0.5550 0.0210  0.5700 0.0182  0.5850 0.0158  0.6000 0.0137  0.6150 0.0119  0.6300 0.0104  0.6450 0.0090  0.6600 0.0079  0.6750 0.0069  0.6900 0.0061  0.7050 0.0054  0.7200 0.0048  0.7350 0.0043  0.7500 0.0039  0.7650 0.0036  0.7800 0.0034  0.7950 0.0033  0.8100 0.0032  0.8250 0.0033  0.8400 0.0033  0.8550 0.0035  0.8700 0.0037  0.8850 0.0040  0.9000 0.0045  0.9150 0.0050  0.9300 0.0056  0.9450 0.0063  0.9600 0.0072  0.9750 0.0082  0.9900 0.0094  1.0050 0.0108  1.0200 0.0124  1.0350 0.0142  1.0500 0.0164  1.0650 0.0188  1.0800 0.0216  1.0950 0.0248  1.1100 0.0284  1.1250 0.0325  1.1400 0.0372  1.1550 0.0424  1.1700 0.0481  1.1850 0.0545  1.2000 0.0616  1.2150 0.0692  1.2300 0.0775  1.2450 0.0863  1.2600 0.0956  1.2750 0.1052  1.2900 0.1149  1.3050 0.1247  1.3200 0.1342  1.3350 0.1432  1.3500 0.1514  1.3650 0.1585  1.3800 0.1642  1.3950 0.1684  1.4100 0.1709  1.4250 0.1716  1.4400 0.1703  1.4550 0.1673  1.4700 0.1625  1.4850 0.1563  1.5000 0.1488  1.5150 0.1404  1.5300 0.1312  1.5450 0.1216  1.5600 0.1118  1.5750 0.1020  1.5900 0.0925  1.6050 0.0834  1.6200 0.0747  1.6350 0.0667  1.6500 0.0592  1.6650 0.0524  1.6800 0.0462  1.6950 0.0406  1.7100 0.0356  1.7250 0.0311  1.7400 0.0271  1.7550 0.0236  1.7700 0.0205  1.7850 0.0178  1.8000 0.0155  1.8150 0.0134  1.8300 0.0116  1.8450 0.0100  1.8600 0.0087  1.8750 0.0075  1.8900 0.0065  1.9050 0.0056  1.9200 0.0048  1.9350 0.0042  1.9500 0.0036  1.9650 0.0031  1.9800 0.0027  1.9950 0.0023  2.0100 0.0020  2.0250 0.0017  2.0400 0.0015  2.0550 0.0013  2.0700 0.0011  2.0850 0.0009  2.1000 0.0008  2.1150 0.0007  2.1300 0.0006  2.1450 0.0005  2.1600 0.0004  2.1750 0.0004  2.1900 0.0003  2.2050 0.0003  2.2200 0.0002  2.2350 0.0002  2.2500 0.0002  2.2650 0.0002  2.2800 0.0001  2.2950 0.0001  2.3100 0.0001  2.3250 0.0001  2.3400 0.0001  2.3550 0.0001  2.3700 0.0001  2.3850 0.0000  2.4000 0.0000  2.4150 0.0000  2.4300 0.0000  2.4450 0.0000  2.4600 0.0000  2.4750 0.0000  2.4900 0.0000  2.5050 0.0000  2.5200 0.0000  2.5350 0.0000  2.5500 0.0000  2.5650 0.0000  2.5800 0.0000  2.5950 0.0000  2.6100 0.0000  2.6250 0.0000  2.6400 0.0000  2.6550 0.0000  2.6700 0.0000  2.6850 0.0000  2.7000 0.0000  2.7150 0.0000  2.7300 0.0000  2.7450 0.0000  2.7600 0.0000  2.7750 0.0000  2.7900 0.0000  2.8050 0.0000  2.8200 0.0000  2.8350 0.0000  2.8500 0.0000  2.8650 0.0000  2.8800 0.0000  2.8950 0.0000  2.9100 0.0000  2.9250 0.0000  2.9400 0.0000  2.9550 0.0000  2.9700 0.0000  2.9850 0.0000  3.0000 0.0000   /

 \color{green}

\setplotsymbol ({\LARGE $\cdot$})

\plot -3.0000 0.0000  -2.9850 0.0000  -2.9700 0.0000  -2.9550 0.0000  -2.9400 0.0000  -2.9250 0.0000  -2.9100 0.0000  -2.8950 0.0000  -2.8800 0.0000  -2.8650 0.0000  -2.8500 0.0000  -2.8350 0.0000  -2.8200 0.0000  -2.8050 0.0000  -2.7900 0.0000  -2.7750 0.0000  -2.7600 0.0000  -2.7450 0.0000  -2.7300 0.0000  -2.7150 0.0000  -2.7000 0.0000  -2.6850 0.0000  -2.6700 0.0000  -2.6550 0.0000  -2.6400 0.0000  -2.6250 0.0000  -2.6100 0.0000  -2.5950 0.0000  -2.5800 0.0000  -2.5650 0.0000  -2.5500 0.0000  -2.5350 0.0000  -2.5200 0.0000  -2.5050 0.0000  -2.4900 0.0000  -2.4750 0.0000  -2.4600 0.0000  -2.4450 0.0000  -2.4300 0.0000  -2.4150 0.0000  -2.4000 0.0000  -2.3850 0.0000  -2.3700 0.0000  -2.3550 0.0000  -2.3400 0.0000  -2.3250 0.0000  -2.3100 0.0000  -2.2950 0.0000  -2.2800 0.0000  -2.2650 0.0000  -2.2500 0.0000  -2.2350 0.0000  -2.2200 0.0000  -2.2050 0.0000  -2.1900 0.0000  -2.1750 0.0000  -2.1600 0.0000  -2.1450 0.0000  -2.1300 0.0000  -2.1150 0.0000  -2.1000 0.0000  -2.0850 0.0000  -2.0700 0.0000  -2.0550 0.0000  -2.0400 0.0000  -2.0250 0.0000  -2.0100 0.0000  -1.9950 0.0000  -1.9800 0.0000  -1.9650 0.0000  -1.9500 0.0000  -1.9350 0.0000  -1.9200 0.0000  -1.9050 0.0000  -1.8900 0.0000  -1.8750 0.0000  -1.8600 0.0000  -1.8450 0.0000  -1.8300 0.0000  -1.8150 0.0000  -1.8000 0.0000  -1.7850 0.0000  -1.7700 0.0000  -1.7550 0.0000  -1.7400 0.0000  -1.7250 0.0000  -1.7100 0.0000  -1.6950 0.0000  -1.6800 0.0000  -1.6650 0.0000  -1.6500 0.0000  -1.6350 0.0000  -1.6200 0.0000  -1.6050 0.0000  -1.5900 0.0000  -1.5750 0.0000  -1.5600 0.0000  -1.5450 0.0000  -1.5300 0.0000  -1.5150 0.0000  -1.5000 0.0000  -1.4850 0.0000  -1.4700 0.0000  -1.4550 0.0000  -1.4400 0.0000  -1.4250 0.0000  -1.4100 0.0000  -1.3950 0.0000  -1.3800 0.0000  -1.3650 0.0000  -1.3500 0.0000  -1.3350 0.0000  -1.3200 0.0000  -1.3050 0.0000  -1.2900 0.0000  -1.2750 0.0000  -1.2600 0.0000  -1.2450 0.0000  -1.2300 0.0000  -1.2150 0.0000  -1.2000 0.0000  -1.1850 0.0000  -1.1700 0.0000  -1.1550 0.0000  -1.1400 0.0000  -1.1250 0.0001  -1.1100 0.0001  -1.0950 0.0001  -1.0800 0.0001  -1.0650 0.0001  -1.0500 0.0001  -1.0350 0.0001  -1.0200 0.0001  -1.0050 0.0002  -0.9900 0.0002  -0.9750 0.0002  -0.9600 0.0003  -0.9450 0.0003  -0.9300 0.0004  -0.9150 0.0004  -0.9000 0.0005  -0.8850 0.0006  -0.8700 0.0006  -0.8550 0.0008  -0.8400 0.0009  -0.8250 0.0010  -0.8100 0.0012  -0.7950 0.0014  -0.7800 0.0016  -0.7650 0.0018  -0.7500 0.0021  -0.7350 0.0025  -0.7200 0.0028  -0.7050 0.0033  -0.6900 0.0038  -0.6750 0.0044  -0.6600 0.0051  -0.6450 0.0059  -0.6300 0.0067  -0.6150 0.0078  -0.6000 0.0089  -0.5850 0.0102  -0.5700 0.0117  -0.5550 0.0134  -0.5400 0.0152  -0.5250 0.0173  -0.5100 0.0195  -0.4950 0.0220  -0.4800 0.0247  -0.4650 0.0276  -0.4500 0.0307  -0.4350 0.0339  -0.4200 0.0372  -0.4050 0.0405  -0.3900 0.0438  -0.3750 0.0470  -0.3600 0.0499  -0.3450 0.0526  -0.3300 0.0548  -0.3150 0.0566  -0.3000 0.0579  -0.2850 0.0587  -0.2700 0.0589  -0.2550 0.0586  -0.2400 0.0578  -0.2250 0.0567  -0.2100 0.0553  -0.1950 0.0537  -0.1800 0.0521  -0.1650 0.0506  -0.1500 0.0494  -0.1350 0.0485  -0.1200 0.0482  -0.1050 0.0485  -0.0900 0.0496  -0.0750 0.0517  -0.0600 0.0548  -0.0450 0.0592  -0.0300 0.0651  -0.0150 0.0725  0.0000 0.0816  0.0150 0.0927  0.0300 0.1056  0.0450 0.1204  0.0600 0.1368  0.0750 0.1543  0.0900 0.1724  0.1050 0.1900  0.1200 0.2062  0.1350 0.2196  0.1500 0.2293  0.1650 0.2342  0.1800 0.2341  0.1950 0.2289  0.2100 0.2191  0.2250 0.2056  0.2400 0.1895  0.2550 0.1719  0.2700 0.1537  0.2850 0.1359  0.3000 0.1190  0.3150 0.1034  0.3300 0.0894  0.3450 0.0769  0.3600 0.0659  0.3750 0.0564  0.3900 0.0482  0.4050 0.0411  0.4200 0.0351  0.4350 0.0300  0.4500 0.0256  0.4650 0.0219  0.4800 0.0187  0.4950 0.0160  0.5100 0.0137  0.5250 0.0117  0.5400 0.0100  0.5550 0.0086  0.5700 0.0074  0.5850 0.0063  0.6000 0.0054  0.6150 0.0047  0.6300 0.0040  0.6450 0.0034  0.6600 0.0029  0.6750 0.0025  0.6900 0.0022  0.7050 0.0019  0.7200 0.0016  0.7350 0.0014  0.7500 0.0012  0.7650 0.0010  0.7800 0.0009  0.7950 0.0008  0.8100 0.0007  0.8250 0.0006  0.8400 0.0005  0.8550 0.0004  0.8700 0.0004  0.8850 0.0003  0.9000 0.0003  0.9150 0.0002  0.9300 0.0002  0.9450 0.0002  0.9600 0.0001  0.9750 0.0001  0.9900 0.0001  1.0050 0.0001  1.0200 0.0001  1.0350 0.0001  1.0500 0.0001  1.0650 0.0001  1.0800 0.0000  1.0950 0.0000  1.1100 0.0000  1.1250 0.0000  1.1400 0.0000  1.1550 0.0000  1.1700 0.0000  1.1850 0.0000  1.2000 0.0000  1.2150 0.0000  1.2300 0.0000  1.2450 0.0000  1.2600 0.0000  1.2750 0.0000  1.2900 0.0000  1.3050 0.0000  1.3200 0.0000  1.3350 0.0000  1.3500 0.0000  1.3650 0.0001  1.3800 0.0001  1.3950 0.0001  1.4100 0.0001  1.4250 0.0001  1.4400 0.0001  1.4550 0.0001  1.4700 0.0002  1.4850 0.0002  1.5000 0.0002  1.5150 0.0002  1.5300 0.0003  1.5450 0.0003  1.5600 0.0004  1.5750 0.0004  1.5900 0.0005  1.6050 0.0006  1.6200 0.0007  1.6350 0.0008  1.6500 0.0009  1.6650 0.0011  1.6800 0.0012  1.6950 0.0014  1.7100 0.0017  1.7250 0.0019  1.7400 0.0022  1.7550 0.0026  1.7700 0.0030  1.7850 0.0035  1.8000 0.0041  1.8150 0.0047  1.8300 0.0055  1.8450 0.0063  1.8600 0.0073  1.8750 0.0085  1.8900 0.0098  1.9050 0.0114  1.9200 0.0131  1.9350 0.0152  1.9500 0.0175  1.9650 0.0201  1.9800 0.0232  1.9950 0.0266  2.0100 0.0305  2.0250 0.0349  2.0400 0.0398  2.0550 0.0454  2.0700 0.0515  2.0850 0.0582  2.1000 0.0656  2.1150 0.0736  2.1300 0.0822  2.1450 0.0912  2.1600 0.1007  2.1750 0.1104  2.1900 0.1202  2.2050 0.1299  2.2200 0.1391  2.2350 0.1477  2.2500 0.1554  2.2650 0.1618  2.2800 0.1667  2.2950 0.1700  2.3100 0.1715  2.3250 0.1711  2.3400 0.1689  2.3550 0.1649  2.3700 0.1593  2.3850 0.1524  2.4000 0.1444  2.4150 0.1355  2.4300 0.1260  2.4450 0.1163  2.4600 0.1065  2.4750 0.0968  2.4900 0.0875  2.5050 0.0786  2.5200 0.0703  2.5350 0.0626  2.5500 0.0554  2.5650 0.0489  2.5800 0.0431  2.5950 0.0378  2.6100 0.0331  2.6250 0.0289  2.6400 0.0252  2.6550 0.0219  2.6700 0.0190  2.6850 0.0165  2.7000 0.0143  2.7150 0.0124  2.7300 0.0107  2.7450 0.0093  2.7600 0.0080  2.7750 0.0069  2.7900 0.0060  2.8050 0.0052  2.8200 0.0044  2.8350 0.0038  2.8500 0.0033  2.8650 0.0028  2.8800 0.0025  2.8950 0.0021  2.9100 0.0018  2.9250 0.0016  2.9400 0.0014  2.9550 0.0012  2.9700 0.0010  2.9850 0.0009  3.0000 0.0007   /

 \color{black}
\put {O} at 0 0 
\setplotarea x from -3.000000 to 3.000000, y from 0 to 0.400000
\axis top /
\axis right /
\axis left /
\axis bottom 
 label {{\Large \hspace{2cm} $f$}} /
\axis bottom 
 ticks out withvalues -30 0 30 / quantity 3 /
\axis left shiftedto x=0 
 label {{\Large $\kappa \,$}} /
\axis left shiftedto x=0 
 ticks out withvalues {} 0.1 {} 0.3 / quantity 5 /

\normalcolor
\endpicture

%% file: Compressgraphs/ScatterW.tex
\beginpicture

 \setcoordinatesystem units <3.000 pt, 0.5 pt>

\color[rgb]{0.147438,0.069532,0.791067}  \put {$0_1$} at  1.098612 1.098612 
\color[rgb]{0.259873,0.164787,0.593611}  \put {$3_1$} at  4.841388 7.833857 
\color[rgb]{0.329163,0.290140,0.412418}  \put {$4_1$} at  10.379345 36.932128 
\color[rgb]{0.303336,0.235363,0.488992}  \put {$5_1$} at  7.836257 19.835862 
\color[rgb]{0.336295,0.293266,0.401883}  \put {$5_2$} at  11.220053 38.220039 
\color[rgb]{0.367999,0.339272,0.324074}  \put {$6_1$} at  15.963186 63.111409 
\color[rgb]{0.378547,0.343357,0.308775}  \put {$6_2$} at  18.022176 66.006972 
\color[rgb]{0.408879,0.381064,0.238564}  \put {$6_3$} at  26.084091 101.045969 
\color[rgb]{0.356717,0.299817,0.372936}  \put {$7_1$} at  14.057661 41.057660 
\color[rgb]{0.380563,0.343405,0.306492}  \put {$7_2$} at  18.450987 66.041590 
\color[rgb]{0.405128,0.378595,0.245173}  \put {$7_3$} at  24.862420 98.175478 
\color[rgb]{0.391214,0.375768,0.262634}  \put {$7_4$} at  20.936786 95.008150 
\color[rgb]{0.405633,0.380050,0.243164}  \put {$7_5$} at  25.022610 99.854747 
\color[rgb]{0.408942,0.381084,0.238473}  \put {$7_6$} at  26.105399 101.069947 
\color[rgb]{0.441151,0.412178,0.170452}  \put {$7_7$} at  41.202547 148.232622 
\color[rgb]{0.401865,0.379073,0.248220}  \put {$8_1$} at  23.861209 98.722783 
\color[rgb]{0.414260,0.382590,0.230937}  \put {$8_2$} at  27.982558 102.871186 
\color[rgb]{0.421303,0.406285,0.199123}  \put {$8_3$} at  30.769117 137.369775 
\color[rgb]{0.406518,0.380326,0.241917}  \put {$8_4$} at  25.306160 100.176825 
\color[rgb]{0.415706,0.382054,0.229741}  \put {$8_5$} at  28.525204 102.225453 
\color[rgb]{0.431273,0.409654,0.184736}  \put {$8_6$} at  35.442990 143.442483 
\color[rgb]{0.443129,0.429960,0.150192}  \put {$8_7$} at  42.529632 189.303377 
\color[rgb]{0.432691,0.408994,0.183769}  \put {$8_8$} at  36.191336 142.224218 
\color[rgb]{0.433896,0.410401,0.180964}  \put {$8_9$} at  36.846627 144.837819 
\color[rgb]{0.436966,0.410755,0.176972}  \put {$8_{10}$} at  38.598808 145.505301 
\color[rgb]{0.434802,0.409123,0.181154}  \put {$8_{11}$} at  37.350823 142.461615 
\color[rgb]{0.430878,0.409018,0.185830}  \put {$8_{12}$} at  35.238567 142.268682 
\color[rgb]{0.444373,0.413785,0.164842}  \put {$8_{13}$} at  43.399692 151.397562 
\color[rgb]{0.447782,0.430635,0.144241}  \put {$8_{14}$} at  45.934845 191.173433 
\color[rgb]{0.425947,0.386570,0.213014}  \put {$8_{15}$} at  32.827044 107.826643 
\color[rgb]{0.460330,0.447641,0.111650}  \put {$8_{16}$} at  57.790914 249.493609 
\color[rgb]{0.465286,0.449123,0.104056}  \put {$8_{17}$} at  64.038477 255.882699 
\color[rgb]{0.477281,0.462492,0.074740}  \put {$8_{18}$} at  86.010107 328.709532 
\color[rgb]{0.348582,0.296885,0.384948}  \put {$8_{19}$} at  12.842841 39.763670 
\color[rgb]{0.376618,0.340847,0.313292}  \put {$8_{20}$} at  17.623039 64.211333 
\color[rgb]{0.399164,0.350045,0.278569}  \put {$8_{21}$} at  23.072029 71.063091 
\color[rgb]{0.377677,0.343128,0.309951}  \put {$3_1^+\#3_1^+$} at  17.840746 65.840746 
\color[rgb]{0.331380,0.246599,0.446079}  \put {$3_1^+\#3_1^-$} at  10.633343 22.633318 
\color[rgb]{0.381962,0.344311,0.304071}  \put {$3_1^+\#4_1$} at  18.755937 66.704062 
\color[rgb]{0.395824,0.348867,0.283707}  \put {$3_1^+\#5_1^+$} at  22.142515 70.142515 
\color[rgb]{0.367623,0.303831,0.356340}  \put {$3_1^+\#5_1^-$} at  15.895092 42.895092 
\color[rgb]{0.383937,0.344980,0.301214}  \put {$3_1^+\#5_2^+$} at  19.196857 67.196857 
\color[rgb]{0.400148,0.378746,0.250371}  \put {$3_1^+\#5_2^-$} at  23.355624 98.348239 
\color[rgb]{0.395607,0.348733,0.284089}  \put {$4_1\#4_1$} at  22.083767 70.039143


 \color{black}
\put {O} at 0 0 
\setplotarea x from 0 to 100, y from 0 to 400
\axis top /
\axis right /
\axis bottom 
 label {{\large \hspace{1cm} $\C{W}$ ($k=1/4$)}} /
\axis bottom 
 ticks withvalues 0 25 {} 75 100 / quantity 5 /
\axis left 
 label {{\large $\C{W}$ ($k=1$)}} /
\axis left 
 ticks out withvalues 0 100 {} 300 400 / quantity 5 /
\setdots <2pt>
\plot -0.04 0 0.4 0 /
\plot 0 -1 0 13 /
\normalcolor
\endpicture

%% file: table-slab-forces.bbl
\begin{thebibliography}{10}

\bibitem{AdCCF83}
C.~Arag{\~a}o~de Carvalho, Caracciolo S., and Fr{\"o}hlich J.
\newblock Polymers and g $|\phi|^4$ theory in four dimensions.
\newblock {\em Nuclear Physics B}, 215(2):209--248, 1983.

\bibitem{BOS07}
M.~Baiesi, E.~Orlandini, and A.L. Stella.
\newblock Ranking knots of random, globular polymer rings.
\newblock {\em Physical Review Letters}, 99(5):58301, 2007.

\bibitem{BF81}
B.~Berg and D.~Foerster.
\newblock Random paths and random surfaces on a digital computer.
\newblock {\em Physics Letters B}, 106(4):323--326, 1981.

\bibitem{D93}
Y.~Diao.
\newblock Minimal knotted polygons on the cubic lattice.
\newblock {\em Journal of Knot Theory and its Ramifications}, 2(4):413--425,
  1993.

\bibitem{DBS89}
C.O. Dietrich-Buchecker and J.P. Sauvage.
\newblock A synthetic molecular trefoil knot.
\newblock {\em Angewandte Chemie International Edition in English},
  28(2):189--192, 1989.

\bibitem{DSB01}
A.~Dobay, P.E. Sottas, J.~Dubochet, and A.~Stasiak.
\newblock Predicting optimal lengths of random knots.
\newblock {\em Letters in Mathematical Physics}, 55(3):239--247, 2001.

\bibitem{GWF06}
A.~Gholami, J.~Wilhelm, and E.~Frey.
\newblock Entropic forces generated by grafted semiflexible polymers.
\newblock {\em Physical Review E}, 74:041803, Oct 2006.

\bibitem{ISDAVS12}
K.~Ishihara, R.~Scharein, Y.~Diao, J.~Arsuaga, M.~Vazquez, and K.~Shimokawa.
\newblock Bounds for the minimum step number of knots confined to slabs in the
  simple cubic lattice.
\newblock {\em Journal of Physics A: Mathematical and Theoretical}, 45:065003,
  2012.

\bibitem{JvR07}
E.J. Janse~van Rensburg.
\newblock Squeezing knots.
\newblock {\em Journal of Statistical Mechanics: Theory and Experiment},
  2007(03):P03001, 2007.

\bibitem{JvROTW08}
E.J. Janse~van Rensburg, E.~Orlandini, M.C. Tesi, and S.G. Whittington.
\newblock Knotting in stretched polygons.
\newblock {\em Journal of Physics A: Mathematical and Theoretical}, 41:015003,
  2008.

\bibitem{JvRR09}
E.J. Janse~van Rensburg and A.~Rechnitzer.
\newblock Generalized atmospheric sampling of self-avoiding walks.
\newblock {\em Journal of Physics A: Mathematical and Theoretical}, 42:335001,
  2009.

\bibitem{JvRR10}
E.J. Janse~van Rensburg and A.~Rechnitzer.
\newblock Generalised atmospheric sampling of knotted polygons.
\newblock {\em Journal of Knot Theory and its Ramifications}, 2010.

\bibitem{JvRR12}
E.J. Janse~van Rensburg and A.~Rechnitzer.
\newblock The compressibility of minimal lattice knots.
\newblock {\em Journal of Statistical Mechanics: Theory and Experiment},
  2012(05):P05003, 2012.

\bibitem{JvRW90}
E.J. Janse~van Rensburg and S.G. Whittington.
\newblock The knot probability in lattice polygons.
\newblock {\em Journal of Physics A: Mathematical and General}, 23:3573, 1990.

\bibitem{JvRW91}
E.J. Janse~van Rensburg and S.G. Whittington.
\newblock The bfacf algorithm and knotted polygons.
\newblock {\em Journal of Physics A: Mathematical and General}, 24(23):5553,
  1991.

\bibitem{KSH07}
N.~Kresge, R.D. Simoni, and R.L. Hill.
\newblock Unwinding the dna topoisomerase story: the work of james c. wang.
\newblock {\em Journal of Biological Chemistry}, 282(22):e17, 2007.

\bibitem{LMZC06}
Z.~Liu, J.K. Mann, E.L. Zechiedrich, and H.S. Chan.
\newblock Topological information embodied in local juxtaposition geometry
  provides a statistical mechanical basis for unknotting by type-2 dna
  topoisomerases.
\newblock {\em Journal of Molecular Biology}, 361(2):268--285, 2006.

\bibitem{MLY11}
R.~Matthews, A.A. Louis, and J.M. Yeomans.
\newblock Confinement of knotted polymers in a slit.
\newblock {\em Molecular Physics}, 109(7-10):1289--1295, 2011.

\bibitem{MHDKK02}
R.~Metzler, A.~Hanke, P.G. Dommersnes, Y.~Kantor, and M.~Kardar.
\newblock Equilibrium shapes of flat knots.
\newblock {\em Physical Review Letters}, 88(18):188101, 2002.

\bibitem{MW86}
J.P.J. Michels and F.W. Wiegel.
\newblock On the topology of a polymer ring.
\newblock {\em Proceedings of the Royal Society of London. A. Mathematical and
  Physical Sciences}, 403(1825):269--284, 1986.

\bibitem{OSV04}
E.~Orlandini, A.L. Stella, and C.~Vanderzande.
\newblock Loose, flat knots in collapsed polymers.
\newblock {\em Journal of Statistical Physics}, 115(1):681--700, 2004.

\bibitem{SIADSV09}
R.~Scharein, K.~Ishihara, J.~Arsuaga, Y.~Diao, K.~Shimokawa, and M.~Vazquez.
\newblock Bounds for the minimum step number of knots in the simple cubic
  lattice.
\newblock {\em Journal of Physics A: Mathematical and Theoretical}, 42:475006,
  2009.

\bibitem{SW93}
S.Y. Shaw and J.C. Wang.
\newblock Knotting of a dna chain during ring closure.
\newblock {\em Science}, 260(5107):533--536, 1993.

\bibitem{SD02}
M.K. Shimamura and T.~Deguchi.
\newblock Knot complexity and the probability of random knotting.
\newblock {\em Physical Review E}, 66:040801, Oct 2002.

\bibitem{SBA12}
S.~Swetman, C.~Brett, and M.P. Allen.
\newblock Phase diagrams of knotted and unknotted ring polymers.
\newblock {\em Physical Review E}, 2012.

\bibitem{T00}
W.R. Taylor.
\newblock A deeply knotted protein structure and how it might fold.
\newblock {\em Nature}, 406(6798):916--919, 2000.

\bibitem{TJvROW94}
M.C. Tesi, E.J. Janse~van Rensburg, E.~Orlandini, and S.G. Whittington.
\newblock Knot probability for lattice polygons in confined geometries.
\newblock {\em Journal of Physics A: Mathematical and General}, 27:347, 1994.

\bibitem{VKK05}
P.~Virnau, Y.~Kantor, and M.~Kardar.
\newblock Knots in globule and coil phases of a model polyethylene.
\newblock {\em Journal of the American Chemical Society}, 127(43):15102--15106,
  2005.

\bibitem{ZKC97}
E.L. Zechiedrich, A.B. Khodursky, and N.R. Cozzarelli.
\newblock Topoisomerase iv, not gyrase, decatenates products of site-specific
  recombination inescherichia coli.
\newblock {\em Genes \& development}, 11(19):2580--2592, 1997.

\end{thebibliography}
